\documentclass[a4paper,11pt]{article}
\usepackage{a4wide}
\usepackage{xcolor}
\usepackage{graphicx}  
\usepackage{amsmath}   
\usepackage[compress]{cite}
\usepackage{amssymb}   
\usepackage{bm} 
\usepackage{dcolumn}
\usepackage{float}
\usepackage{color}
\usepackage{mathrsfs}
\usepackage{amsfonts}
\usepackage{varioref}
\numberwithin{equation}{section}
\usepackage{tabularx}
\usepackage[caption=false]{subfig}
\usepackage[export]{adjustbox}
\usepackage{multirow}
\RequirePackage[colorlinks,citecolor=blue,urlcolor=magenta,linkcolor=blue]{hyperref}
\input epsf

\usepackage[bottom=0.9in,top=0.5in,right=0.5in,left=0.5in]{geometry}

\labelformat{subfigure}{Fig.~\thefigure#1}
\hypersetup{
    colorlinks,
    linkcolor={green!50!black},
    citecolor={magenta},
    urlcolor={blue!80!black}
}

	\title{\bf Probing Hairy Kerr Black Holes through Quasi-Periodic Oscillations I: A study based on the kinematic models }
	\author{ \textbf{Anirban Dasgupta$^{a}$\thanks{522ph1007@nitrkl.ac.in}, \textbf{Supragyan Priyadarshinee$^{a, b}$}\thanks{supragyanpriyadarshinee52@gmail.com},Indrani Banerjee$^{a}$}\thanks{banerjeein@nitrkl.ac.in}, \textbf{Subhash Mahapatra$^{a}$}\thanks{mahapatrasub@nitrkl.ac.in}
			\\\\\textit{{\small $^a$ Department of Physics and Astronomy, National Institute of Technology Rourkela, }}\\
			\textit{{\small Rourkela - 769008, India}}\\
			\textit{{\small $^b$ Department of Physics, IISER Mohali,
Sector 81, }}\\
			\textit{{\small Knowledge City, Mohali, Punjab, 140306, India}}
			}
	
	\date{}

\begin{document}
\maketitle

	\begin{abstract}
	
Black holes endowed with nontrivial scalar or matter fields, known as hairy black holes, possess additional parameters beyond mass, charge, and spin, leading to richer phenomenology. Constraining their parameter space is therefore essential, particularly in view of current electromagnetic and gravitational-wave observations. In this work, we study rotating hairy black hole solutions inspired by the gravitational decoupling method, which satisfy the Einstein field equations with a conserved energy–momentum tensor obeying the strong energy conditions. We explore in detail the horizon structure of such black holes and report for the first time certain unique features, not observed in Kerr BHs. 
We examine how the hair parameters influence the fundamental frequencies governing particle motion in the hairy Kerr spacetime and compare these effects with the Kerr case. Since theoretical models of high-frequency quasi-periodic oscillations (HFQPOs) are directly linked to these fundamental frequencies, this provides a powerful observational probe of hairy black holes. By confronting several kinematic HFQPO models with observations from six black hole sources, we report that for most sources, the HFQPO data with the current level of precision cannot distinguish between the Kerr and the hairy Kerr scenario. However, based on the agreement with previous spin estimates 
our analysis offers a systematic framework to assess the relative suitability of different HFQPO models for each source. Notably, even with current observational precision, we report that both the Relativistic Precession Model and the Tidal Disruption Model seem to be unsuitable for the sources GRO J1655-40 and GRS 1915+105. 
The broader implications of these findings are discussed.

	\end{abstract}

\section{Introduction}
Black holes, among the most mysterious objects in the Universe and remarkable predictions of general relativity, have attracted significant attention in recent decades. They naturally integrate key aspects of gravitational theory, thermodynamics, and quantum physics. Recent observational breakthroughs have further accelerated research in this area. Electromagnetic observations from satellites such as RXTE, Chandra, and NuSTAR, along with direct detections of gravitational waves by the LIGO–VIRGO collaboration \cite{LIGOScientific:2016aoc,LIGOScientific:2018mvr,LIGOScientific:2017vwq}, and the imaging of black hole shadows by the Event Horizon Telescope (EHT) \cite{EHT:2019ds,EventHorizonTelescope:2022wkp}, have enabled tests of gravity in the strong-field regime near the event horizon. These observations remain consistent with the predictions of general relativity.

Black hole solutions with nontrivial scalar or matter fields, commonly referred to as hairy black holes, have received considerable attention in recent years. Such solutions often arise in modified gravity theories, effective field models, or Einstein gravity coupled to additional fields with appropriate self-interactions. Unlike the no-hair results of classical general relativity \cite{ruffini1971introducing, Bekenstein:1971hc, Bekenstein:1995un, Sudarsky:1995zg}, these configurations can possess additional parameters beyond mass, charge, and angular momentum, thereby enriching the phenomenology of compact objects. Understanding the allowed parameter space of hairy black holes is therefore essential, particularly in light of growing observational constraints from gravitational-wave detections and electromagnetic observations of strong-gravity environments. For a recent discussion and comprehensive reviews on the existence of scalar hair in asymptotically flat spacetimes, see~\cite{Herdeiro:2015waa, Hertog:2006rr}. Note that the no-hair conjecture lacks the rigor of a strict mathematical theorem, and a variety of counterexamples have been discovered over the years \cite{bocharova1970exact, Bekenstein:1974sf, Bekenstein:1975ts, Bronnikov:1978mx, torii1999toward, zloshchastiev2005coexistence, berti2013numerical, garfinkle1991charged, brodbeck1996instability, volkov1995number, bizon1991n, bizon1991stability, zhou1991nonlinear, bizon1990colored,  Mahapatra:2020wym, Priyadarshinee:2021rch, Priyadarshinee:2023cmi}.

An interesting geometric framework for constructing hairy black hole solutions is gravitational decoupling, most notably implemented through the minimal geometric deformation (MGD) approach \cite{Ovalle:2017fgl, Ovalle:2018gic, Ovalle:2020kpd, Ovalle:2018umz, Ovalle:2017wqi}. In this method, a known seed solution of Einstein's equations is deformed by coupling it to an additional source sector via a controlled decoupling parameter. This technique has been successfully employed across a broad range of applications, including anisotropic stellar configurations \cite{Ovalle:2017fgl, Torres-Sanchez:2019wjv}, wormholes \cite{Panyasiripan:2024kyu}, regular black holes \cite{Casadio:2015gea, Misyura:2024fho, Hua:2025qwu}, and a variety of hairy black hole geometries \cite{Ovalle:2020kpd, Mahapatra:2022xea}. A key strength of gravitational decoupling is that it enables the generation of new solutions without requiring an explicit matter Lagrangian for the additional sector.

Within the context of black holes, the decoupling parameter dictates the strength of the extra ``hair'' sector and thereby characterizes deviations from standard vacuum solutions such as Schwarzschild or Kerr. However, not all deformations yield physically viable spacetimes. Ensuring regularity outside the horizon, the existence of a well-defined event horizon, compliance with reasonable energy conditions, and thermodynamic consistency is essential. With increasingly precise observational probes, including gravitational-wave detections and black hole shadow measurements, it has become crucial to determine the regions of parameter space compatible with astrophysical constraints.

The extended gravitational decoupling framework \cite{Sharif:2020arn, Ovalle:2018gic, ovalle2018black, Contreras:2021yxe, Ovalle:2020kpd} achieves a systematic modification of spherically symmetric backgrounds by adding new sources in such a way that Einstein's equations decouple into separate sectors, allowing the deformed metric to be solved independently. Physical viability additionally requires the deformed geometry to satisfy the strong energy condition, remain free of extra singularities, and possess a well-defined horizon. Under these criteria, new static, spherically symmetric, asymptotically flat metrics were constructed in \cite{Ovalle:2020kpd}, and their rotating counterparts lead to hairy Kerr black holes characterized by deviation parameters $\alpha$ and $l$. The combination $l_{0}=\alpha l$ plays a crucial role by modifying the thermodynamics and observational signatures of these objects \cite{Mahapatra:2022xea}. These solutions have attracted significant theoretical and observational interest due to their improved dynamical stability and potential astrophysical relevance \cite{Ovalle:2020kpd, Contreras:2021yxe, PhysRevD.103.124052}. Observational constraints, particularly from the Event Horizon Telescope (EHT), have been used to set limits on the deformation parameter $l$ and $\alpha$ \cite{Afrin:2021imp}. These hairy Kerr solutions have therefore found successful applications in black hole thermodynamics \cite{Mahapatra:2022xea, 2023arXiv230301413V}, gravitational lensing and shadow analysis \cite{Islam:2021dyk}, quasi-normal modes (QNMs) \cite{Yang:2025hqke}, gravitational-wave phenomenology, accretion disk physics \cite{Meng:2023htc, Meng:2025ivb, Ali:2025dje,Zi:2023omh}, black hole spectroscopy \cite{Guimaraes:2025jsh}, and parameter estimation using the EHT constraints \cite{Afrin:2021imp, Vagnozzi:2022moj}. For other related works on hairy black holes, see~\cite{wang2025exploring, Vertogradov:2024qpfe, Li:2023htz, Li:2022hkq, Heydarzade:2023dof, Al-Badawi:2024iax, 2025arXiv251117137M}.
The present work explores the horizon structure, the time-like geodesics and the fundamental frequencies associated with the hairy Kerr spacetime and reports some unique and interesting features which has not been previously reported to the best of our knowledge. Thereafter, we test the viability of the hairy Kerr spacetime in explaining the high-frequency quasi-periodic oscillations observed in certain black hole sources.

One of the most significant observational phenomena that characterizes strong gravitational interactions in the near-horizon region of black holes is the occurrence of Quasi-Periodic Oscillations (QPOs).  These QPOs frequently appear as sharp peaks in the power-density spectra of certain black hole and neutron star sources and are commonly observed in their X-ray fluxes. 
Based on their observed frequencies, QPOs are conventionally classified into two categories: low-frequency QPOs (LFQPOs), with frequencies in the $\rm mHz$ range, and high-frequency QPOs (HFQPOs), which typically occur at a few hundred hertz in stellar-mass black hole (BH) systems~\cite{vanderKlis:2000ca,Maselli:2014fca,Torok:2004xs,Abramowicz:2011xu,Aschenbach:2004kj,Torok:2011qy}. Observational confirmation of HFQPOs was first achieved with NASA’s Rossi X-Ray Timing Explorer satellite \cite{2006csxs.book..623T}. 
In low-mass X-ray binaries (LMXRBs), HFQPOs correspond to characteristic timescales of $0.1$--$1~\rm ms$, which are commensurate with the dynamical timescales of accreting matter in the innermost regions ($r < 10 R_{\rm g}$) of compact objects~\cite{1971SvA....15..377S,1973SvA....16..941S,PhysRevLett8217}. 
Such millisecond dynamical timescales for stellar-mass BHs \cite{1971SvA....15..377S,1973SvA....16..941S} correspond to Fourier-domain frequencies of a few hundred hertz, consistent with the observed HFQPOs \cite{2006csxs.book.....L,vanderKlis:2000ca}. This agreement underscores the physical relevance of HFQPOs as diagnostics of accretion dynamics and strong-field gravity in the vicinity of BHs \cite{vanderKlis:2000ca}.
One of the most prominent observational features of HFQPOs is the appearance of pairs of frequency peaks in simple rational ratios, most prominently the 3:2 ratio observed in several black hole X-ray binaries. This has led to the development of resonance-based models, in which the observed frequencies are interpreted as arising from nonlinear resonances between the orbital and the epicyclic modes in the accretion flow. Both parametric and forced resonance mechanisms, as well as more general multi-resonance scenarios, have been extensively explored in the literature.  Numerical simulations also support the relevance of such rational ratios in HFQPO phenomenology \cite{Remillard:2006fc,Smith:2020npb,2001A&A...374L..19A,Abramowicz:2004rr,
Stuchlik:2013esa}.
While the 3:2 ratio is the most commonly observed feature, more complex frequency structures have also been reported. In particular, the source GRS~1915+105 exhibits multiple HFQPO frequencies, indicating a richer phenomenology that may involve more intricate resonance mechanisms or multiple oscillation modes. These observations further highlight the importance of considering a broad range of theoretical models in order to interpret the HFQPO data.
At the same time, an alternative class of models interpret QPOs in terms of the fundamental frequencies of orbital motion without explicitly invoking resonance conditions. In this work, we adopt this kinematic approach, which allows for a direct connection between the observed frequencies and the underlying spacetime geometry. This will be soon accompanied by a follow-up paper
where the phenomenological implications of the resonance models based on the QPO-based
observations will be discussed in detail.

Since the HFQPO frequencies are closely associated with the orbital and epicyclic frequencies of matter near the compact object, they exhibit an inverse scaling with the mass of the central object. Consequently, HFQPOs occur at $\rm kHz$ frequencies in neutron-star systems, whereas in supermassive BHs—for example, Sgr A*, they are shifted to the $\rm mHz$ range. Owing to their intrinsic connection to the motion of accreting matter in the immediate vicinity of compact objects, HFQPOs are believed to encode crucial information about gravity in the strong-field, high-curvature regime \cite{PhysRevLett8217}.  Their investigation therefore offers a distinctive observational probe of near-horizon physics and the fundamental properties of BH spacetimes and several theoretical studies have delved into this field \cite{Allahyari:2021bsq,Jiang:2021ajk,Bambi:2012pa,Bambi:2013fea,Dasgupta:2025qwq,Dasgupta:2025fuh,Banerjee:2022ffu,Azreg-Ainou:2020bfl,Ghasemi-Nodehi:2020oiz,Rayimbaev:2022mrk,Shaymatov:2022enf,Alloqulov:2025cta,Shaymatov:2023jfa}.
In the present work we test the hairy Kerr spacetime with the available HFQPO data of the six BH sources assuming the kinematic models for the HFQPOs. A similar study has been done previously \cite{Liu:2023ggze} where only one of the available kinematic models was considered and the parameter space seemingly has not be exhaustively explored.
The present work analyses the parameter space more comprehensively and therefore offers to bridge this gap in the literature.

The paper is organized as follows: In \autoref{S2} we provide an overview of the rotating hairy Kerr BH spacetime and discuss its horizon structure. We investigate the nature of time-like geodesics and the fundamental frequencies associated with the hairy Kerr spacetime in \autoref{S3}. \autoref{error-analysis} is dedicated to test the viability of the hairy Kerr spacetime in explaining the observed HFQPOs of BHs in the context of the available kinematic models. A comparison between the different HFQPO models is performed to test the suitability of each of these models source-wise.
We summarize the constraints obtained on the deformation parameter and the length parameter, discuss their implications, and
conclude with some scope for future work in \autoref{S6}. We work with mostly positive metric convention and consider geometrized units, i.e., $G=c=1$.
\\

\section{A brief review of the rotating hairy solutions}
\label{S2}
In this section, we briefly review the rotating hairy black hole (BH) solution which is a solution of Einstein field equations with a conserved energy-momentum (EM) tensor that satisfies the strong energy conditions (SECs). This conserved energy-momentum tensor may encompass one or more fundamental fields (which may be of scalar, vector, or tensor origin) arising due to the presence of dark-matter and/or dark-energy, whose nature is unknown. This EM tensor needs to fulfill only certain minimal conditions, namely, (i) should be well-behaved everywhere outside the event horizon, (ii) should fulfill the strong-energy conditions outside the event horizon, and (iii) the resultant hairy metric should possess a well-defined event horizon. The rotating hairy black hole solution can be obtained by first applying the extended gravitational decoupling method \cite{Ovalle:2020kpd, Contreras_2021} to generate the static hairy black hole background \cite{Ovalle:2018gic, Ovalle:2017fgl}, and subsequently employing the Newman–Janis algorithm \cite{Mahapatra:2022xea}. The static seed metric for this construction is given by
\begin{align}
ds^2=-\bigg(1-\frac{2\mathcal{M}}{r} + \alpha e^{\frac{-r}{\mathcal{M} - \alpha l/2}}\bigg)dt^2 + \bigg(1-\frac{2\mathcal{M}}{r} + \alpha e^{\frac{-r}{\mathcal{M} - \alpha l/2}}\bigg)^{-1}dr^2 + r^2 d\Omega^2 \,,
\label{metfunclSEC}
\end{align}
where $\alpha$ is the deviation parameter measuring deviation from the Schwarzschild metric, $l_0=\alpha l$ corresponds to the primary hair, and $\mathcal{M}$ is associated with the ADM mass of the black hole. This is the mass that will be observed
by an asymptotic observer at infinity and is related to the usual Schwarzschild mass $M$ by the relation $\mathcal{M} = M +\alpha l/2$. 



  It is important to note that a Lagrangian description of the hairy gravitational theory under consideration is not known at present. 
In this framework, one works directly at the level of the Einstein field equations by introducing an additional effective energy-momentum tensor $T_{\mu\nu}$ that deforms a known seed solution in a controlled manner \cite{Ovalle:2018gic, Ovalle:2017fgl, Ovalle:2020kpd, Contreras_2021}. This procedure allows one to systematically generate new solutions by decoupling the gravitational sources, without requiring an explicit derivation from a fundamental Lagrangian. This feature is inherent to the gravitational decoupling method and has been widely adopted in the literature to explore phenomenological extensions of General Relativity
\cite{Vagnozzi:2022moj,Afrin:2021imp,Islam:2021dyk,Yang:2025hqk,Meng:2023htc, Meng:2025ivb, Ali:2025dje,Zi:2023omh,Guimaraes:2025jsh}. The lack of precise knowledge of the matter fields giving rise to the hairy BH solution is attributed mainly to the yet unknown nature of the microphysics it is invoked to explain, e.g. the dark sector. Consequently, an action principle giving rise to the hairy BH solution has not yet been constructed. Only minimal conditions are imposed on the matter sector which we mentioned above, that ensures the possibility to encompass a large class of matter fields based on the physics it is required to explain.

To constrain the parameters of the hairy black hole and test their physical validity, it is crucial to obtain their rotating counterparts. Since the black hole spin is crucial in astrophysical processes, it makes rotating black holes important as they provide prototypes of black holes that are usually observed in nature. The axisymmetric and stationary counterpart of the above spherically symmetric static solution in  Boyer-Lindquist coordinates reads
\cite{Mahapatra:2022xea,Contreras:2021yxe},
\begin{align}
ds^2= g_{tt}(r,\theta)dt^2 + 2g_{t\phi}(r,\theta) dtd\phi + g_{\phi\phi}(r,\theta)d\phi^2 + g_{rr}(r,\theta) dr^2 +g_{\theta\theta}(r,\theta)d\theta^2 \,,
\label{hairyrotmetric}
\end{align}
where
\begin{align}
g_{tt}(r,\theta)&= -\frac{r^2 \big(1-2\mathcal{M}/r + \alpha e^{-r/(\mathcal{M}-\alpha l/2)}\big)+ a^2 \cos^2\theta}{\Sigma}\,, \nonumber \\
g_{t\phi}(r,\theta)&=-\frac{ar^2 \sin^2\theta}{\Sigma}\bigg(\frac{2\mathcal{M}}{r}-\alpha e^{-r/(\mathcal{M}-\alpha l/2)}\bigg)\,,\nonumber \\
g_{\phi\phi}(r,\theta)&=\Sigma \sin^2\theta\bigg[1 + a^2 \sin^2\theta\frac{r^2(1+2\mathcal{M}/r-\alpha e^{-r/(\mathcal{M}-\alpha l/2)}) +a^2 \cos^2\theta }{\Sigma^2}\bigg]\,,\nonumber \\
g_{rr}(r,\theta)&=\frac{\Sigma}{a^2+r^2\big(1-2\mathcal{M}/r+ \alpha e^{-r/(\mathcal{M}-\alpha l/2)}\big)}\,, \nonumber\\
g_{\theta\theta}(r,\theta)&=\Sigma \,,
\label{rotatinghairmet}
\end{align}
with $\Sigma=r^2+a^2\cos^2{\theta}$. Notice that all metric coefficients, except the $g_{\theta\theta}$ term, receive corrections due to the deviation and the hair parameter. As expected, in the limit $\alpha \rightarrow 0$ (vanishing deviation), this new axisymmetric hairy black hole solution reduces to the Kerr geometry and approaches the Schwarzschild geometry with mass $\mathcal{M}$ when the spin $a$ also vanishes. Note that in order to ensure asymptotic flatness of \autoref{rotatinghairmet}, $\alpha l\leq 2$ is necessary, and one can show that the Kerr solution is again recovered when $\alpha l=2$. 
This hairy black hole solution (\autoref{hairyrotmetric}) results from the matter energy-momentum tensor that satisfies the strong energy condition and is the prototype of non-Kerr black holes with additional hairy parameters $\{\alpha, l \}$. 
Note that when $\alpha$ is non-zero but $l$ vanishes, the resultant static or stationary BH solution has no GR counterpart. We will show below that such a geometry originates from an effective EM tensor that violates the strong energy condition. Consequently, the $l = 0$ branch will not be considered when discussing observational constraints. 

The EM tensor associated with the above-mentioned rotating hairy BH solution is given by
\begin{align}
    T^{\mu\nu}=\Tilde{\rho}e^\mu_t e^\nu_t + \Tilde{p_r}e^\mu_r e^\nu_r + \Tilde{p_\theta}e^\mu_\theta e^\nu_\theta + \Tilde{p_\phi}e^\mu_\phi e^\nu_\phi\,,
\end{align}
where
\begin{eqnarray}
& & e^\mu_t =    \frac{(r^2+a^2,0,0,a)}{\sqrt{\Sigma\Delta}}\,,~~~ e^\mu_r =    \frac{\sqrt{\Delta}(0,1,0,0)}{\sqrt{\Sigma}}\,,  \nonumber \\
& & e^\mu_\theta =    \frac{(0,0,1,0)}{\sqrt{\Sigma}}\,,~~~e^\mu_\phi =    \frac{(a\sin^2\theta,0,0,1)}{\sqrt{\Sigma}\sin\theta}\,,
\end{eqnarray}
are the tetrads with $\Delta=a^2+r^2\big(1-2\mathcal{M}/r+ \alpha e^{-r/(\mathcal{M}-\alpha l/2)})$  and $\tilde{\rho}$ is the energy density while $\tilde{p_r},\tilde{p_\theta},\tilde{p_\phi}$ are the pressures. The energy density and pressures giving rise to the hairy BH solution are given by \cite{Mahapatra:2022xea},
\begin{align}
\tilde{\rho}&=   \frac{\alpha r^2e^{-\frac{r}{\mathcal{M}-\frac{\alpha l}{2}}}}{\kappa^2\Sigma^2 (\mathcal{M}-\frac{\alpha l}{2})} \lbrace r-(\mathcal{M}-\frac{\alpha l}{2})\rbrace\,, \nonumber \\
\tilde{p}_r&= -\tilde{\rho}\,, \nonumber \\
\tilde{p}_\theta&= \frac{\alpha r^2e^{-\frac{r}{\mathcal{M}-\frac{\alpha l}{2}}}}{\kappa^2\Sigma^2 (\mathcal{M}-\frac{\alpha l}{2})} \lbrace r-(\mathcal{M}-\frac{\alpha l}{2})\rbrace + \frac{\alpha e^{-\frac{r}{\mathcal{M}-\frac{\alpha l}{2}}}}{2\kappa^2 ({\mathcal{M}-\frac{\alpha l}{2}})^2\Sigma}\lbrace 2({\mathcal{M}-\frac{\alpha l}{2}})^2 -4 ({\mathcal{M}-\frac{\alpha l}{2}}) r +r^2 \rbrace\,, \nonumber \\
\tilde{p}_\phi&= \tilde{p}_\theta\,.
\label{edp}
\end{align}
 One may note that the present framework does not provide a clear idea of the origin of the hair. However, the parameters $\alpha$ and $l$ can be interpreted as effective quantities encoding the overall influence of additional matter sectors or modified gravitational interactions, without committing to a specific fundamental theory. Similar effective parametrizations are commonly employed in astrophysical tests of gravity, where the primary goal is to constrain possible deviations from General Relativity using observational data.

In order to fulfill the strong energy conditions, one needs to ensure that the energy density and pressures discussed above satisfy,
\begin{align}
\tilde{\rho}+\tilde{p}_r+2\tilde{p}_\theta\geq 0,~~~~~\tilde{\rho}+\tilde{p}_r\geq 0, ~~~~~\tilde{\rho}+\tilde{p}_\theta\geq 0\,.
\end{align}
In order for the energy density and pressures in \autoref{edp} to satisfy the strong energy conditions we need to ensure that: (a) $\tilde{\rho}\geq 0$ and $\tilde{p}_\theta\geq 0$. Throughout this paper we will consider motion in the equatorial plane, such that positivity of $\tilde{\rho}$ and $\tilde{p}_\theta$ respectively requires,
\begin{align}
    r&>\mathcal{M}-\frac{\alpha l}{2}\,, \\
    r&>2(\mathcal{M}-\frac{\alpha l}{2})\,.
    \label{SEC}
\end{align}
 By analyzing the geodesic equation for $\theta$ and by computing the curvature invariants like the Ricci scalar, the quadratic Ricci scalar, the Kretschmann scalar, etc. it is straightforward to verify that $\theta=\pi/2$ is indeed a solution of the geodesic equations and that the hairy spacetime geometry is regular at $\theta=\pi/2$.
In what follows we will scale the length parameter $l$, the angular momentum of the black hole $a$ and the distance $r$ with the ADM mass $\mathcal{M}$, such that $l\equiv \frac{l}{\mathcal{M}}$, $a\equiv \frac{a}{\mathcal{M}}$
and $r\equiv \frac{r}{\mathcal{M}}$. We need to ensure that the strong energy conditions are satisfied everywhere outside the event horizon which requires $r\geq r_{SEC}$ where $r_{SEC}=2(1-\frac{\alpha l}{2})$. Note that $r_{SEC}$ does not depend on the rotation of the black hole.

The event horizon is obtained by solving for the roots of $g^{rr}=\Delta=0$, which mostly yield two solutions, but sometimes up to four solutions, depending on the choice of $l$, $\alpha$, and $a$. The horizon structure of the rotating hairy BH has been discussed earlier \cite{Afrin:2021imp,Islam:2021dyk,Mahapatra:2022xea}, here we highlight some of the additional features of the horizon that have not been mentioned before, to the best of our knowledge.  
Analysis of the horizon structure requires one to plot $\Delta$ with the radial distance $r$. Further, one needs to ensure that the outer horizon $r_{EH}>r_{SEC}$, based on which three scenarios emerge, which we discuss below:

\subsection{Case $l>e^{-2}$:}
Just like the Kerr BH, a hairy black hole with a given $(l,\alpha)$ has the largest horizon when its spin is zero. As spin increases the outer horizon decreases while the inner horizon increases until they coincide at $a_{max}$, which is the extremal BH scenario. Thus, for a given $(l, \alpha)$ if $r_{SEC}>r_{EH}$ for the non-rotating case then the SEC is violated for all higher spin BHs. 
\begin{itemize}
    \item For a given $l$ (with $l>e^{-2}$) as we increase $\alpha$, $r_{SEC}$ decreases and the critical $\alpha$ (denoted by $\alpha_{crit}$) is obtained when $r_{SEC}$ coincides with the horizon radius of the static BH. Thus, 
    \begin{align}
     \alpha_{crit}=\frac{2}{l}-e^2
     \label{alpha}
    \end{align}
which is obtained by solving $\Delta(r_{SEC})=0$ for $\alpha$ and $l$ assuming $a=0$.
For a given $l>e^{-2}$, $\alpha_{crit}$ vanishes when $l\simeq 2/e^2$. Thus when $l>2/e^2$, SEC is satisfied for all allowed $\alpha$, i.e., between $0\lesssim \alpha \lesssim \alpha_{max}=\frac{2}{l}$.

\item For a given $l$, if $\alpha<\alpha_{crit} $ then the SEC is not satisfied. Hence, we will confine ourselves only to the cases where $\alpha\geq \alpha_{crit} $. At $\alpha= \alpha_{crit}$ only the static BH satisfies the SEC. Thus, BHs with $|a|>0$ will not be considered in this case. Note that for a given $(l,\alpha)$, $0\leq |a|\leq a_{max}$, where $a_{max}$ corresponds to the extremal BH which is obtained by solving $\Delta(l,\alpha,a,r)=\Delta'(l,\alpha,a,r)=0$.

\begin{figure}[t!]
		\centering
		\subfloat[\label{Fig1a}]%
{\includegraphics[width=0.45\linewidth]{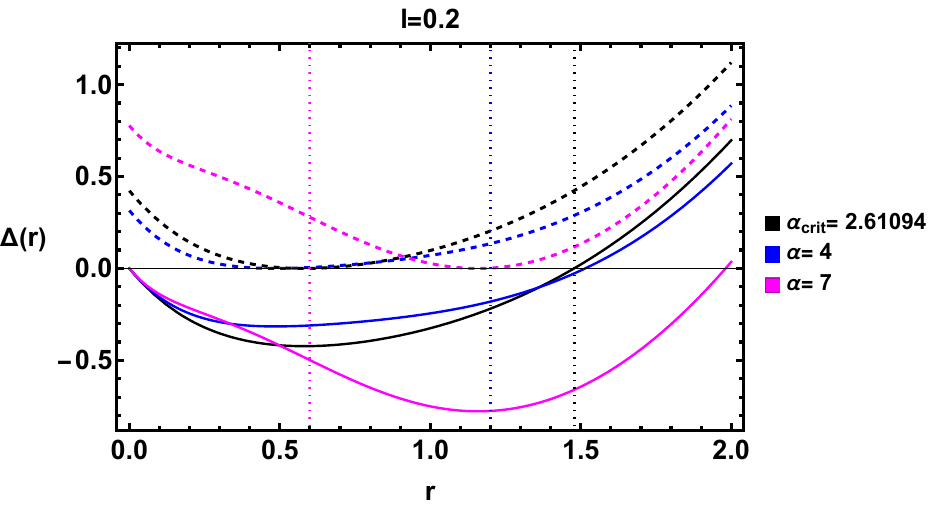}}
             \hfill
		\subfloat[\label{Fig1b}]%
        {\includegraphics[width=0.41\linewidth]{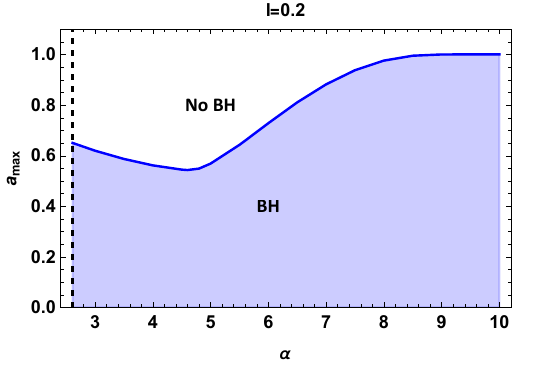}}
		\caption{In the figure above, we consider the case \( l = 0.2 \). 
Figure~(a) shows the variation of \( \Delta(r) \) for different values of \( \alpha \): 
\( \alpha = \alpha_{crit} \), \( \alpha = 4 \) (which lies between \( \alpha_{crit} \) and \( \alpha_{coin} \)), 
and \( \alpha = 7 \) (which is greater than \( \alpha_{coin} \)). 
The solid curves correspond to the non-rotating case \( a = 0 \), while the dashed curves represent the extremal cases \( a = a_{max} \), 
for which \( \Delta(r) \) develops a coincident root. 
For each value of \( \alpha \), the vertical dotted line marks the position \( r = r_{SEC} \), 
indicating the radius at which the strong energy condition begins to fail. Figure~(b) shows, for the same choice \( l = 0.2 \), the maximum spin parameter \( a_{SEC} \) allowed by the SEC 
as a function of \( \alpha \), for the interval \( \alpha_{crit} \leq \alpha \leq \alpha_{max} \). }
        \label{Fig:rvsdeltalpt2}
	\end{figure}

\begin{figure}[hbt!]
		\centering
		\subfloat[]
        {\includegraphics[width=0.5\linewidth]{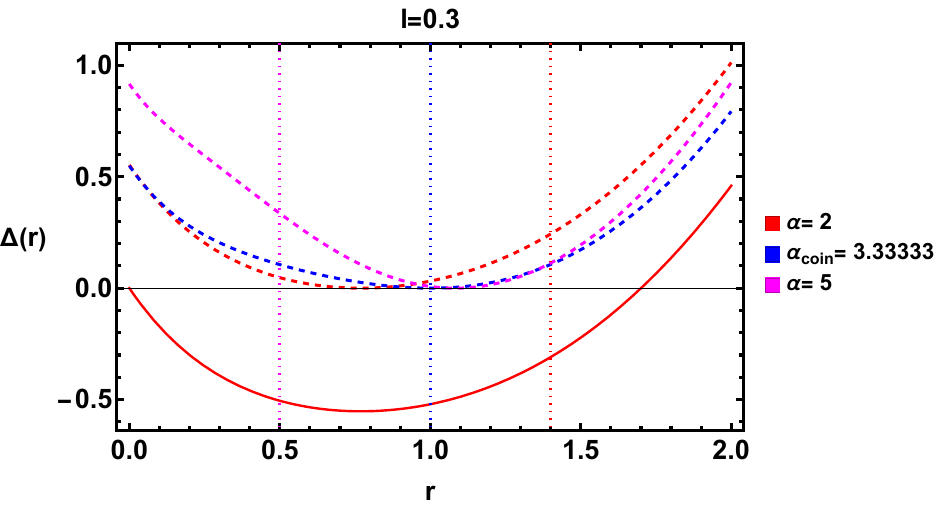}\label{Figa:rvsdeltalpt3}}
		\hfill
		\subfloat[]
        {\includegraphics[width=0.41\linewidth]{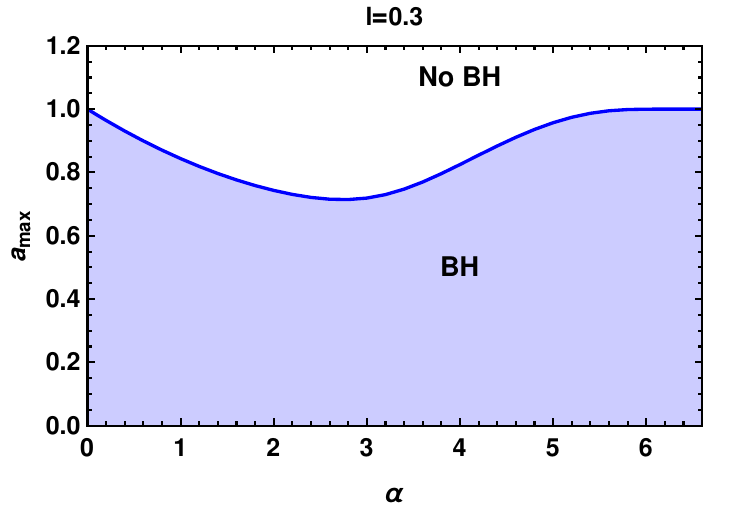}\label{Figb:alphavsamaxpt3}}
		\caption{In the figure above, we consider the case \( l = 0.3 \). 
Figure~(a) illustrates the behavior of \( \Delta(r) \) for several representative values of \( \alpha \), ranging from $\alpha<\alpha_{coin}$ ($\alpha=2$), $\alpha=\alpha_{coin}$, and $\alpha>\alpha_{coin}$ ($\alpha=5$). 
Solid curves denote the non-rotating configurations \( a = 0 \), while the dotted curves correspond to the extremal cases \( a = a_{\max} \), 
where \( \Delta(r) \) develops a coincident root. 
For each \( \alpha \), the vertical dotted line marks the radius \( r = r_{SEC} \) at which the strong energy condition ceases to hold. Figure~(b) shows, for the same choice \( l = 0.3 \), the maximum spin parameter allowed by the strong energy condition as a function of \( \alpha \), 
for the range \( 0 \leq \alpha \leq \alpha_{\max} \). }
        \label{Fig:rvsdeltalpt3}
	\end{figure}
   
\item When $\alpha_{crit}\leq \alpha \leq \alpha_{coin}$, where $\alpha_{coin}=\frac{1}{l}$, BHs with any allowed spin between $0\leq a \leq a_{max}$ do not satisfy the SEC. The spin up to which SEC holds good is given by $a_{SEC}$,
\begin{align}
a_{SEC}=\sqrt{2r_{SEC}-r^2_{SEC}(1+\alpha e^{-2})   }
\end{align}
which is obtained by solving $\Delta(r_{SEC})=0$ for $a$, for fixed $(l,\alpha)$.

\item When $\alpha=\alpha_{coin}$, $r_{SEC}$ corresponds to the horizon radius of the extremal BH, i.e., when one solves $\Delta(l,\alpha,a,r_{SEC})=\Delta'(l,\alpha, a,r_{SEC})=0$, one arrives at $\alpha_{coin}=\frac{1}{l}$ and $a_{SEC}=a_{max}$. Thereafter, as we consider $\alpha>\alpha_{coin}$, BHs with all allowed values of spin satisfy the SEC. Note that we cannot have $\alpha>\alpha_{max}=\frac{2}{l}$, otherwise asymptotic flatness would not be ensured.

\item Thus, to summarize, for values of $l>e^{-2}$, when $\alpha_{crit}\leq \alpha\leq \alpha_{coin}$ the spin range is $0\leq |a|\leq a_{SEC}$, while when $\alpha_{coin}<\alpha\leq \alpha_{max}$ the spin range is $0\leq |a|\leq a_{max}$ and $\Delta$ always has two real roots. 
\end{itemize}

~\autoref{Fig:rvsdeltalpt2} and \autoref{Fig:rvsdeltalpt3} show the variation of $\Delta$ with $r$ for various choices of $\alpha$ and $a$ for $l=0.2$ and $l=0.3$, respectively. In these figures, corresponding to each choice of $\alpha$, the solid lines are used to represent the $a=0$ cases while the dotted lines are used to denote the $a=a_{max}$ (corresponding to a given $\alpha, l$) cases. The $\Delta$ corresponding to $\alpha_{crit}$ is plotted with black color. For $l=0.3$, $\alpha_{crit}<0$ hence it is not shown in \ref{Figa:rvsdeltalpt3} and in these kind of cases all values of $\alpha$ between 0 to $\alpha_{max}$ are allowed. The vertical dotted lines in these figures denote $r_{SEC}$ corresponding to a given choice of $\alpha$ and $l$.

For \( l = 0.2 \), the critical value of the deformation parameter is 
\( \alpha_{crit} \simeq 2.6109 \), at which the outer horizon of the non-rotating 
black hole coincides with \( r_{SEC} \). Equivalently, at this critical point the zero of 
\( \Delta(r) \) aligns exactly with \( r_{SEC} \), as indicated by the black vertical 
dotted line. Clearly for $\alpha=\alpha_{crit}$, the SEC is immediately violated once rotation is 
introduced (\( |a|>0 \)), which is evident in \ref{Fig1a}, where 
the black dashed curve (corresponding to \( a = a_{max} \)) has coincident roots 
at radii smaller than \( r_{SEC} \). The blue curves in \ref{Fig1a} correspond to \( \alpha = 4 \), 
a value between \( \alpha_{crit} \) and \( \alpha_{coin} \). Here the SEC holds for 
the static case (\( a=0 \)) but fails for the extremal rotating case, as indicated by 
the blue vertical dotted line lying between the zero crossings of the blue solid 
and blue dashed curves. In contrast, for \( \alpha = 7 > \alpha_{coin} \) 
(shown by the magenta curves in ~\ref{Fig1a}), the SEC is satisfied for all 
allowed spins \( 0 \leq a \leq a_{max} \), since the magenta vertical dotted line 
marking \( r_{SEC} \) lies to the left of the zero crossings of both corresponding 
\(\Delta(r)\) curves.

In \autoref{Fig:rvsdeltalpt3} (for \( l = 0.3 \)), the SEC is satisfied for 
\( 0 \le a \le a_{SEC} \) when \( \alpha = 2 \), but it fails at \( a = a_{max} \). 
At \( \alpha_{coin} \simeq 3.33 \), the SEC becomes satisfied across the entire 
spin range \( 0 \le a \le a_{max} \); the blue dotted vertical line marks the 
radius where the corresponding \( \Delta(r) \) curve develops a coincident root. 
Since \( r_{SEC} \le r_{EH} \) at \( a = a_{max} \) implies the same for all lower 
spins, the \( a = 0 \) cases are omitted for \( \alpha \ge \alpha_{coin} \). For 
\( \alpha > \alpha_{coin} \) (e.g., \( \alpha = 5 \) in \autoref{Figa:rvsdeltalpt3}), the SEC 
is satisfied for all allowed spins, as illustrated in \autoref{Figa:rvsdeltalpt3}.


In \ref{Fig1b} and \ref{Figb:alphavsamaxpt3}, we show the maximum 
allowed spin as a function of \( \alpha \) within the admissible range 
\( \alpha_{crit} \le \alpha \le \alpha_{max} \) for \( l = 0.2 \) and \( l = 0.3 \), 
respectively. In \ref{Fig1b}, \( \alpha_{crit} \) is indicated by the 
black dashed vertical line, whereas it does not appear in 
\ref{Figb:alphavsamaxpt3} because it lies at a negative value of \( \alpha \), and 
\( \alpha < 0 \) is not allowed~\cite{Ovalle:2020kpd}. For any fixed \( l \) and \( \alpha \), increasing 
the spin beyond \( a_{max} \) makes the horizon imaginary, resulting in a naked 
singularity. Since we are interested only in black hole solutions (the blue shaded 
regions in ~\ref{Fig1b} and \ref{Figb:alphavsamaxpt3}), we will only focus in the regime $0\leq |a|\leq a_{max}$ for a given $l, \alpha$ in subsequent analysis.


\subsection{Case $l= e^{-2}$:}
In the case $l > e^{-2}$, we observed that for $\alpha_{crit} \leq \alpha \leq \alpha_{coin}$, the extremal black holes do not satisfy the SEC. However, as $l$ is decreased to the limiting value $l = e^{-2}$, the parameter $\alpha_{coin}$ coincides with $\alpha_{crit}$. The implications of this coincidence, and its impact on the SEC behavior and the structure of extremal solutions, are discussed below.
\begin{itemize}
    \item In this case, $\alpha_{crit} = \alpha_{coin} = e^{2}$, which implies that only the region $\alpha > e^{2}$ should be considered for the SEC to be satisfied.
    \item For completeness, we note that when $\alpha<e^2$ and $0< a < a_{max}$, the function $\Delta(r)$ has two real roots, which coincide when $a=a_{max}$.  
    \item When $\alpha =e^2$, the non-rotating BH also happens to be the extremal BH, hence $|a|>0$ should not be considered.
    \item For this value of $l$, when $\alpha=e^2$ and $a=0$, the event horizon corresponds to the inflection point of $\Delta$ and coincides with $r_{SEC}$, i.e., $\Delta(r_{EH})=\Delta'(r_{EH})=\Delta''(r_{EH})=0$ with $r_{EH}=r_{SEC}=2(1-\frac{\alpha l}{2})$.
    \item When $\alpha\geq e^2$, the extremal BH condition is achieved at a higher value of spin, and for all spins between $0\leq a\leq a_{max}$, the strong energy condition holds, meaning that $r_{SEC}<r_{EH}$. Therefore, the SEC is satisfied throughout the entire allowed spin range, unlike in the previous case. 
    \item For $e^{2} < \alpha < 8.15$ and $a = a_{\max}$, $\Delta$ admits three real roots, with the two inner roots becoming coincident when $\alpha \simeq 8.15$. For $\alpha$ above this value, $\Delta(a_{\max})$ possesses only one real root. In this range of $\alpha$, and for certain choices of the spin parameter, $\Delta$ can have four real roots. However, we do not discuss these additional roots in detail, as our main interest lies in the outer root, which always remains greater than $r_{SEC}$.
   \item When $\alpha\gtrsim 9.5$ and $ 0\leq a\leq a_{max}$, $\Delta$ again exhibits two real roots. As $\alpha$ increases further, the hairy black hole begins to resemble the horizon structure of the Kerr black hole.
    \item Once again, we cannot increase $\alpha$ beyond $\alpha_{max}=2e^2$.
    Therefore, in this case the allowed range of the deviation parameter is $e^2\leq \alpha\leq 2e^2$, and for every value of $\alpha$ within this interval, the entire range of spin is permitted.
     \end{itemize}

\begin{figure}[hbt!]
		\centering
		\subfloat[]{\includegraphics[width=0.5\linewidth]{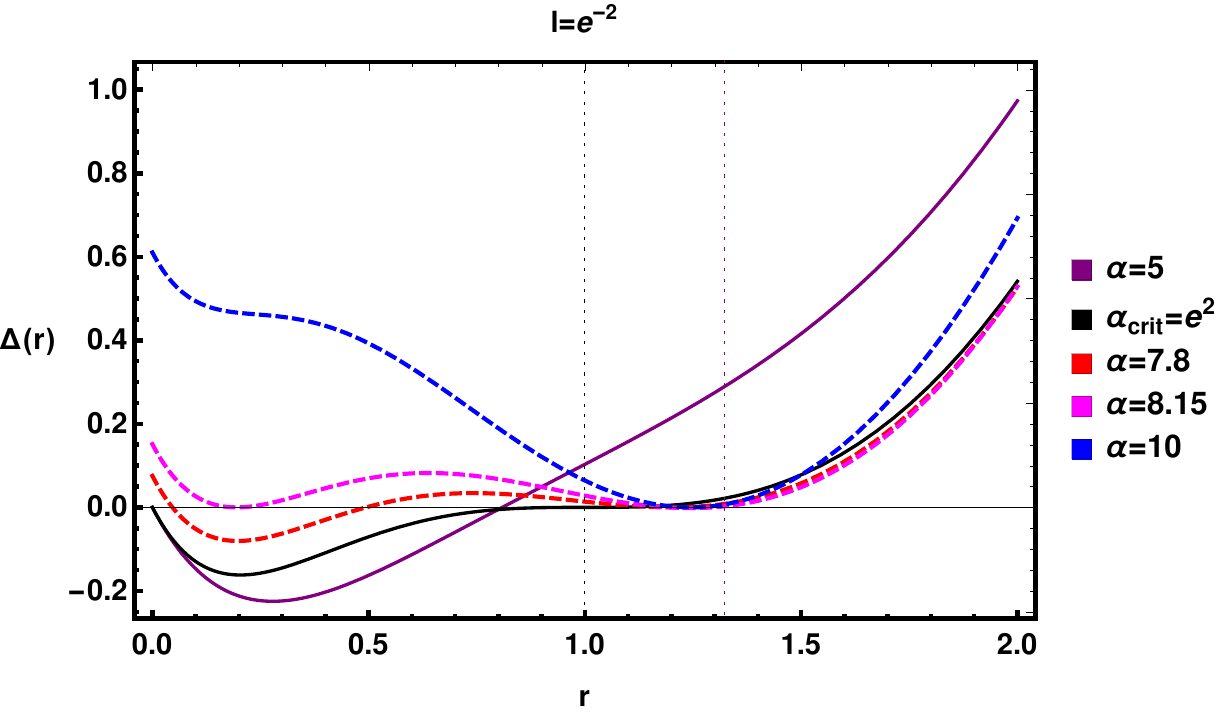}\label{Figa:rvrdeltalptesq}}
		\hfil
		\subfloat[]{\includegraphics[width=0.41\linewidth]{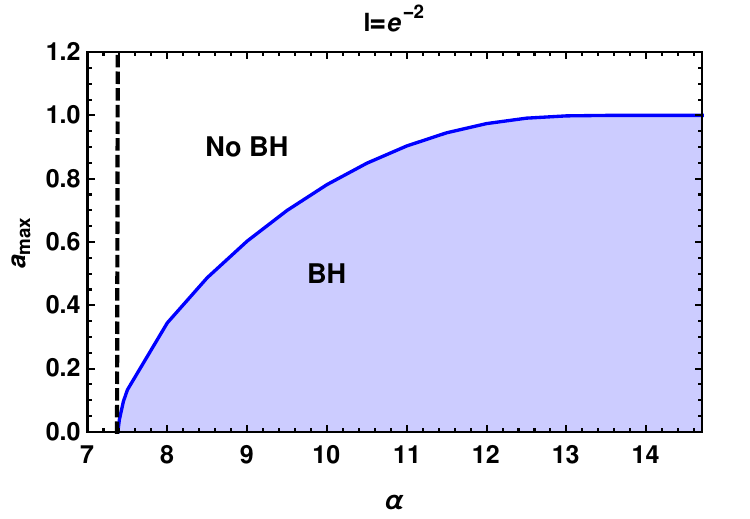}\label{Figa:alphavsamaxptesq}}
		\caption{In the figure above, we consider the case \( l = e^{-2} \). 
Figure~(a) shows the behavior of \( \Delta(r) \) for several values of \( \alpha \), 
including \( \alpha < \alpha_{crit} \) (\( \alpha = 5 \)), the critical case 
\( \alpha = \alpha_{crit} = e^{2} \), and values \( \alpha > \alpha_{crit} \) 
(\( \alpha = 7.8,\, 8.15,\, 10 \)). Solid curves represent the non-rotating 
case \( a = 0 \), while dashed curves correspond to the extremal configurations 
\( a = a_{max} \). The vertical 
dotted lines indicate \( r = r_{SEC} \) for each choice of \( \alpha \). Figure~(b) displays the corresponding dependence of the maximum allowed spin 
\( a_{max} \) on \( \alpha \) within the interval 
\( \alpha_{crit} \leq \alpha \leq \alpha_{max} \). }
\end{figure}

The above features are most clearly demonstrated in ~\ref{Figa:rvrdeltalptesq}, which shows the behavior of \( \Delta(r) \) for \( l = e^{-2} \) and different values of \( \alpha \). As before, the solid and dotted curves correspond to 
\( a = 0 \) and \( a = a_{max} \) respectively, and the vertical dotted lines mark 
the associated \( r_{SEC} \). For this choice of \( l \), the critical value is 
\( \alpha_{crit} = e^2 \), for which \( r_{EH} = r_{SEC} \) and \( a_{max} = 0 \). 
At this point, \( \Delta(r) \) develops an inflection at the horizon, as shown by the 
black solid curve. When \( \alpha < \alpha_{crit} \) (e.g., \( \alpha = 5 \)), the 
SEC fails for all spins, since the corresponding \( r_{SEC} \) lies outside the outer horizon already at \( a = 0 \), and the horizon radius decreases with spin. For \( \alpha > \alpha_{crit} \), the allowed spin increases from zero, and the SEC  is satisfied for every \( 0 \le a \le a_{max} \). In the interval 
\( e^2 < \alpha \lesssim 8.15 \), \( \Delta(r) = 0 \) admits four real roots for \( a < a_{max} \), while the extremal geometry (\( a = a_{max} \)) exhibits three roots with a double outer root. Near \( \alpha \simeq 8.15 \), the two inner roots merge for \( a < a_{max} \), and for \( a = a_{max} \) the outer roots also become 
coincident, as seen from the magenta dashed curve. For \( \alpha > 8.15 \), the structure simplifies: \( \Delta(r) = 0 \) yields two real roots for \( a < a_{max} \) 
and a single degenerate root at extremality, resembling the Kerr limit. This trend  is expected, since increasing \( \alpha \) toward \( \alpha_{max} = 2/l \) restores 
asymptotic flatness.

\ref{Figa:alphavsamaxptesq} shows how \( a_{max} \) varies with \( \alpha \) in the range \( \alpha_{crit} \le \alpha \le \alpha_{max} \) for \( l = e^{-2} \), where  \( \alpha_{crit} \) is indicated by the black dashed line. Consistent with the  previous analysis, \( a_{max} = 0 \) at the critical value 
\( \alpha = \alpha_{crit} = e^{2} \), meaning that the non-rotating black hole is already extremal at this point.
    
    
\subsection{Case $l<  e^{-2}$:}
When \( l < e^{-2} \) and \(\alpha\) lies below a critical threshold $\alpha_{crit}$, the SEC is violated since \( r_{EH} < r_{SEC} \). This behavior is illustrated for \( l = 0.1 \) in \ref{Fig4a}. For \(\alpha < \alpha_{crit}\) (with \(\alpha_{crit} \simeq 11.484\) for this choice of \(l\)), the function \( \Delta(r) \) admits two real roots, but both lie inside \( r_{SEC} \). This is seen for \(\alpha = 10\) and \(a = 0\) (purple solid curve), where the outer horizon remains smaller than the SEC threshold (purple vertical line). Since a violation of the SEC already occurs for \(a = 0\), increasing the spin cannot restore it, and therefore other spin values are not shown for \(\alpha = 10\). At the critical value \(\alpha = \alpha_{crit} \simeq 11.484\), $\Delta(r)$ at \(a = 0\) develops three real roots (black solid curve), with the outermost horizon becoming a coincident root and shifting beyond \(r_{SEC}\) (black dotted vertical line). This implies that:


\begin{figure}[hbt!]
		\centering
		\subfloat[]{\includegraphics[width=0.5\linewidth]{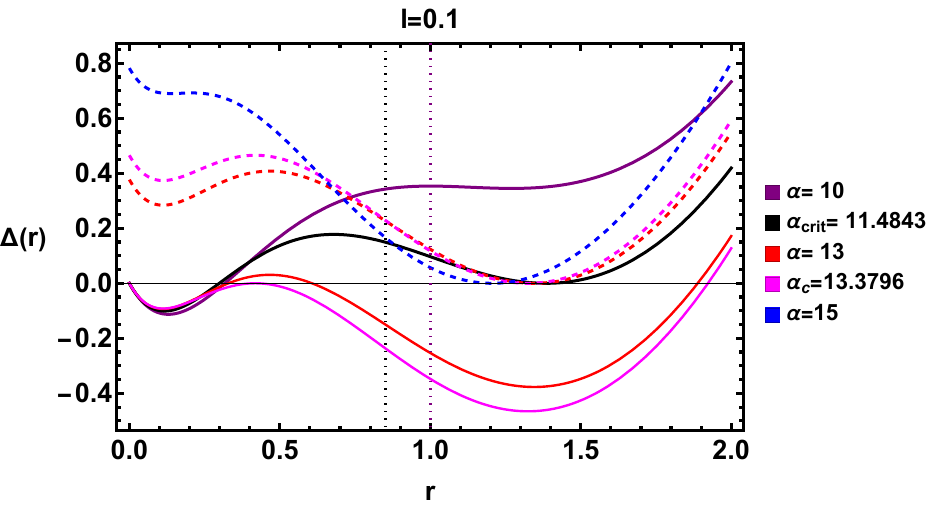}\label{Fig4a}}
		\hfil
		\subfloat[]{\includegraphics[width=0.41\linewidth]{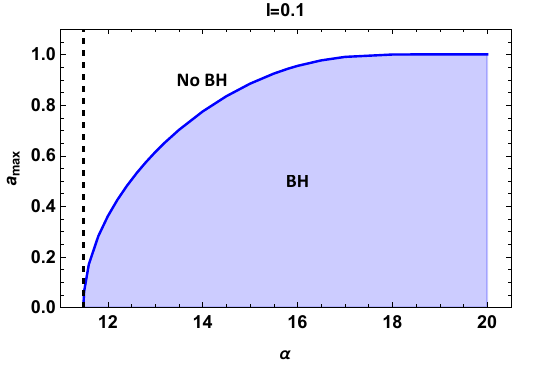}\label{Fig4b}}
		\caption{In the figure above, we consider the case \( l = 0.1 < e^{-2} \). 
Figure~(a) shows the behavior of \( \Delta(r) \) for several values of \( \alpha \), 
including \( \alpha < \alpha_{crit} \) (\( \alpha = 10 \)), the critical case 
\( \alpha = \alpha_{crit}\), and values \( \alpha > \alpha_{crit} \) 
(\( \alpha = 13,\, \alpha_c,\, 15 \)). Solid curves represent the non-rotating 
case \( a = 0 \), while dashed curves correspond to the extremal configurations 
\( a = a_{max} \). The vertical 
dotted lines indicate \( r = r_{SEC} \) for each choice of \( \alpha \). Figure~(b) displays the corresponding dependence of the maximum allowed spin 
\( a_{max} \) on \( \alpha \) within the interval 
\( \alpha_{crit} \leq \alpha \leq \alpha_{max} \). }
	\end{figure}

\begin{itemize}
\item Once again, the critical $\alpha$ corresponds to the case when the non-rotating BH is also the extremal BH, but unlike the previous case, now $r_{EH}>r_{SEC}$. Note that, this BH has two inner horizons with radii less than $r_{SEC}$, which are not important from an observational point of view.

\item  The critical $\alpha_{crit}$ is obtained by solving $\Delta(r_{EH})=\Delta'(r_{EH})=0$ with $a=0$ which gives us the radii of the coincident roots of $\Delta$,
    \begin{align}
        r_{h_{\pm}}=1\pm \sqrt{\alpha l -1}\,.
        \label{13}
    \end{align}
    The critical $\alpha$ for a given $l<e^{-2}$ is obtained by substituting $r_{h_{+}}$ in $\Delta=0$ with $a=0$ and solving for $\alpha$. This requires solving the following equation:
    \begin{align}
        \alpha e^{-\frac{r_{h_{+}}}{1-\frac{\alpha l}{2}}}=\frac{2}{r_{h_{+}}}-1\,,
        \label{14}
    \end{align}
    for a given $l$.
    
\item For $\alpha_{crit}<\alpha<\alpha_{c}$ and for certain spins between $0\leq a< a_{max}$, $\Delta$ has four real roots with the largest root $r_{EH}>r_{SEC}$. Such a situation corresponds to $\alpha=13$ and $a=0$ (the red solid curve in \ref{Fig4a}). The red dashed curve in \ref{Fig4a} is associated with $a=a_{max}$ (extremal BH case) for $\alpha=13$, which does not have 4 real roots of $\Delta(r)$. The inner roots are however not relevant astrophysically. The important thing to note is that, BHs with any allowed spin with $\alpha>\alpha_{crit}$ satisfies the SEC anywhere outside the outer horizon, i.e., $r_{EH}>r_{SEC}$. Here $\alpha_c$ is obtained when $r_{h_{-}}$ in \autoref{13} is substituted in $\Delta=0$ with $a=0$, i.e., using $r_{h_{-}}$ in \autoref{14} instead of $r_{h_{+}}$ and solving for $\alpha$ for a given $l<e^{-2}$.

\item When $\alpha=\alpha_c\simeq 13.3796$ (for $l=0.1)$ the inner roots of $\Delta(r)$ coincide for $a=0$, which is evident from the solid magenta curve in \ref{Fig4a}. For higher spin values, but $a<a_{max}$, there are two real roots for $\Delta$, while a single coincident root arises for $a=a_{max}$ as illustrated by the magenta dashed curve in \ref{Fig4a}.
    
\item  As $\alpha$ is gradually increased beyond $\alpha_c$ (e.g. $\alpha=15$ and $a=a_{max}$, denoted by the blue dashed curve in \ref{Fig4a}), the larger spins first begin to mimic the Kerr-like behavior of $\Delta$ (two real roots) followed by smaller spins. 
    
\item Overall, for $l<e^{-2}$, when the deviation parameter is between $\alpha_{crit}\leq \alpha\leq \alpha_{max}$, the SEC is satisfied for any $r$ outside the event horizon for any allowed range of spin. Finally as $\alpha\to \alpha_{max}=\frac{2}{l}$ we recover the Kerr scenario.
\end{itemize}

In \ref{Fig4b}, we plot the variation of $a_{max}$ with $\alpha$ for $l=0.1$. Here, $\alpha_{crit}$ is shown with the black dashed vertical line. It shows that at $\alpha_{crit}$, the non-rotating BH is also the extremal BH.
The above discussion elucidates that the non-trivial features of hairy Kerr BH become apparent for smaller values of $0<l\leq e^{-2}$ with $\alpha\gtrsim \alpha_{crit}$ but much less than $\alpha_{max}$. We also note that for a fixed $l$, one may in principle consider $\alpha<\alpha_{crit}$, provided the region of interest remains outside the radius $r_{SEC}$ associated with that parameter set (see \autoref{SEC}).\\


We end this section by pointing out that the case \( l = 0 \) is ruled out, since the resulting geometry admits only two horizons, both lying well inside \( r_{SEC} = 2 \), and therefore fails to satisfy the SEC at all points outside the outer horizon. Thus, to bring out the non-trivial horizon structures of the hairy Kerr BH we need both non-vanishing $l$ and $\alpha$. 

\section{Epicyclic frequencies for hairy black holes}
\label{S3}
The stationary, axisymmetric spacetime described by the hairy Kerr metric has the general form:
\begin{eqnarray}
ds^2=g_{tt}dt^2+2g_{t{\phi}}dtd{\phi}+g_{\phi\phi}d{\phi}^2+g_{rr}dr^2+g_{\theta\theta}d{\theta}^2\,,
\end{eqnarray}
where $g_{\nu\mu}=g_{\nu\mu}(r,{\theta})$ and $g_{\nu\mu}(r,{\theta})=g_{\nu\mu}(r,-{\theta})$. Here, we consider the motion of massive test particles around such a spacetime. When we study the Lagrangian we get two conserved quantities here. The first one is the energy $E$ per unit mass, and the second one is the angular momentum $L$ per unit mass.
Now, the Lagrangian is
\begin{eqnarray}
\mathcal{L}=g_{tt}\dot{t}^2+2g_{t\phi}\dot{t}\dot{\phi}+g_{\phi\phi}\dot{\phi}^2+g_{rr}\dot{r}^2+g_{\theta\theta}\dot{\theta}^2\,.
\end{eqnarray}
From the Euler-Lagrangian equation, we get
\begin{eqnarray}
\frac{d}{d\lambda} \Big( \frac{\partial \mathcal{L}}{\partial \dot{t}} \Big) - \frac{\partial \mathcal{L}}{\partial t} = 0\,.
\end{eqnarray}
Since $\frac{\partial\mathcal{L}}{\partial t}=0$, as $\mathcal{L}$ is independent of $t$, we have $\frac{\partial\mathcal{L}}{\partial\dot{t}}=-2E=\text{constant}$. This leads to
\begin{eqnarray}
g_{tt}\dot{t}+g_{t\phi}\dot{\phi}=-E = p_t\,.
\end{eqnarray}
Now, $\mathcal{L}$ is also independent of $\phi$, which implies $\frac{\partial\mathcal{L}}{\partial\dot{\phi}}=-2L=\text{constant}$. From this, we get
\begin{eqnarray}
g_{\phi\phi}\dot{\phi}+g_{t\phi}\dot{t}=L = p_\phi\,.
\end{eqnarray}
From the above two conserved quantities, we can determine the potential under whose influence the test particle is rotating,
\begin{eqnarray}
U(r,\theta)=g^{tt}-2\Big(\frac{L}{E}\Big)g^{t\phi}+\Big(\frac{L}{E}\Big)^{2}g^{\phi\phi}\,.
\end{eqnarray}
From $g_{\mu\nu}u^{\mu}u^{\nu}=-1$, we also have
\begin{eqnarray}
    g_{rr} \dot{r}^2 + g_{\theta\theta} \dot{\theta}^2 + E^2 U(r, \theta) = -1\,.
    \label{3.7}
\end{eqnarray}
Since we restrict the test particle motion to the equatorial plane, the above equation reduces to $\dot{r}^2=V(r)$, where the effective potential is given by $V(r)=-\frac{1+E^{2}U(r_c,\frac{\pi}{2})}{g_{rr}}$.

The marginally stable circular orbit can be obtained from $V(r)=0$, $V^{\prime}(r)=0$, and $V^{\prime\prime}(r)=0$. This is the last stable orbit where the particle can move in a circular path. After that, it will fall radially towards the black hole. As the particle moves along a circular orbit around the black hole, it acquires a corresponding angular velocity $\Omega=\frac{u^{\phi}}{u^t}$, where $u^{\mu} = \dot{x}^{\mu} = \frac{dx^{\mu}}{d\tau}$. From the radial Euler-Lagrangian equation and by considering circular, equatorial geodesics, we obtain
\begin{eqnarray}
\frac{\partial g_{tt}}{\partial r}+2\Omega\frac{\partial g_{t\phi}}{\partial r}+\Omega^{2}\frac{\partial g_{\phi\phi}}{\partial r}=0\,.
\end{eqnarray}
We get $\Omega$ by solving the above quadratic equation,
\begin{eqnarray}
\Omega=\frac{-\partial_{r}g_{t\phi}\pm\sqrt{(\partial_{r}g_{t\phi})^{2}-(\partial _{r}g_{\phi\phi})(\partial_{r}g_{tt})}}{\partial _{r}g_{\phi\phi}}\,,
\end{eqnarray}
where $\Omega=2\pi f_\phi$, with $f_\phi$ being the frequency with which the test particle is rotating around the black hole.

\begin{figure}[htpb!]
     \centering
     \subfloat[]{
        \includegraphics[scale=0.43]{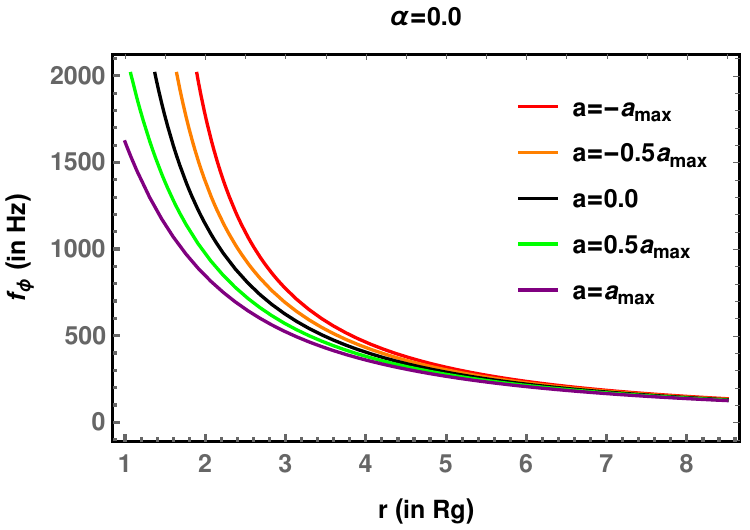}
       \label{Fig5a} 
     }\hfill
     	\subfloat[]{
		\includegraphics[scale=0.43]{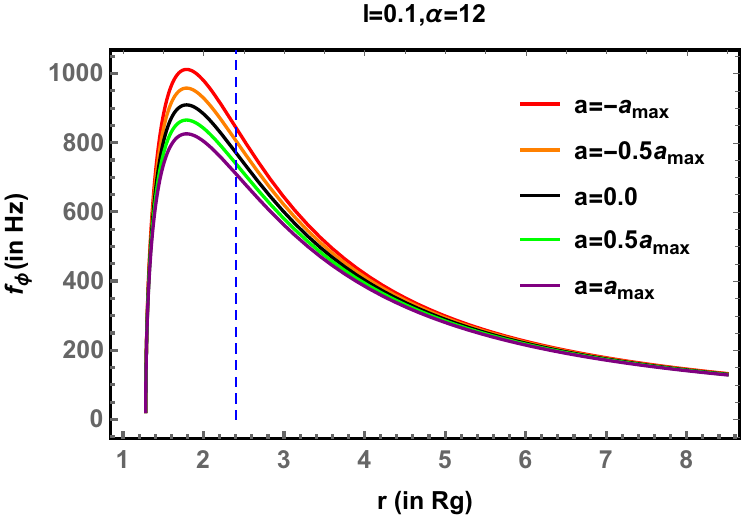}\label{Fig5b}
        
	}
	\subfloat[]{
		\includegraphics[scale=0.43]{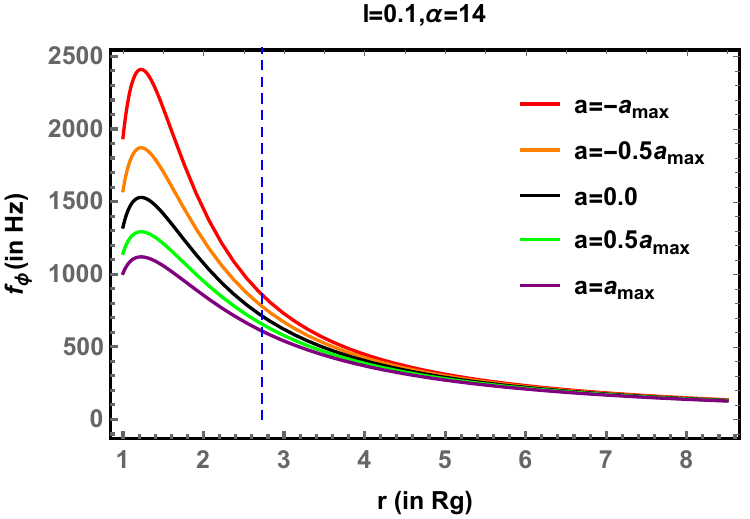}\label{Fig5c}
	}
	\subfloat[]{
		\includegraphics[scale=0.43]{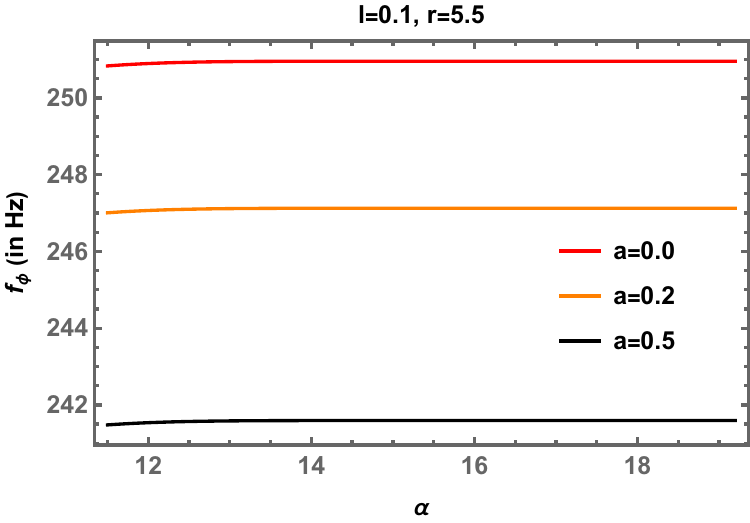}\label{Fig5d}
	} \hfill
   \subfloat[]{
		\includegraphics[scale=0.43]{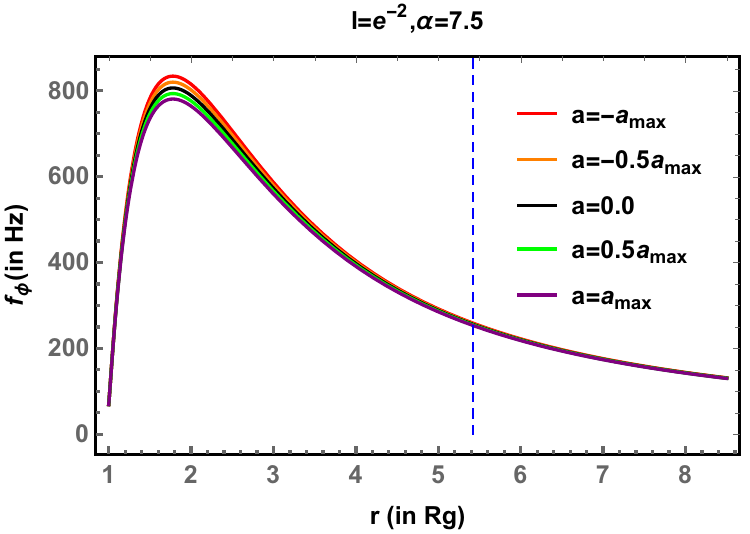}\label{Fig5e}
	}
     \subfloat[]{
		\includegraphics[scale=0.43]{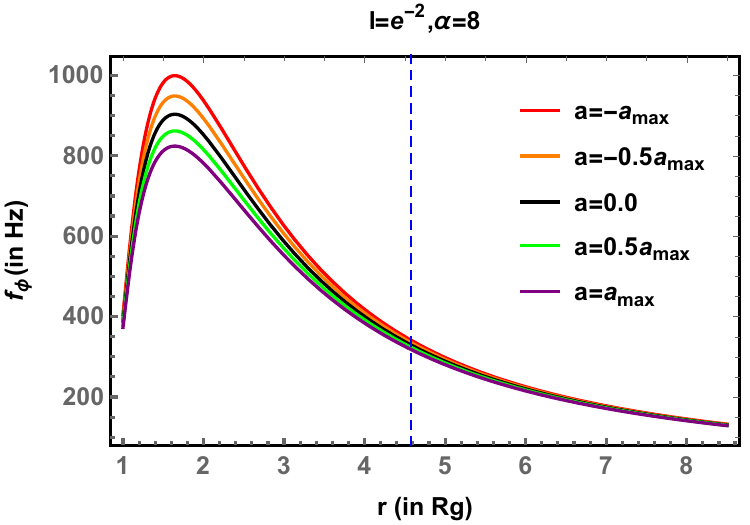}\label{Fig5f}
	}
     \subfloat[]{
		\includegraphics[scale=0.43]{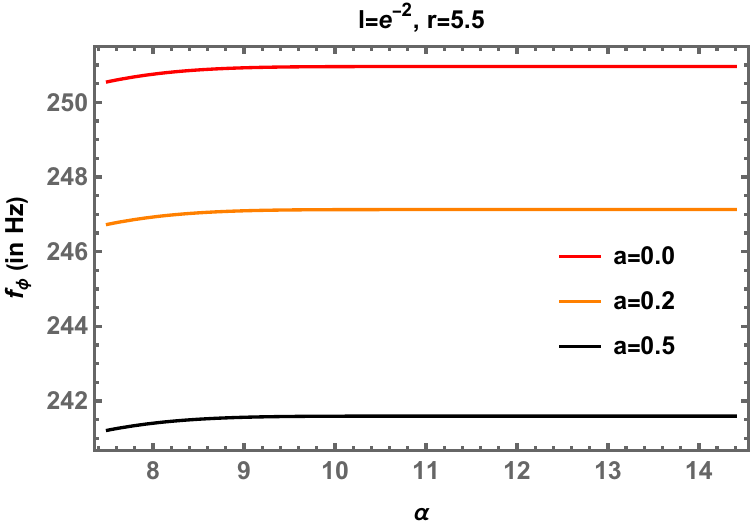}\label{Fig5g}
	}\hfill

    \subfloat[]{
        \includegraphics[scale=0.43]{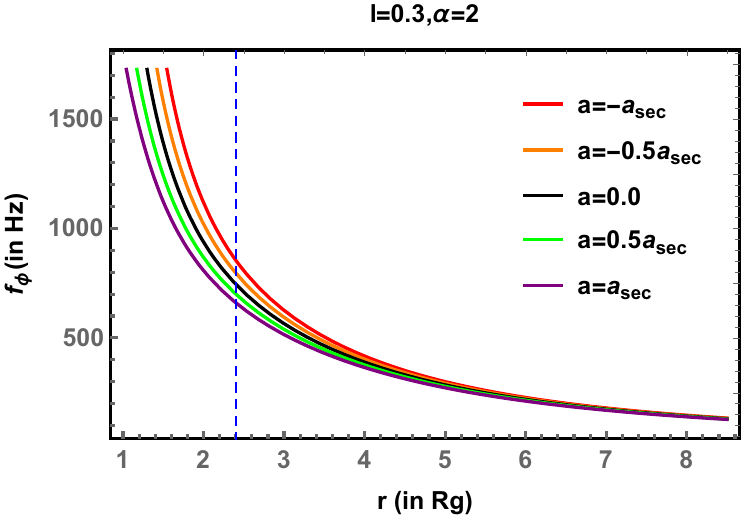}\label{Fig5h}
    }
    \subfloat[]{
        \includegraphics[scale=0.43]{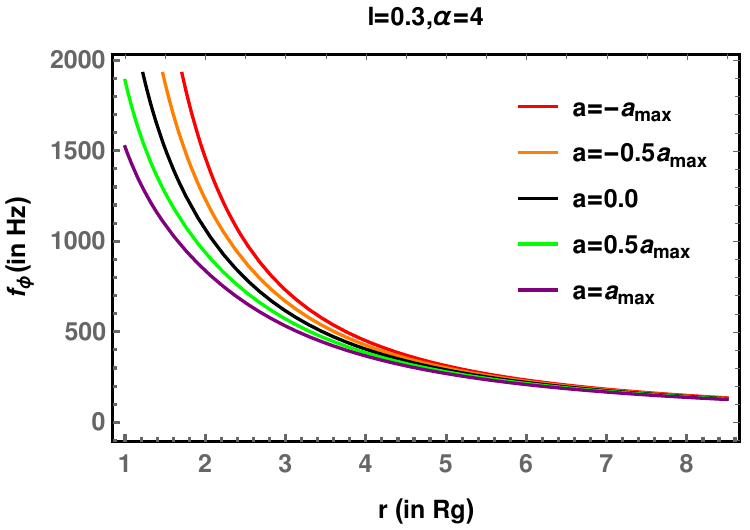}\label{Fig5i}
        
    }
    \subfloat[]{
        \includegraphics[scale=0.43]{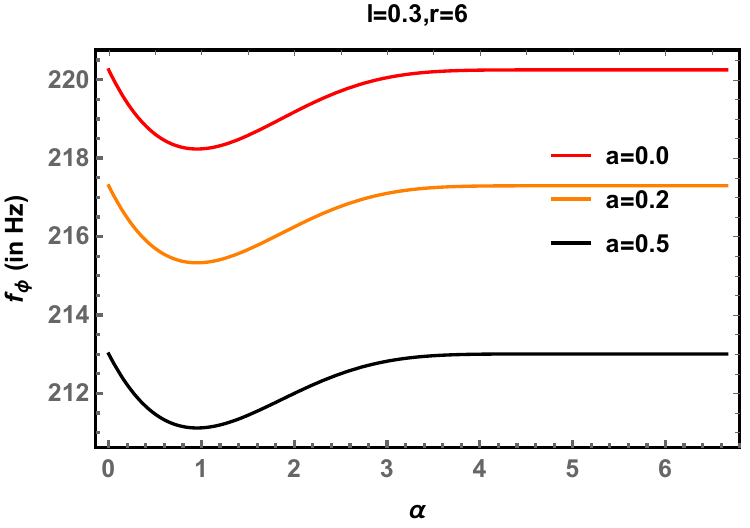}\label{Fig5j}
    }
    \caption{{ The above figures show the radial variation of $f_\phi$ with different values of $\alpha$ for a fixed $l$ but different allowed values of spin. We show vatiation of $f_\phi$ (a) with $r$ for $ \alpha=0$ (Kerr scenario), (b) with $r$ for $l=0.1, \alpha=12$, and (c) with $r$ for $l=0.1, \alpha=14$ ,(d) with $\alpha$ for $l=0.1$, $r=5.5$, (e) with $r$ for $l=e^{-2}, \alpha=7.5$, (f) with $r$ for $l=e^{-2}, \alpha=8$, (g) with $\alpha$ for $l=e^{-2}, r=5.5$, (h) with $r$ for $l=0.3, \alpha=2$, (i) with $r$ for $l=0.3, \alpha=4$, (j) with $\alpha$ for $l=0.3, r=6$. In each case we have considered variation with respect to spins from $-a_{max}$ to $a_{max}$/ from $-a_{sec}$ to $a_{sec}$. The above figure is plotted for a $M=10 M_\odot$ BH.}}
    \label{fphi1}
\end{figure}

In \autoref{fphi1}, we plot the variation of $f_\phi$ with $r$ for different choices of $l$, $\alpha$ and $a$. Here we mainly show variation with respect to $\alpha$ and $a$ for a fixed $l$.
In each row, we consider a fixed $l$ but allow variation with respect to $\alpha$. Panel~(a) shows the Kerr case with $\alpha=0$, where $f_\phi$ is plotted from $a=-a_{max}$ to $a=a_{max}$. In panels (b)-(c), we fix $l=0.1$ ($l<e^{-2}$) and vary $\alpha$ as $12$ and $14$, respectively. Next, we fix $l=e^{-2}$ and vary $\alpha$ in panels (e)-(f). Finally, we consider $l=0.3$ ($l>e^{-2}$) and vary $\alpha$ in panels (h)-(i) of \autoref{fphi1}. 

From \autoref{fphi1} we note that for a fixed $l$ and $\alpha$, $f_\phi$ increases upto the $r_{ms}$ as one moves closer to the black hole and also upon decreasing the black hole spin. Such a trend is also observed in the Kerr case.
The vertical dotted line marks the innermost stable circular orbit radius ($r_{ms}$) for the maximum spin value, which is the smallest $r_{ms}$ among all spin values for a given $l$ and $\alpha$. Note that, in panels (a) and (i), this minimum $r_{ms}\sim 1$.  The hairy Kerr BH differs significantly from the Kerr BH when $l\leq e^{-2}$ and $\alpha$ is close to $\alpha_{crit}$ (panels (b), (c), (e), and (f)) when $f_\phi$ exhibits a maxima close to the BH and eventually drops to zero. However, such a maxima occurs inside the $r_{ms}$, which marks the last stable circular orbit. Hence, as long as we confine ourselves to $r\gtrsim r_{ms}$, the behavior of $f_\phi$ is similar to the Kerr case.  For $l\leq e^{-2}$ a clear departure from the Kerr behavior is observed in the magnitude of $f_\phi$. For such values of $l$ and when $\alpha$ is close to $\alpha_{crit}$, the hairy Kerr BH exhibits much smaller $f_\phi$ than the corresponding Kerr BH with similar spin. Note however, that for such values of $l$ and $\alpha$, $a_{max}$ is much smaller than $a_{max}$ of the corresponding Kerr BH which is $\sim  1$ (see \autoref{S2}). The Kerr-like behavior of $f_\phi$ is restored for a given $l$ as $\alpha$ approaches $\alpha_{max}$. 
We further note from panels (b)-(c) and (e)-(f) of \autoref{fphi1} that for a fixed $l$ and $a$, $f_\phi$ seems to increase with $\alpha$. This behavior is prevalent near the horizon radius, but for $r\gtrsim r_{ms} $ for a given $l$ and $a$, $f_\phi$ does not change with $\alpha$, which is evident from \ref{Fig5d} and \ref{Fig5g}. For $l>e^{-2}$ and $\alpha<\alpha_{coin}$ (e.g. $l=0.3$ and $\alpha=2$ \ref{Fig5h}, see \autoref{S2}), $a$ can only be as high as $a_{sec}$ which is less than $a_{max}$. In this case, the values of $f_\phi$ are lower than in the Kerr spacetime.
Since, $r_{ms}$ increases with decreasing spin, $r_{ms}$ of these hairy BHs are generally larger than the $r_{ms}$ of hairy BHs with  $l>e^{-2}$ and $\alpha>\alpha_{coin}$ (e.g. $l=0.3$ and $\alpha=4$, \ref{Fig5i}) where $a$ can go upto $a_{max}$. For $l=0.3, \alpha=4$, $f_\phi$ rises almost up to the Kerr case after which it saturates up to $\alpha=6.66$, which is the maximum value $\alpha$ can attain when $l=0.3$.
For hairy BHs with $l>e^{-2}$, when $r$ is close to $r_{ms}$ (e.g. $r\sim 6$), $f_\phi$ initially decreases with $\alpha$, then increases and finally reaches a constant value much before $\alpha$ reaches $\alpha_{max}$ as is evident from \ref{Fig5j}. However, when $r\gtrsim r_{ms}  + 3 R_g$, for a given $l$ and $a$, $f_\phi$ does not change with $\alpha$.

\begin{figure}[htpb!]
    \subfloat[\label{Fig6a}]{
        \includegraphics[scale=0.43]{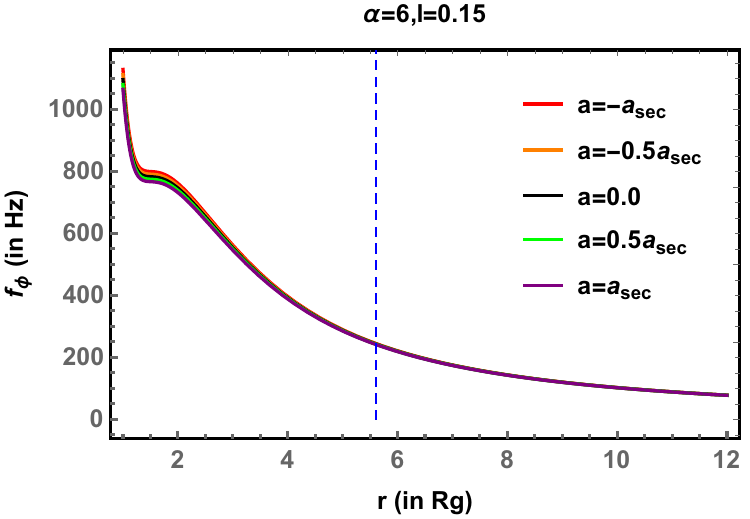}
    }
    \subfloat[\label{Fig6b}]{
        \includegraphics[scale=0.43]{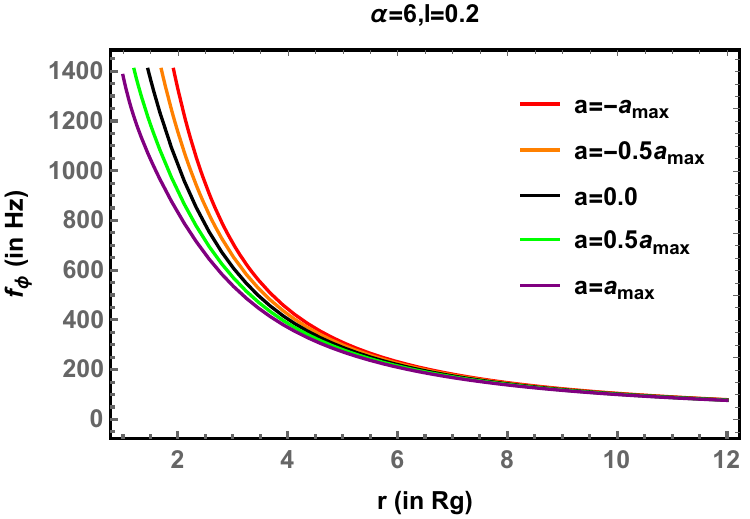}
    }
    \subfloat[\label{Fig6c}]{
        \includegraphics[scale=0.43]{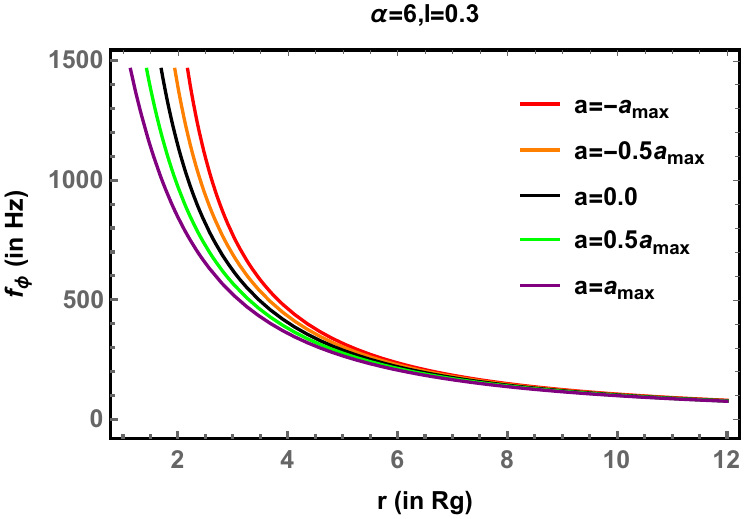}
    }\hfill
    \subfloat[\label{Fig6d}]{
        \includegraphics[scale=0.43]{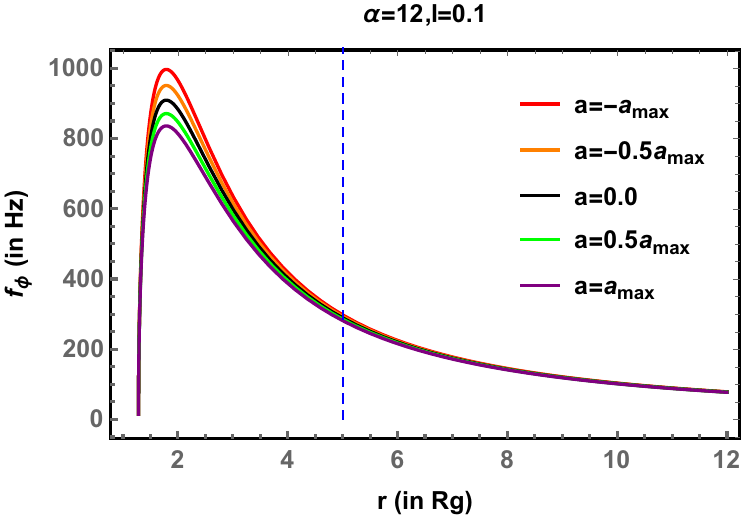}
    }
    \subfloat[\label{Fig6e}]{
        \includegraphics[scale=0.43]{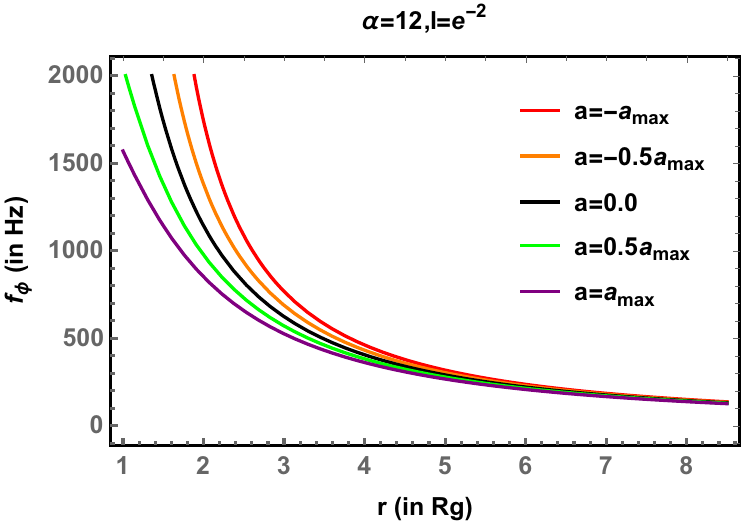}
    }
    \subfloat[\label{Fig6f}]{
        \includegraphics[scale=0.43]{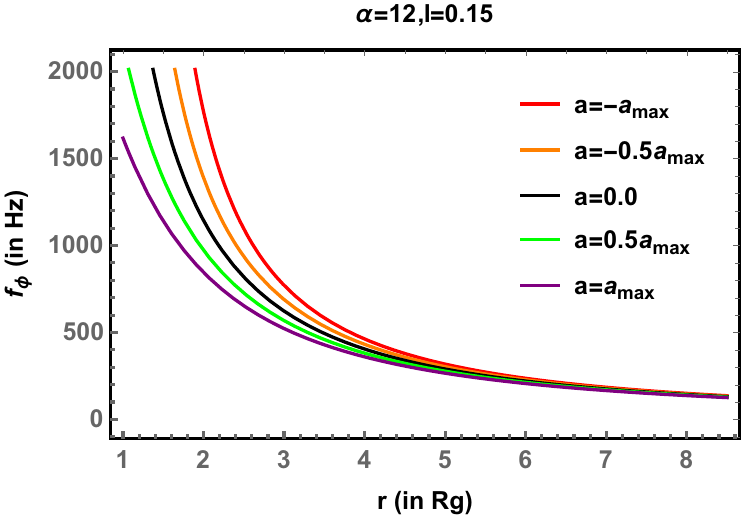}
    }

    \caption{The above figures show the variation of $f_\phi$ with $r$ for (a) $\alpha=6, l=0.15$, (b)$\alpha=6, l=0.2$, (c)  $\alpha=6, l=0.3$, (d) $\alpha=12, l=0.1$, (e)  $\alpha=12, l=0.13533$, (f)  $\alpha=12, l=0.15$. In each case we have considered variation with respect to spins from $-a_{max}$ to $a_{max}$/$-a_{sec}$ to $a_{sec}$. The above figures are made for a $M=10 M_\odot$ BH.}
    \label{fphi2}
\end{figure}

In \autoref{fphi2}, we further examine the variation of $f_\phi$ with $r$ for fixed $\alpha$ but varying $l$ for different choices of spin. Each row corresponds to a fixed value of $\alpha$, increasing from top to bottom, while $l$ is varied. 
In \ref{Fig6a}, \ref{Fig6b} and \ref{Fig6c} we fix the deformation parameter to $\alpha = 6$ and consider $l = 0.15$, $l=0.2$ and $l=0.3$ respectively. Similarly, in \ref{Fig6d}, \ref{Fig6e} and \ref{Fig6f} the deformation parameter is fixed to $\alpha = 12$ and $l$ is varied to 0.1, $e^{-2}$ and 0.15 respectively. For $\alpha=6 ~\& ~l=0.15$, the spin can go only as high as $a_{sec}$. Thus, for hairy BHs with $\alpha=6 ~\& ~l=0.15$, the $r_{ms}$ values are larger than their corresponding hairy counterparts with same $\alpha$ but larger $l$.
The $r_{ms}$ values of the hairy Kerr BH with maximum achievable prograde spin (which is also the smallest $r_{ms}$) is represented by the blue dashed line in \autoref{fphi2}. For \ref{Fig6b}, \ref{Fig6c}, \ref{Fig6e} and \ref{Fig6f}, the smallest $r_{ms}\sim 1$. The behavior is similar for lower values of $\alpha$.
It is clear from \autoref{fphi1} and \autoref{fphi2} that $f_{\phi}$
is more sensitive to $\alpha$ than $l$. $f_\phi$ varies with $l$ for a given $\alpha$ only when $\alpha\neq 0$ and $\alpha l<<2$.

To get the epicyclic frequencies, we have to perturb the stable circular equatorial orbit both radially and vertically. We can show the perturbation by the following equations
\begin{eqnarray}
r(t)= r_c+\delta r ;\hspace{2mm}  \delta r \sim e^{i\omega_r t} ; \hspace{2mm}  \delta r<<r\,,\\ 
\theta(t)=\frac{\pi}{2}+\delta\theta; \hspace{2mm}    \delta\theta \sim e^{i\omega_{\theta}t};\hspace{2mm}         \delta\theta<< \frac{\pi}{2}\,.
\end{eqnarray}
As a first approximation, we assume the two oscillations to be decoupled. From (\ref{3.7}), we get
\begin{eqnarray}
& & -g_{rr}(u^{0}2\pi f_{r}\delta r)^2-g_{\theta\theta}(u^{0} 2\pi f_{\theta})^2+E^{2}\Big[U(r_c,\frac{\pi}{2})+\frac{1}{2}\frac{\partial^2 U}{\partial r^2}(r_c,\frac{\pi}{2})\delta r^2+\frac{1}{2}\frac{\partial^2 U}{\partial\theta^2}(r_c,\frac{\pi}{2})\delta\theta^2\Big]=0\,.
\end{eqnarray}
Equating the coefficients of $\delta r^2$ and $\delta\theta^2$ to zero, we get the epicyclic frequencies as follows
\begin{eqnarray}
f_{r}^2=\frac{c^6}{G^2M^2}\Big[\frac{(g_{tt}+g_{t\phi}\Omega)^2}{2(2\pi)^{2}g_{rr}}\Big(\frac{\partial^2{U}}{\partial r^2}\Big)_{r_c,\frac{\pi}{2}}\Big]\,,
\label{3.13}
\end{eqnarray}
\begin{eqnarray}
f_{\theta}^2=\frac{c^6}{G^2M^2}\Big[\frac{(g_{tt}+g_{t\phi}\Omega)^2}{2(2\pi)^{2}g_{\theta\theta}}\Big(\frac{\partial^2{U}}{\partial\theta^2}\Big)_{r_c,\frac{\pi}{2}}\Big]\,,
\label{3.14}
\end{eqnarray}
where the factor $\frac{c^6}{G^2M^2}$ is used in order to match the dimension of frequency squared.\\

\begin{figure}[htpb!]
     \centering
     \subfloat[\label{Fig7a}]{
        \includegraphics[scale=0.43]{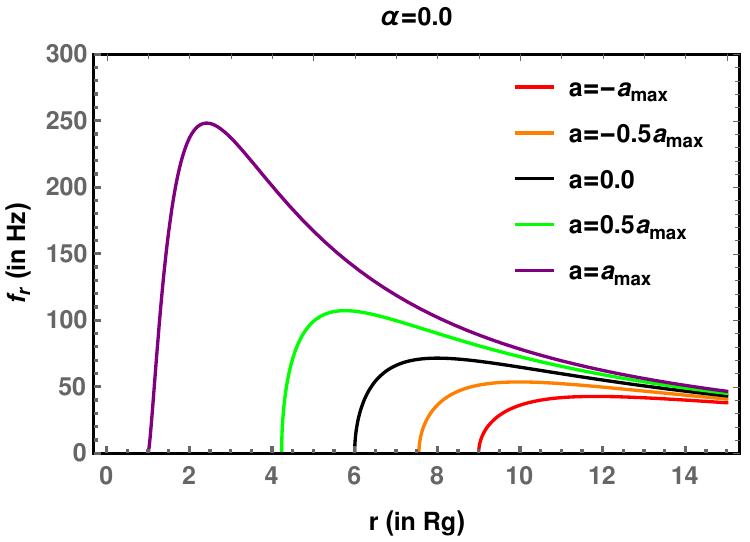}
     }\hfill

	\subfloat[\label{Fig7b}]{
		\includegraphics[scale=0.43]{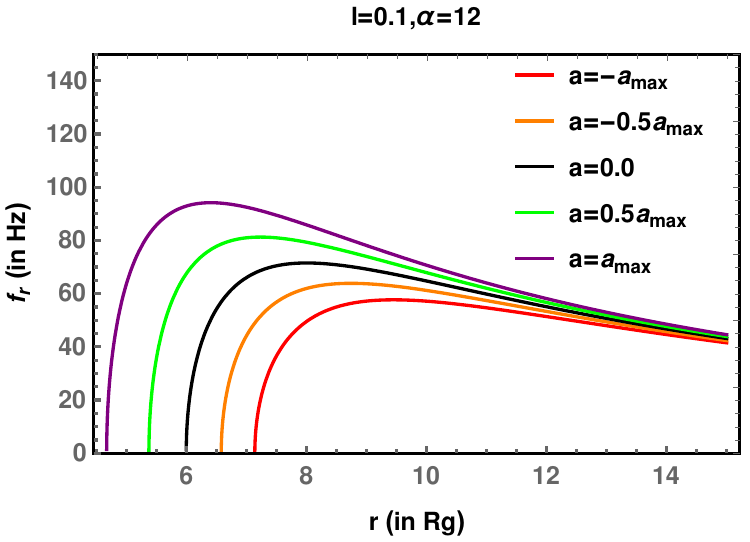}
	}
	\subfloat[\label{Fig7c}]{
		\includegraphics[scale=0.43]{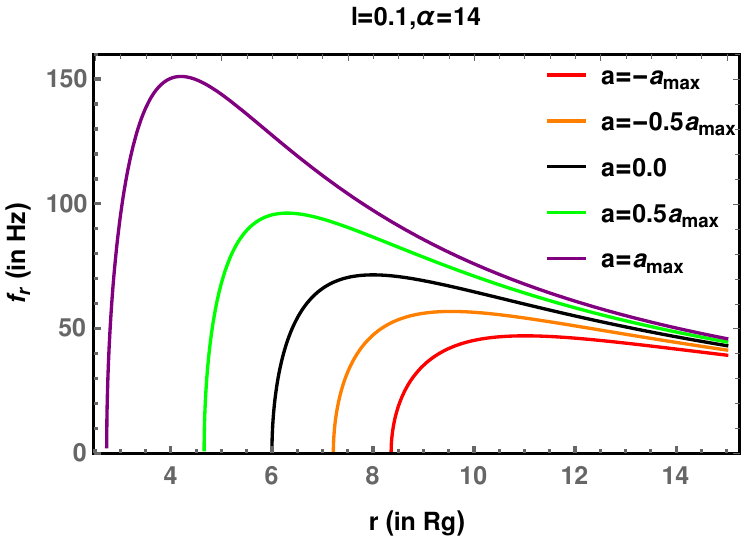}
	}
	\subfloat[\label{Fig7d}]{
		\includegraphics[scale=0.43]{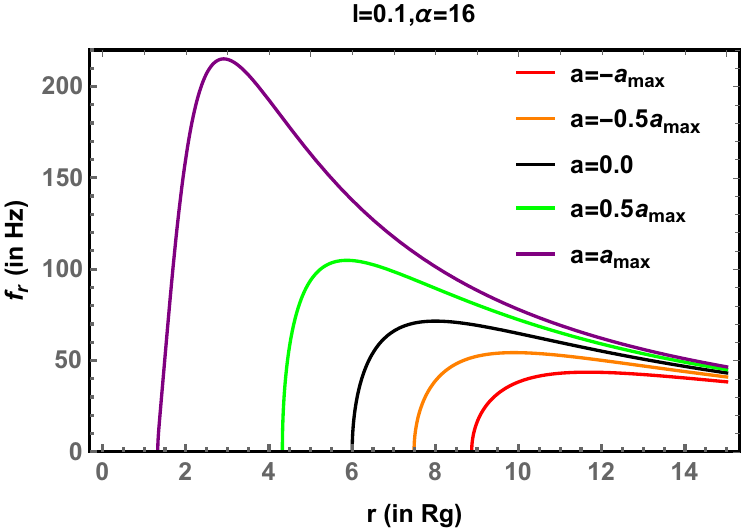}
	}\hfill

    \subfloat[\label{Fig7h}]{
        \includegraphics[scale=0.43]{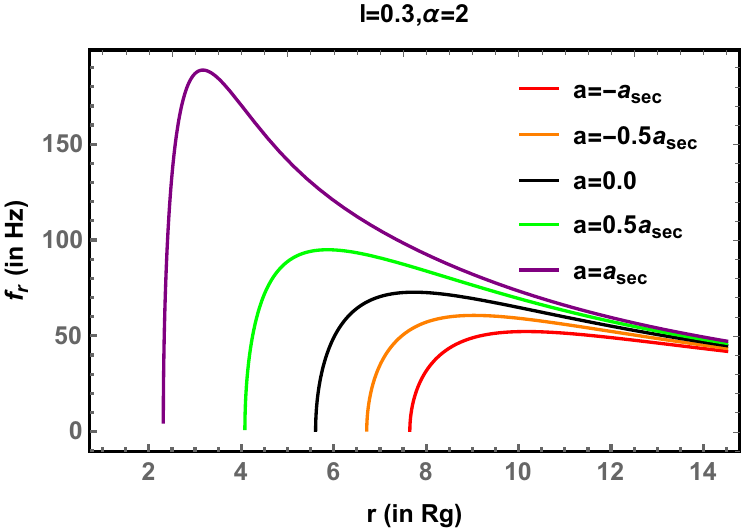}
    }
    \subfloat[\label{Fig7i}]{
        \includegraphics[scale=0.43]{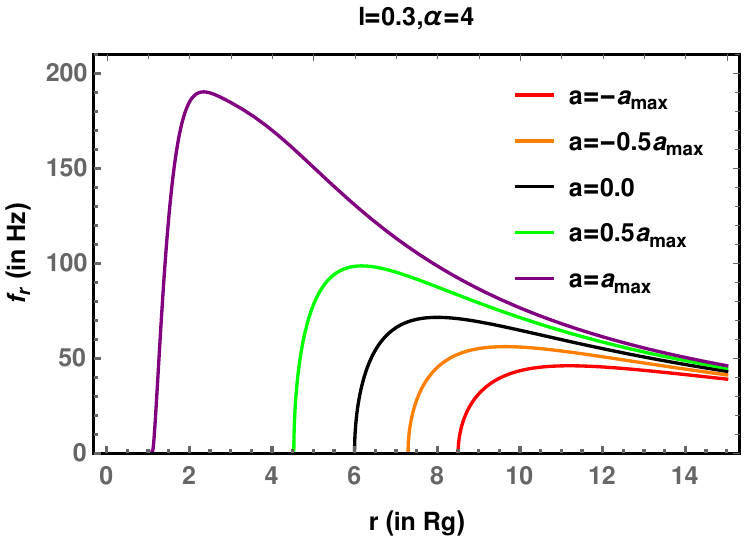}
    }
    \subfloat[\label{Fig7j}]{
        \includegraphics[scale=0.43]{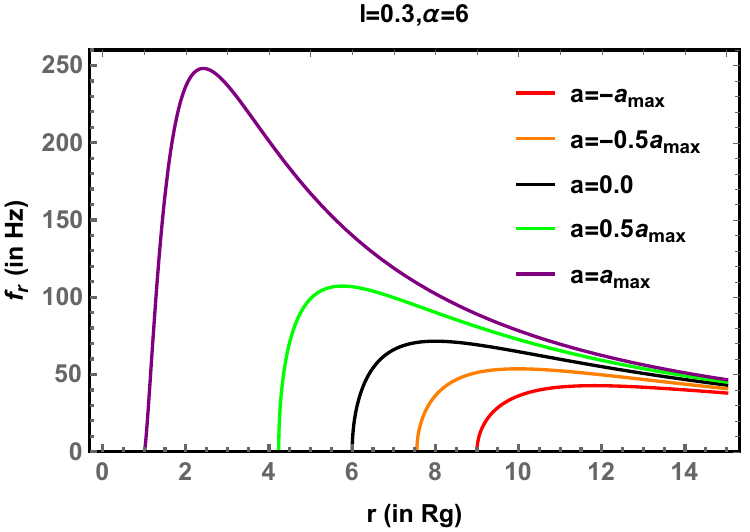}
    }

    \caption{The above figures illustrate the dependence of $f_r$ on $\alpha$ for a fixed $l$.
    They show the radial variation of $f_r$ for (a) $ \alpha=0$ (Kerr scenario), (b) $l=0.1, \alpha=12$, (c) $l=0.1, \alpha=14$, (d) $l=0.1,\alpha=16$, (e) $l=0.3, \alpha=2$, (f) $l=0.3, \alpha=4$, and (g) $l=0.3, \alpha=6$. In each subfigure, we have shown variation with respect to spins in the entire allowed range. The above plot is made for a $M=10 M_\odot$ BH. }
    
    \label{fr1}
\end{figure}

\begin{figure}[htpb!]
	\subfloat[\label{Fig8a}]{
		\includegraphics[scale=0.43]{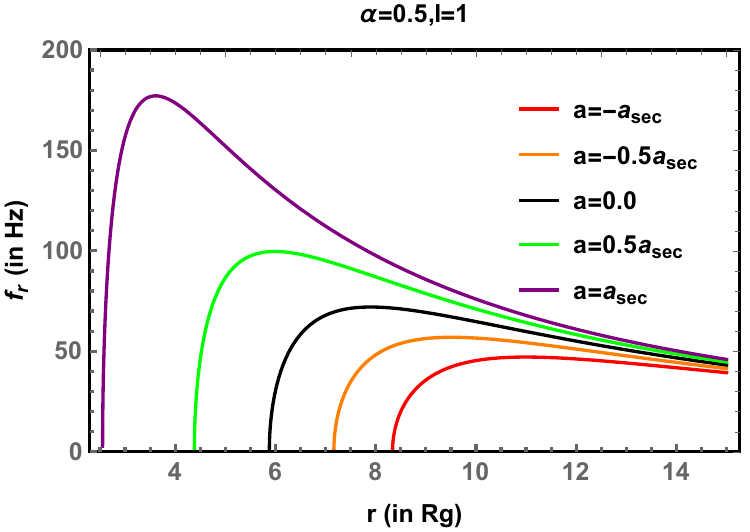}
	}
	\subfloat[\label{Fig8b}]{
		\includegraphics[scale=0.43]{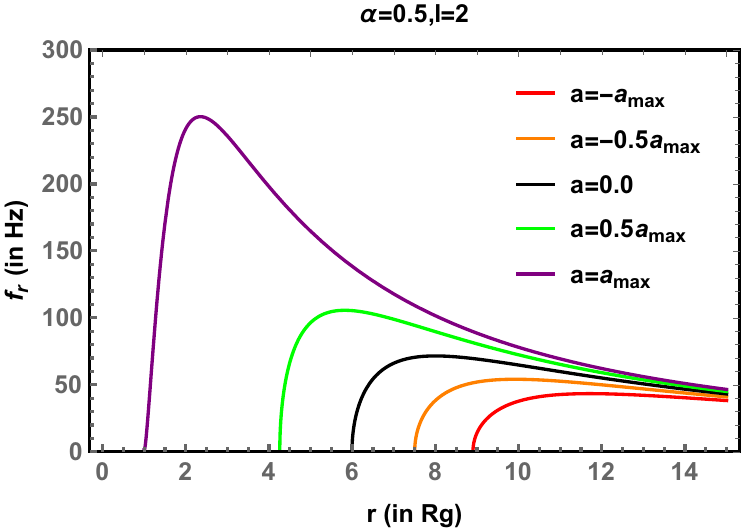}
	}
	\subfloat[\label{Fig8c}]{
		\includegraphics[scale=0.43]{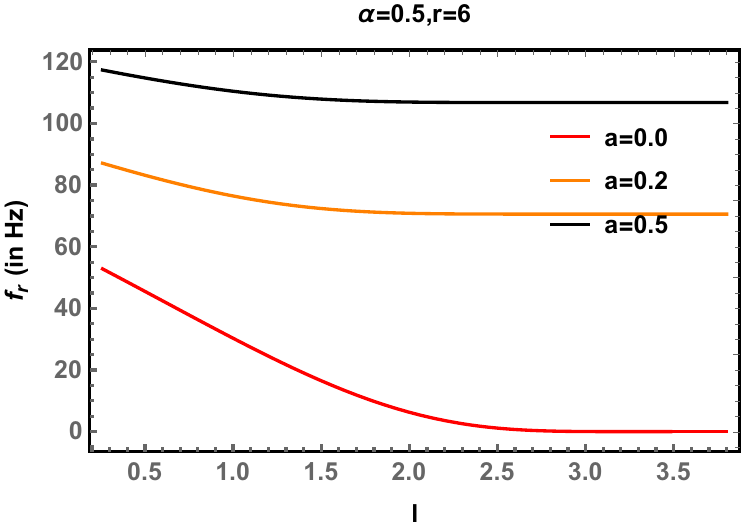}
	}\hfill

      \subfloat[\label{Fig8g}]{
		\includegraphics[scale=0.43]{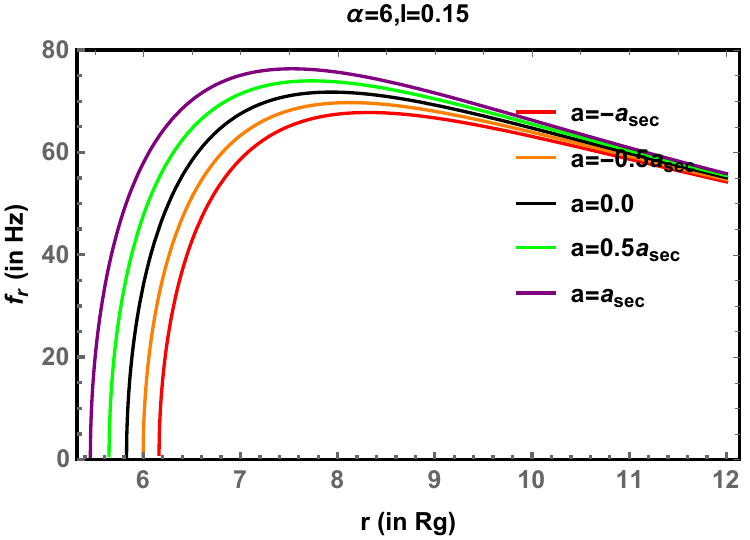}
	}
     \subfloat[\label{Fig8h}]{
		\includegraphics[scale=0.43]{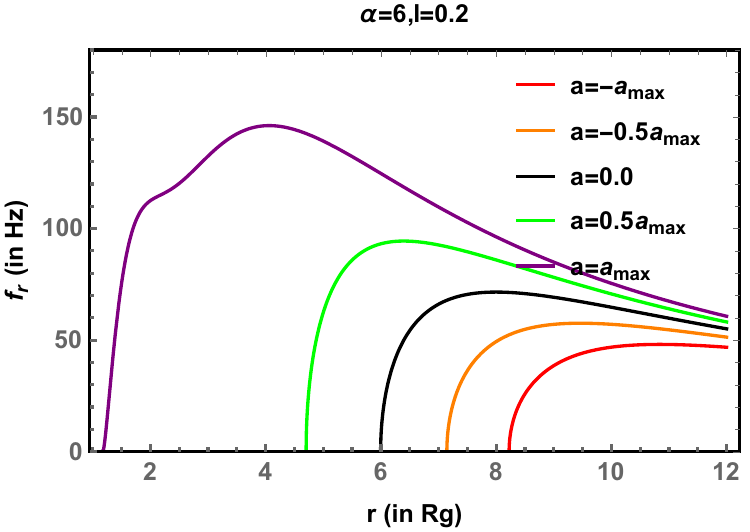}
	}
     \subfloat[\label{Fig8i}]{
		\includegraphics[scale=0.43]{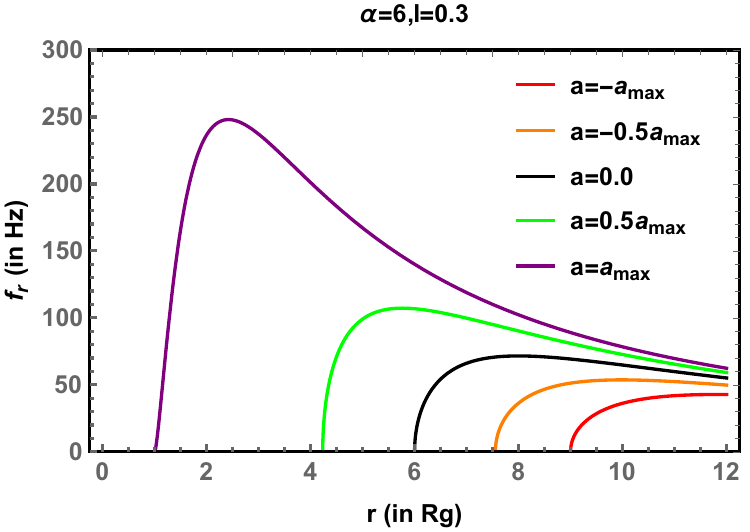}
	}\hfill
    \subfloat[\label{Fig8j}]{
        \includegraphics[scale=0.43]{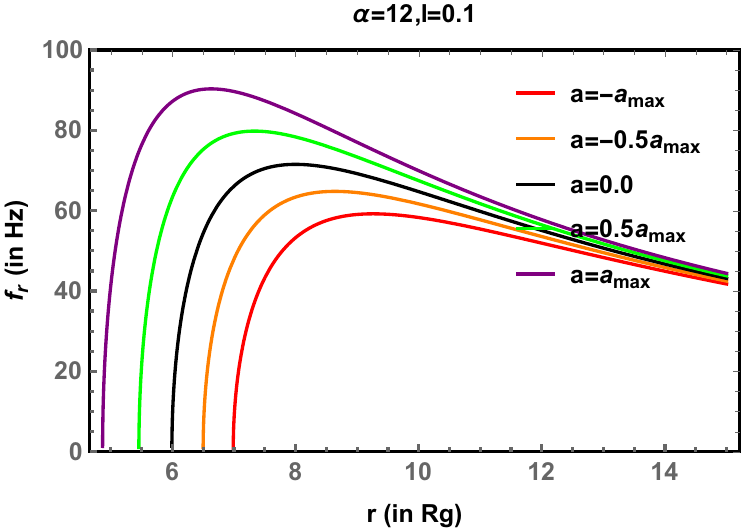}
    }
    \subfloat[\label{Fig8k}]{
        \includegraphics[scale=0.43]{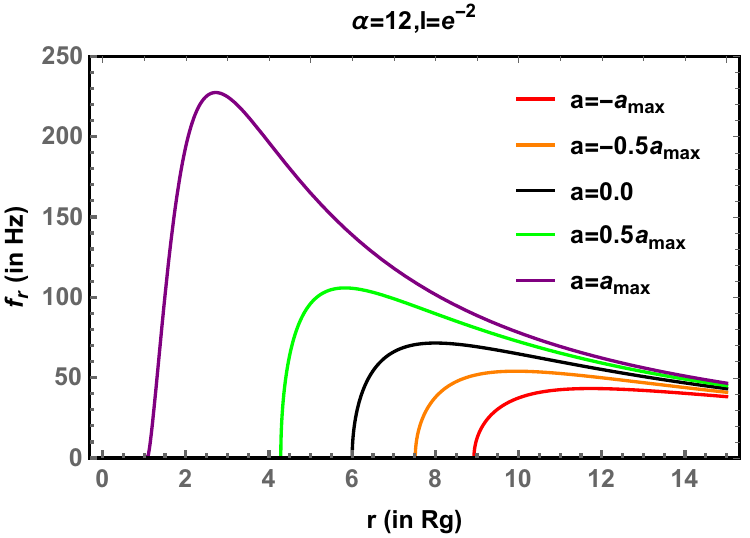}
    }
    \subfloat[\label{Fig8l}]{
        \includegraphics[scale=0.43]{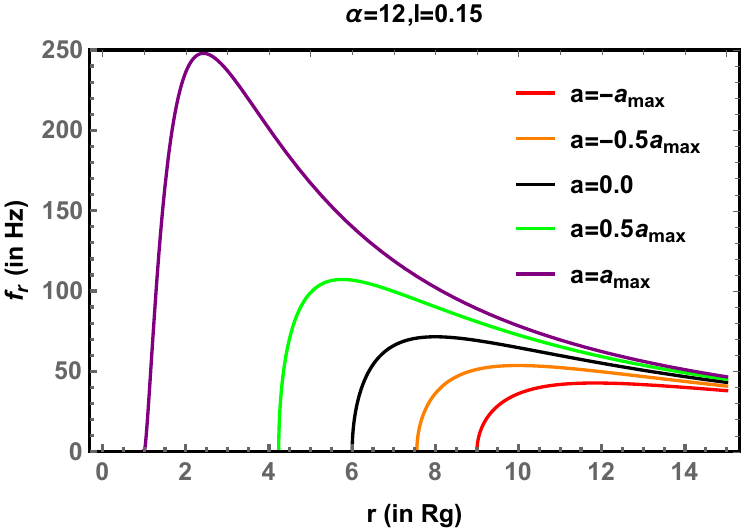}
    }
    \caption{ The above figures illustrate the dependence of $f_r$ on $l$ for a fixed $\alpha$. In particular they show the variation of $f_r$ (a) with $r$, for $\alpha=0.5, l=1$, (b) with $r$, for $\alpha=0.5, l=2$, (c)  with $l$, for $\alpha=0.5$ at $r=6$, (d) with $r$, for $\alpha=6, l=0.15$, (e) with $r$, for $\alpha=6, l=0.2$, (f) with $r$, for $\alpha=6, l=0.3$, (g) with $r$, for $\alpha=12, l=0.1$, (h) with $r$, for $\alpha=12, l=e^{-2}$, (i) with $r$, for $\alpha=12, l=0.15$,  for different spins  of a $M=10 M_\odot$ BH.  In each subfigure, we have shown variation with respect to spins in the entire allowed range.}
    \label{fr2}
\end{figure}

 In  \autoref{fr1}, we have shown the variation of $f_r$ with $r$, keeping the length parameter $l$ fixed and varying the deformation parameter $\alpha$ and the spin in the allowed range. \ref{Fig7a} corresponds to the Kerr case as when $\alpha=0$, the metric reduces to Kerr metric, which is provided for comparison purposes. From the figure we note that similar to the Kerr scenario, $f_r$ increases with decreasing distance from the BH, attains a maximum, and then drops to zero at the $r_{ms}$. For a given $\alpha$ and $l$, $f_r$ increases with spin and is maximum for prograde maximum spin (+$a_{max}$) and minimum for retrograde maximum spin (-$a_{max}$), as observed in the Kerr BH.
In \ref{Fig7b}, \ref{Fig7c} and \ref{Fig7d} we have fixed the length parameter $l$ to $0.1$ and varied the deformation parameter $\alpha$ to $12$, $14$, $16$ respectively. For all these $\alpha$ values, $f_r$ are less compared to the Kerr case but increase with increasing $\alpha$ and would resemble the Kerr case as $\alpha \to 2/l$. 
The same holds true for $l=e^{-2}$ and hence we have not shown this case. In \ref{Fig7h}, \ref{Fig7i}, and \ref{Fig7j}, we have fixed $l$ to $0.3$ and varied $\alpha$ to $2$, $4$ and $6$ respectively. In case of $\alpha=2$, the spin value can rise up to $a_{SEC}$, beyond which the strong energy condition is not satisfied. For $\alpha=2$, $f_r$ values decrease from Kerr case but for $\alpha=4$, it rises up and eventually it will saturate to the Kerr case for maximal $\alpha$. 
From \autoref{fr1}, we note that for a given $l$, $f_r$ increases with $\alpha$ as the spin becomes closer to $a_{max}$, while for lower spin, $f_r$ is not that much sensitive to $\alpha$.

In  \autoref{fr2}, we have shown the variation of $f_r$ with $r$, keeping the deformation parameter $\alpha$ fixed and varying the length parameter $l$ in the allowed spin range. In \ref{Fig8a} and \ref{Fig8b}, we have fixed $\alpha$ value to $0.5$ and varied $l$ values to $1$ and $2$. For lower spins and near the $r_{ms}$, $f_r$ decreases with $l$ and eventually saturates at higher $l$ (\ref{Fig8c}). 
From \ref{Fig8g}-\ref{Fig8i} and \ref{Fig8j}-\ref{Fig8l}, $f_r$ seems to increase with increase in $l$ for a fixed $\alpha$.
For $l=1$, the spin can be as high as $a_{SEC}$ and $f_r$ is lower than the Kerr case. Thus, the maximum achievable $f_r$ outside $r_{ms}$ is higher for hairy BHs with a fixed $\alpha$ but larger $l$. This may be attributed to the increase in $a_{max}$ as $l$ approaches $l_{max}=2/\alpha$ for a given $\alpha$.
With increasing $l$ and as $l\to 2/\alpha$ the Kerr scenario is recovered. 
From \autoref{fr1} and \autoref{fr2} we note that in general, the hairy Kerr BH with non-trivial $\alpha, l$ and satisfying the SEC has lower values of radial epicyclic frequency as compared to the Kerr case. 

\begin{figure}[htpb]
     \centering
     \subfloat[\label{Fig9a}]{
        \includegraphics[scale=0.43]{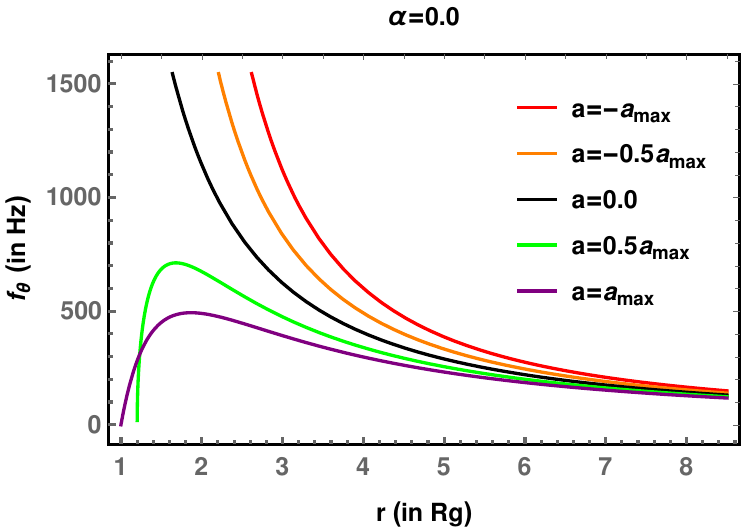}
     }\hfill
	\subfloat[\label{Fig9b}]{
		\includegraphics[scale=0.43]{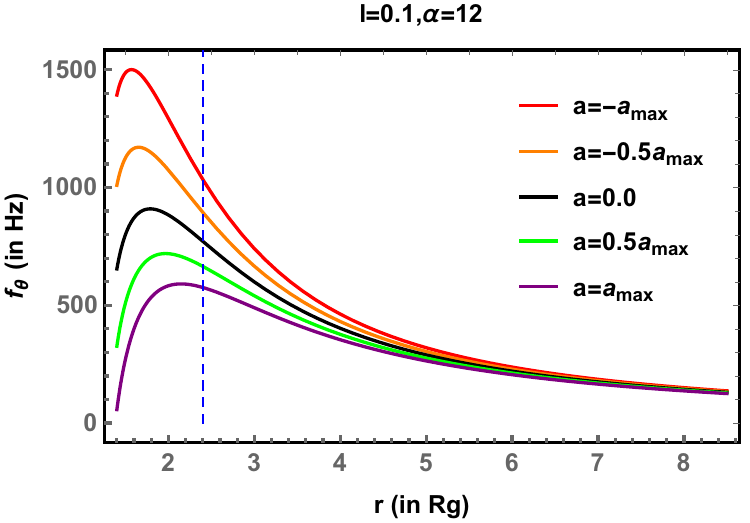}
	}
	\subfloat[\label{Fig9c}]{
		\includegraphics[scale=0.43]{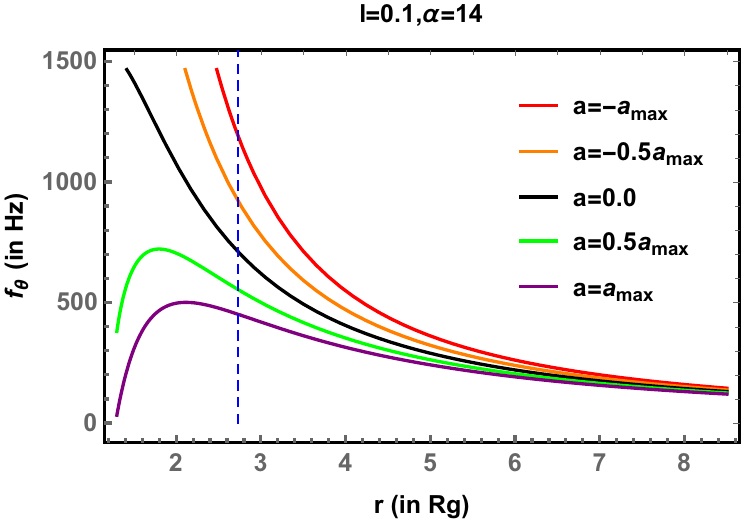}
	}
	\subfloat[\label{Fig9d}]{
		\includegraphics[scale=0.43]{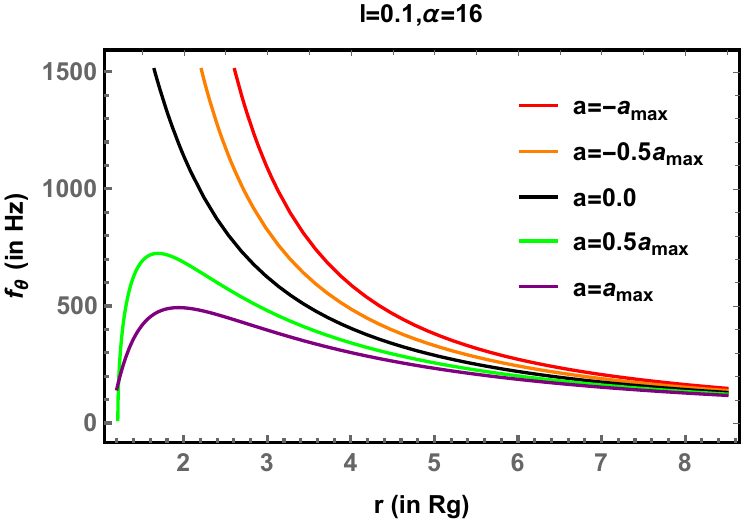}
	}\hfill

    \subfloat[\label{Fig9h}]{
        \includegraphics[scale=0.43]{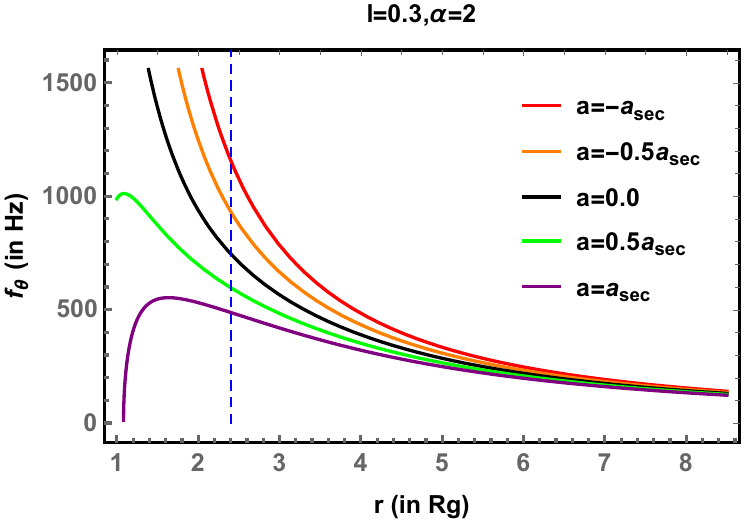}
    }
    \subfloat[\label{Fig9i}]{
        \includegraphics[scale=0.43]{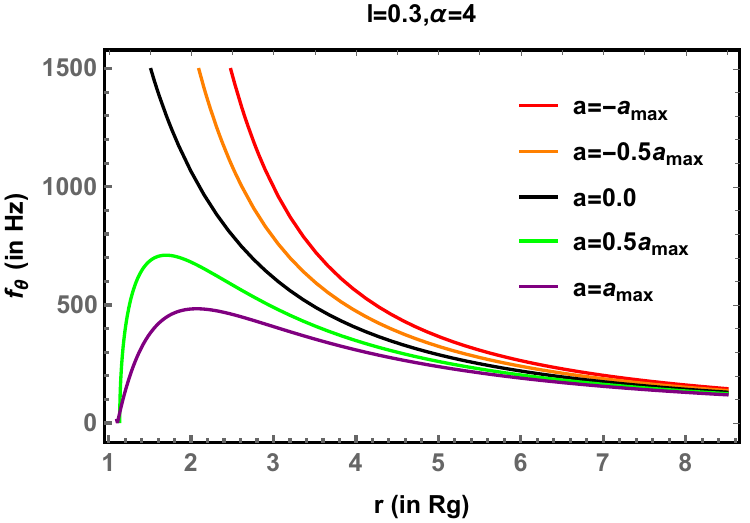}
    }
    \subfloat[\label{Fig9j}]{
        \includegraphics[scale=0.43]{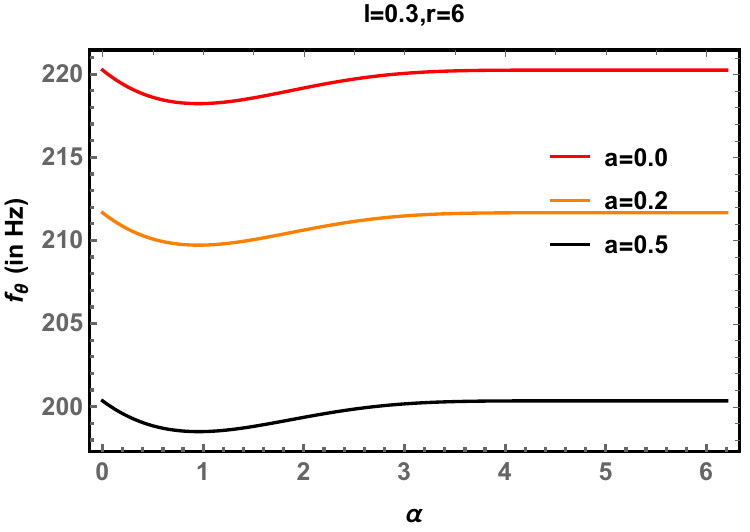}
    }

    \caption{{
    The above figures illustrate the dependence of $f_\theta$ on $\alpha$ for a fixed $l$.
    In particular, they show the variation of $f_\theta$ (a) with $r$ for $ \alpha=0$ (Kerr scenario), (b) with $r$ for $l=0.1, \alpha=12$, (c) with $r$ for $l=0.1, \alpha=14$, (d) with $r$ for $l=0.1$, $\alpha=16$, (e) with $r$ for $l=0.3, \alpha=2$, (f) with $r$ for $l=0.3, \alpha=4$, and (g) with $\alpha$ for $l=0.3$ at $r=6$, for different spins of a $M=10 M_\odot$ BH. In each subfigure, we have shown variation with respect to spins in the entire allowed range.}}
    
    \label{ftheta1}
\end{figure}

Finally in \autoref{ftheta1} we have shown the variation of $f_\theta$ with $r$, keeping the length parameter $l$ fixed and varying the deformation parameter $\alpha$ in the allowed spin range. Here \ref{Fig9a} corresponds to the Kerr case where $a_{max}\sim 1$ is the maximum spin that the Kerr black hole can attain. We have shown the variation of $f_\theta$ for $a=-a_{max}$, $a=-0.5 a_{max}$, $a=0$, $a=0.5 a_{max}$, $a=a_{max}$. We note that $f_\theta$ monotonically increases with decreasing radial distance from the BH for the Schwarzschild BH or retrograde spins while for intermediate to maximal prograde spins $f_\theta$ attains a maximum and then drops to zero inside the $r_{ms}$. However, for $r>r_{ms}$, $f_\theta$ is monotonically increasing with decreasing $r$. Also, $f_\theta$ decreases with an increase in spin of the BH.
 In \ref{Fig9b}-\ref{Fig9d}, we fix the length parameter $l$ to $0.1$ while in \ref{Fig9h}-\ref{Fig9i} we fix $l=0.3$; while varing $\alpha$ for each chosen value of $l$. 
The behaviour for $l=e^{-2}$ is similar to that of $l=0.1$, and hence we do not present this case.
Here the vertical dotted line represents the innermost stable circular orbit radius $r_{ms}$ for the maximum spin value for a given $l,\alpha$, which is also the smallest $r_{ms}$ radius compared to other spin values. In case of \ref{Fig9d} and \ref{Fig9i}, the smallest $r_{ms}$ is almost $1$. From \autoref{ftheta1}, we observe that the behavior of $f_\theta$ as a function of $r$ closely resembles the Kerr case. However, for hairy Kerr black holes with nonzero $l$ and $\alpha$, and satisfying $l\alpha < 2$, the corresponding $r_{ms}$ is larger than that of their Kerr counterparts. This is mainly because the extremal spin $a_{max}$ of hairy Kerr BHs is smaller than that of their extremal Kerr counterparts (see \autoref{S2}). As a consequence, for $r>r_{ms}$, the hairy Kerr BHs generally have smaller $f_\theta$ compared to Kerr BHs with similar spin. For a given $l$ and $a$, and at a particular radius, $f_\theta$ exhibits minimal variation with respect to $\alpha$.
This behavior can be better understood from \ref{Fig9j} where at $r=6 ~R_g$ (i.e., near $r_{ms}$), $f_\theta$ first decreases slightly, then marginally increases and finally saturates to a constant value much before $\alpha \sim \alpha_{max}$.

\begin{figure}[htpb]
	\subfloat[\label{Fig10a}]{
		\includegraphics[scale=0.43]{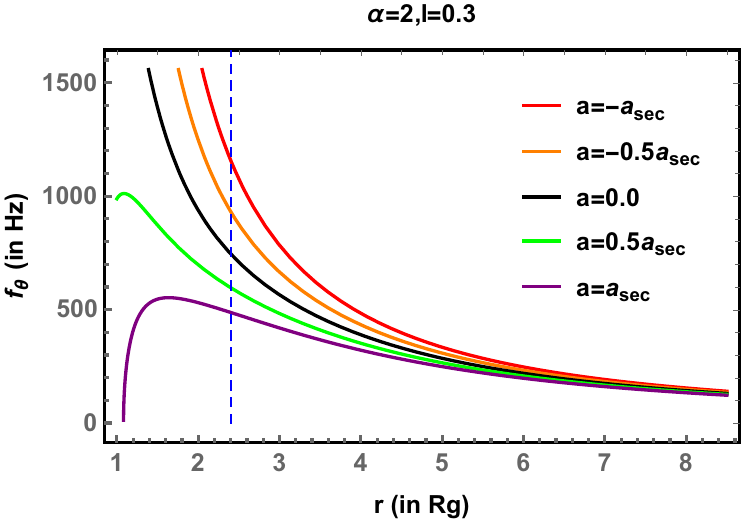}
	}
     \subfloat[\label{Fig10b}]{
		\includegraphics[scale=0.43]{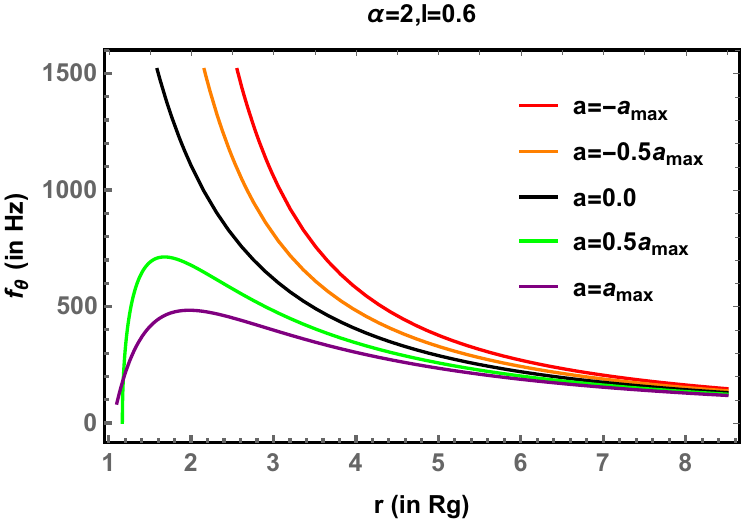}
	}
     \subfloat[\label{Fig10c}]{
		\includegraphics[scale=0.43]{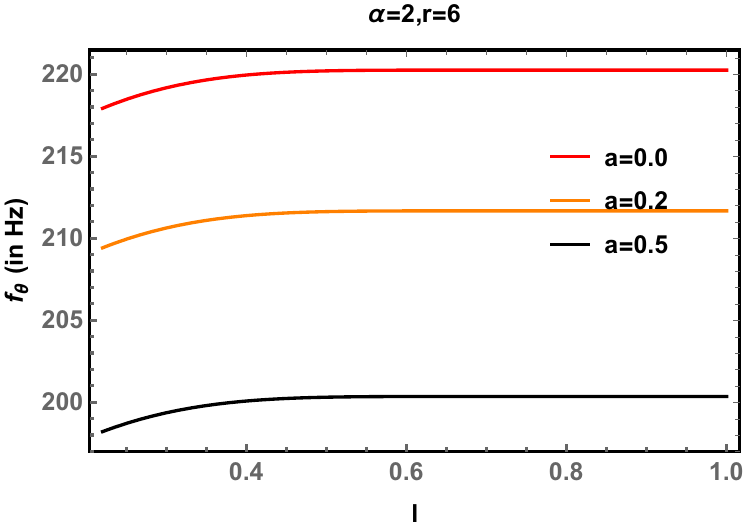}
	}\hfill

    \subfloat[\label{Fig10d}]{
		\includegraphics[scale=0.43]{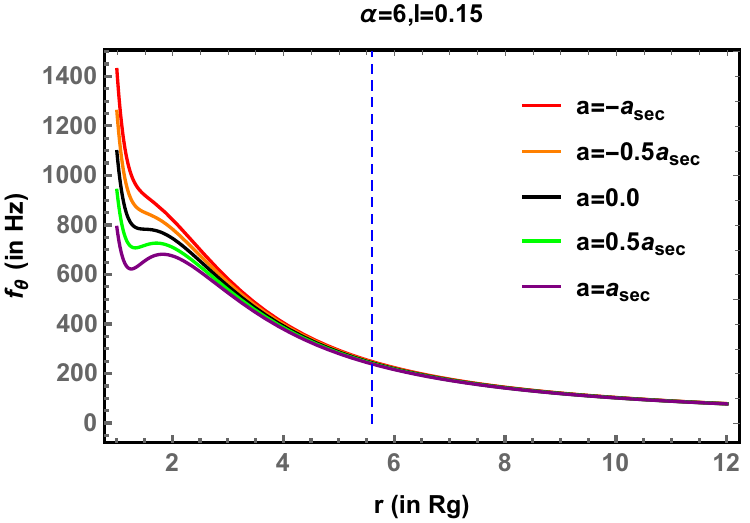}
	}
     \subfloat[\label{Fig10e}]{
		\includegraphics[scale=0.43]{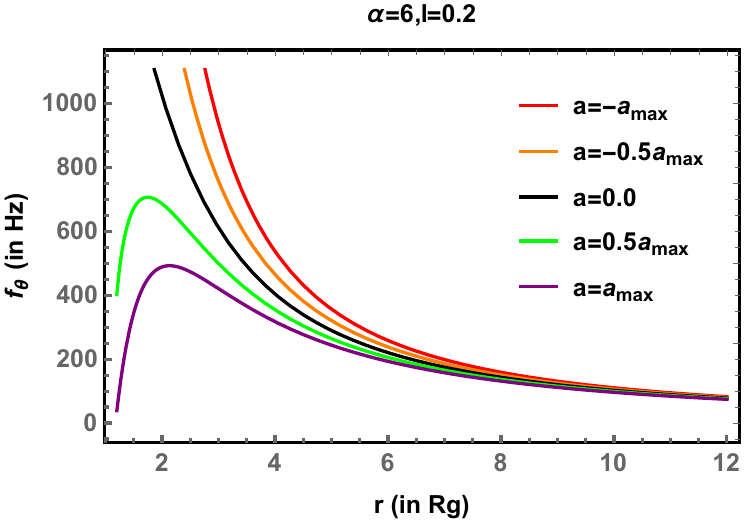}
	}
     \subfloat[\label{Fig10f}]{
		\includegraphics[scale=0.43]{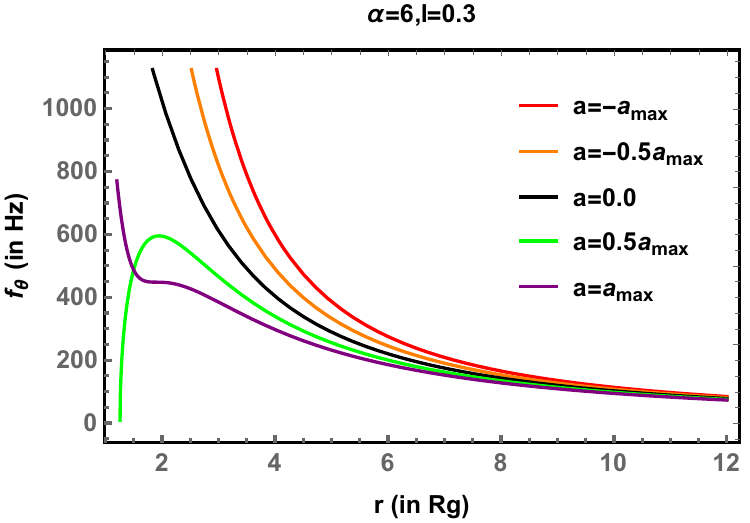}
	}\hfill
    
    \subfloat[\label{Fig10g}]{
        \includegraphics[scale=0.43]{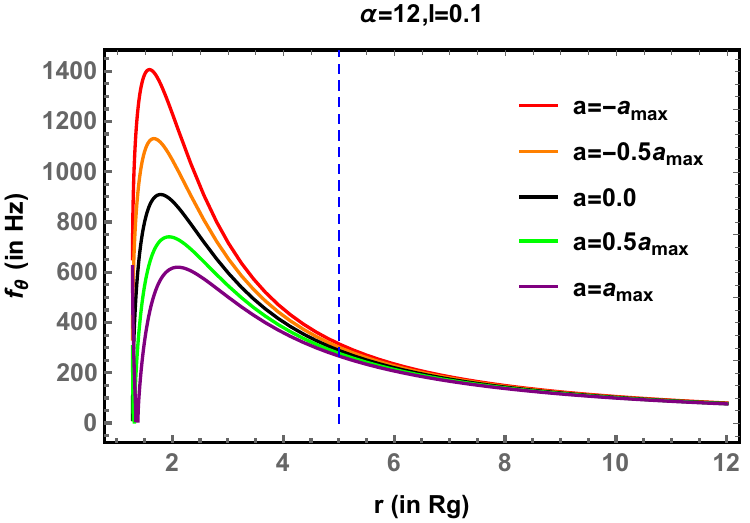}
    }
    \subfloat[\label{Fig10h}]{
        \includegraphics[scale=0.43]{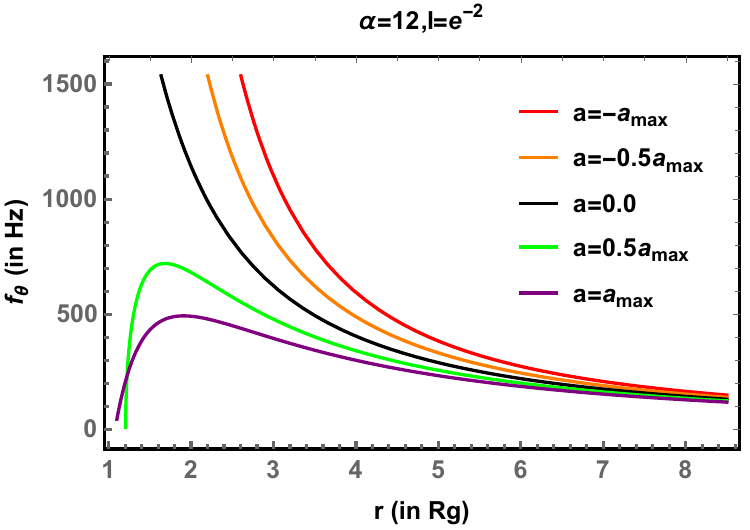}
    }
    \subfloat[\label{Fig10i}]{
        \includegraphics[scale=0.43]{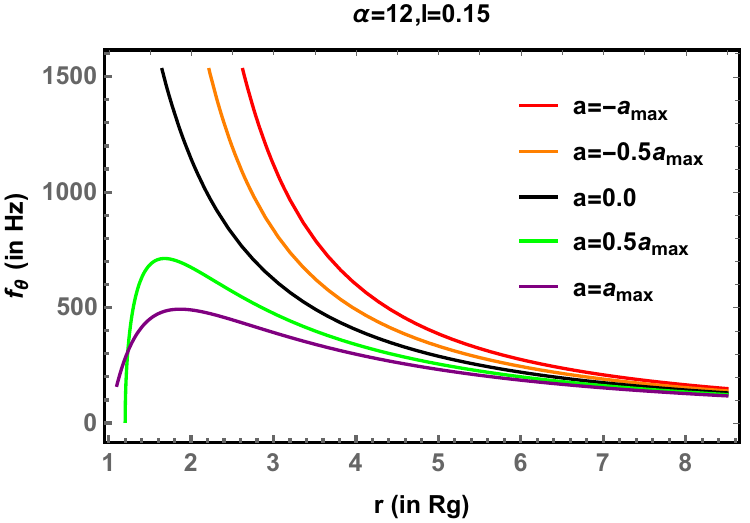}
    }

    \caption{
    The above figures illustrate the dependence of $f_\theta$ on $l$ for a fixed $\alpha$.
    In particular, they show the variation of $f_\theta$ (a) with $r$, for $\alpha=2, l=0.3$, (b) with $r$, for $\alpha=2, l=0.6$, (c) with $l$, for $\alpha=2$ at $r=6$, (d) with $r$, for $\alpha=6, l=0.15$, (e) with $r$, for $\alpha=6, l=0.2$, (f) with $r$, for $\alpha=6, l=0.3$, (g) with $r$, for $\alpha=12, l=0.1$, (h) with $r$, for $\alpha=12, l=e^{-2}$, (i) with $r$, for $\alpha=12, l=0.15$, for different spins for a $M=10 M_\odot$ BH. In each subfigure, we have shown variation with respect to spins in the entire allowed range.}
    
    \label{ftheta2}
\end{figure}

To understand the sensitivity of $f_\theta$ on $l$ we consider \autoref{ftheta2}. Here, we have shown the variation of $f_\theta$ with $r$, keeping the deformation parameter $\alpha$ fixed and varying the length parameter $l$ in the allowed range of spin.  We note that for a given $l$, when $\alpha l<1$ (e.g. \ref{Fig10a}, \ref{Fig10d}), the hairy BH has lower $f_\theta$  compared to the Kerr case as long as we are outside $r>r_{ms}$, for reasons discussed earlier. But as $l$ increases for a given $\alpha$, the Kerr scenario is recovered. Moreover, $f_\theta$ is almost insensitive to variation with respect to $l$ for a given $\alpha$ (as elucidated in \ref{Fig10c} ), as $l\to 2/\alpha$. We find that the behavior for $\alpha \lesssim 2$ is qualitatively similar to the $\alpha \sim 2$ case, hence those panels are not shown separately. Overall, departure from the Kerr scenario is observed in the behavior of $f_\theta$ with $r$ if the particle is in the vicinity of $r_{ms}$ and the hairy BH has $\alpha l<1$.

 It is important to note that the hairy BH solution we consider is fully analytic, which allows for a systematic and controlled study of their properties. Given the paramount importance of hairy black hole solutions in the context of the no-hair theorem, and their potential relevance for astrophysical applications, it is both natural and worthwhile to investigate their phenomenological implications in detail. In this work, we have carried out a comprehensive analysis across the full parameter space of the solution, and tried to constrain it using QPOs. Our primary goal is to explore whether hairy deviations can leave observable imprints in QPO data.
We now turn to the kinematic models proposed in the literature to explain the high-frequency QPOs (HFQPOs) observed in black hole and neutron star systems. HFQPOs typically appear in commensurable pairs, and notably, the ratio between the twin-peak HFQPOs is often found to be $3:2$. In these models, the predicted QPO frequencies are expressed as linear combinations of the three fundamental frequencies, $f_r$, $f_\theta$, and $f_\phi$, introduced earlier. Although most of these frameworks are designed primarily to account for HFQPOs, the Relativistic Precession Model (RPM) uniquely provides a unified mechanism capable of explaining both HFQPOs and low-frequency QPOs (LFQPOs) simultaneously. We investigate the viability of these models and the hairy Kerr BH scenario in explaining the observed the HFQPOs in BH sources.

\newpage
\section{Assessment of Model-Predicted HFQPOs in Light of Observational Data}
\label{error-analysis}
In this section, we compare the HFQPO frequencies predicted by different kinematic models with the corresponding observational measurements. To evaluate how well each model fits the data, we perform an error analysis. The $\chi^2$ minimization technique is employed to quantify the discrepancy between the theoretical estimates and the observed QPO frequencies. Since there are two parameters, $l$ and $\alpha$, that quantify the deviation of the hairy black holes from the Kerr scenario, the $\chi^2$ function is defined as
\begin{eqnarray}
\chi_i^{2}(l,\alpha) =\frac{[f_{up1,i} - f_{1}(l,\alpha, a_{min}, M_{min}, r_{min})]^2}{\sigma_{f_{up1},i}^2} 
+  \frac{[f_{up2,i} - f_{2}(l,\alpha, a_{min}, M_{min}, r_{min})]^2}{\sigma_{f_{up2},i}^2}\,.
\label{chi}
\end{eqnarray}

\begin{table}[htpb]
    \centering
\begin{adjustbox}{width=0.9\textwidth}
    \begin{tabularx}{\textwidth}{|X|X|X|X|X|}
    \hline
    \textbf{Source} & Mass \boldmath$(M_{\boldsymbol{\odot}})$ & $f_{up1} \pm \Delta f_{up1}$ \textbf{(Hz)} & $f_{up2} \pm \Delta f_{up2}$ \textbf{(Hz)} & $f_{up3} \pm \Delta f_{up3}$ \textbf{(Hz)}\\
    \hline
    \hline
    GRO J1655-40 & $5.4 \pm 0.3$ \cite{beer2002quiescent} & $441 \pm 2$ \cite{motta2014precise} & $298 \pm 4$ \cite{motta2014precise} & $17.3 \pm 0.1$ \cite{motta2014precise} \\
    \hline
    XTE J1550-564 & $9.1 \pm 0.61$\cite{Orosz:2011ki} & $276 \pm 3$ & $184 \pm 5$ & - \\
    \hline
    GRS 1915+105 & $12.4^{+2.0}_{-1.8}$ \cite{Reid:2014ywa} & $168 \pm 3$ & $113 \pm 5$ & - \\
    \hline
    H 1743+322 & $8.0 - 14.07$ \cite{Pei:2016kka,Bhattacharjee:2017rbl,Petri:2008jc} & $242 \pm 3$ & $166 \pm 5$ & - \\
    \hline
    Sgr A* & $(3.5-4.9)\times 10^6$\cite{Ghez:2008ms,Gillessen:2008qv} & $(1.445 \pm 0.16)\times10^{-3}$\cite{Torok:2004xs,Stuchlik:2008fy} & $(0.886 \pm 0.04)\times10^{-3}$\cite{Torok:2004xs,Stuchlik:2008fy} & - \\
    \hline
    XTE J1859+226 & $7.85 \pm 0.46$\cite{motta2022black} & $227.5^{+2.1}_{-2.4}$\cite{motta2022black} & $128.6^{+1.6}_{-1.8}$\cite{motta2022black} & $3.65 \pm 0.01$ \\
    \hline
    \end{tabularx}
\end{adjustbox}
\caption{BH Sources where High-Frequency Quasi-Periodic Oscillations (HFQPOs) are observed.}
\label{t1}
\end{table}

This expression quantifies the deviation between the model predicted and the observed HFQPO frequencies. 
\autoref{t1} enlists the BH sources which exhibits HFQPOs in their power spectrum.
In this context, $f_{\mathrm{up1},i}$ and $f_{\mathrm{up2},i}$ denote the upper and lower HFQPO frequencies for the $i$-th source, as listed in  \autoref{t1}. The corresponding uncertainties, $\sigma_{f_{\mathrm{up1},i}}$ and $\sigma_{f_{\mathrm{up2},i}}$, represent the observational errors in these measured frequencies, also reported in  \autoref{t1}. The theoretical frequencies $f_1$ and $f_2$ for each model are calculated from $f_\phi$, $f_\theta$, and $f_r$ following the definitions provided for Relativistic Precession model (RPM) where $f_1=f_\phi$, $f_2=f_\phi-f_r$, $f_3=f_\phi-f_\theta$ and Tidal Disruption model (TDM) where $f_1=f_\phi+f_r$, $f_2=f_\phi$.\\

The steps of our analysis are outlined below:
\begin{itemize}
    \item  First, we choose one of the models between RPM and TDM, in which the frequencies $f_{1}$, $f_{2}$, and, when applicable, $f_{3}$ depend on $l$, $\alpha$, $a$, $M$, and the emission radius $r_{\mathrm{cm}}$ associated with the QPO generation.
    \item We then choose a black hole source from  \autoref{t1} whose mass range is known, enabling us to compute the theoretical frequencies $f_{1}$, $f_{2}$, and, when applicable, $f_{3}$ across that mass interval.

    \item We fix $l$ and  $\alpha$, with $\alpha l\leq 2$, such that the strong energy condition is satisfied. For this fixed $l$ and $\alpha$, the spin parameter $a$ is allowed to vary within 
    $-a_{\max} \leq a \leq a_{\max}$ or $-a_{\sec} \leq a \leq a_{\sec}$ (see \autoref{S2}), guaranteeing that 
    the event horizon remains real and positive.

    \item The black hole mass $M$ is varied within 
    $(M_{0} - \Delta M) \leq M \leq (M_{0} + \Delta M)$, where $M_{0}$ and $\Delta M$ denote 
    the central mass and its observational uncertainty as listed in \autoref{t1}.

    \item Our main task is to evaluate the frequencies $f_{1}$ and $f_{2}$. 
    For each value of $M$, we allow the emission radius $r_{\mathrm{cm}}$ to vary within the range  
    $r_{\mathrm{ms}}(a,l,\alpha) \leq r_{\mathrm{cm}} \leq r_{\mathrm{ms}}(a,l,\alpha) + 20 R_{g}$, where $R_{g} = GM/c^{2}$ is the gravitational radius and $r_{\mathrm{ms}}$ is the radius of the marginally stable orbit.
    \item For every combination of $M$, $a$, and $r_{\mathrm{cm}}$ at the chosen value of $l$, $\alpha$, 
    we compute the corresponding $\chi^{2}_{i}$ for the selected source. 
    The set of parameters $(M, a, r_{\mathrm{cm}})$ that yields the minimum $\chi^{2}_{i}$ provides the 
    best-fit estimates of the mass, spin, and emission radius for that particular $l$ and $\alpha$. 
    These optimal values are denoted as $M_{\min}$, $a_{\min}$, and $r_{\mathrm{em},\min}$  in \autoref{chi}.

    \item We then repeat this procedure for multiple choices of $l$, $\alpha$, allowing us to obtain the variation 
    of $\chi^{2}$ as a function of $l$, $\alpha$ for the selected black hole source.

    \item The same analysis is performed for all black hole sources listed in  \autoref{t1} for the chosen model.

    \item Finally, the procedure is repeated for the other model. The minimum value of $\chi^{2}$ identifies the most favored $l$, $\alpha$, together with the corresponding 
    values of $M_{\min}$ and $a_{\min}$ for each black hole source within the framework of that model.

\end{itemize}
We now provide an overview of the kinematic models proposed to explain HFQPOs observed in the power spectra of certain BHs. By comparing the model predicted frequencies with the observed QPO data, we assess the effectiveness of each model. \autoref{t1} lists several BH sources that exhibit HFQPOs, along with their independently measured masses.  The characteristic frequencies associated with HFQPOs scale inversely with the mass of the black hole (see \autoref{S3}). As a result, stellar mass BHs show HFQPOs in the range of a few hundred hertz, whereas supermassive BHs such as Sgr A* produce oscillations in the millihertz range. From \autoref{t1}, it is clear that HFQPOs often appear as pairs of commensurate frequencies, typically denoted by $f_{\mathrm{up1}}$ and $f_{\mathrm{up2}}$, with a characteristic ratio close to $3{:}2$. Since these frequencies correspond to combinations of the three fundamental frequencies $(f_{r}, f_{\theta}, f_{\phi})$, they depend mainly on the spacetime geometry near the BH and are far less sensitive to the detailed accretion physics. Consequently, HFQPOs provide a relatively clean probe of the underlying metric compared to other observational diagnostics such as the continuum spectrum or the iron line profile.

 Kinematic models interpret QPOs as a direct consequence of the motion of matter in the accretion disk around compact objects. They mainly relate the QPOs to the motion of accreting blobs rather than the forcing or dynamical terms, e.g. the complicated magnetohydrodynamic or radiative processes within the disk, that brings about such a motion \cite{Yagi:2016jm}.
The oscillations are therefore attributed to fundamental dynamical motions - such as orbital rotation, radial epicyclic oscillations, and vertical epicyclic oscillations - as well as relativistic precession effects induced by strong gravitational fields. In their simplest formulation, these models assume neutral test particle motion governed purely by the gravitational field. However, accretion environments are composed of ionized plasma and are often threaded by magnetic fields associated with the accretion disk. In such settings, the dynamics of particles may be influenced not only by gravity but also by electromagnetic interactions, which can modify the effective motion and shift the characteristic frequencies. Extensions of kinematic models that incorporate these effects have been explored in the literature \cite{Stuchlik:2020rls}, leading to a richer phenomenology of QPOs.
At the same time, the inclusion of electromagnetic interactions introduces additional parameters, such as the magnetic field strength and the charge-to-mass ratio of the particles, which can lead to degeneracies with purely gravitational modifications of the spacetime. In the present work, our focus is on isolating the role of the underlying geometry, in particular the impact of the hairy parameters $(\alpha, l)$ on the fundamental frequencies. For this reason, we restrict our analysis to the standard and widely used framework of neutral test particle motion, which also facilitates direct comparison with a large body of existing QPO-based studies.\\
 Resonance-based interpretations, including linear and nonlinear resonance scenarios, make use of the same fundamental orbital and epicyclic frequencies that underlie kinematic models, but attribute the observed QPOs to resonant interactions between these modes \cite{2016A&A...586A.130S}. In this sense, kinematic and resonance models are closely related at the level of the underlying frequency structure, differing primarily in their physical interpretation of the observed signals.
We acknowledge that these frequencies play a central role in models based on oscillating accretion tori, where global disk oscillation modes can give rise to QPOs \cite{2003MNRAS.344L..37R,Rezzolla:2003zy,Abramowicz:2003xyz,kluzniak2001strong}, further emphasizing the fundamental importance of epicyclic motion in strong gravity environments. 
In the present work however, we adopt a kinematic approach which mainly associate the QPOs to the motion of accreting blobs rather than the forcing or dynamical terms that brings about such a motion. Thus, the observed QPOs are directly associated with the fundamental frequencies, as in the relativistic precession and the tidal disruption models, allowing for a direct mapping between observations and spacetime parameters. At the same time, we fully recognize the importance of resonance-based interpretations. A dedicated and more detailed investigation of resonance models in the same spacetime background is currently in progress and will be presented in a forthcoming companion work.

\subsection{Relativistic Precession Model}
\label{S4.2}
The Relativistic Precession Model (RPM) is one of the most influential kinematic frameworks 
used to interpret HFQPOs in systems containing compact objects such as black holes and neutron stars\cite{Stella:1998mqe,stella1997lense,stella1999correlations}. Originally developed to explain the 
twin-peak HFQPOs observed in neutron star binaries, the model was later extended to black hole 
sources once similar oscillatory features were detected in their power spectra.
In the RPM framework, the observed QPO frequencies are directly associated with the relativistic 
orbital motion of matter in the inner regions of the accretion disk. The model identifies three 
fundamental frequencies arising from geodesic motion in strong gravity:

\begin{itemize}
    \item \textbf{Azimuthal (orbital) frequency} $f_{\phi}$: the basic frequency corresponding to 
    circular motion of matter around the compact object.

    \item \textbf{Periastron precession frequency} $f_{\phi} - f_{r}$: produced by the relativistic 
    advance of the periastron due to spacetime curvature. This,  alongside the orbital frequency provides a natural 
    explanation for the observed twin HFQPO peaks.

    \item \textbf{Nodal (Lense Thirring) precession frequency} $f_{\phi} - f_{\theta}$: arising 
    from the frame-dragging effect induced by the spin of the compact object. This frequency 
    is used to account for low-frequency QPOs such as that observed in GRO~J1655--40.
\end{itemize}

A central assumption of the RPM is that all QPOs originate from the same radial location in the 
accretion disk, typically close to the innermost stable circular orbit (ISCO). Under this assumption, the three frequencies above provide a self-consistent and physically motivated scheme for interpreting both high- and low-frequency QPOs within a unified model. The strength of this model lies in its ability to connect the observed QPO frequencies to the geometry 
of the spacetime. Since the model depends only on the metric parameters and the emission radius, it serves as a 
powerful probe of strong-gravity effects, particularly periastron advance and frame-dragging in the vicinity of compact objects. Now, we discuss our result related to the RPM.\\
\\
\textbf{1.\underline{\ GRO J1655-40}}\\
\vspace*{0.2cm}
\begin{figure}[hbt!]
		\centering
		\subfloat[]{\includegraphics[width=0.41\linewidth]{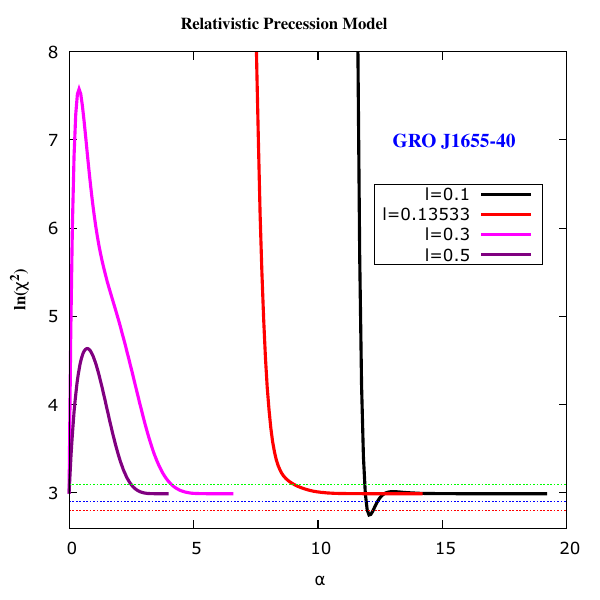}\label{RPMBH1All1}}
		\hfil
		\subfloat[]{\includegraphics[width=0.41\linewidth]{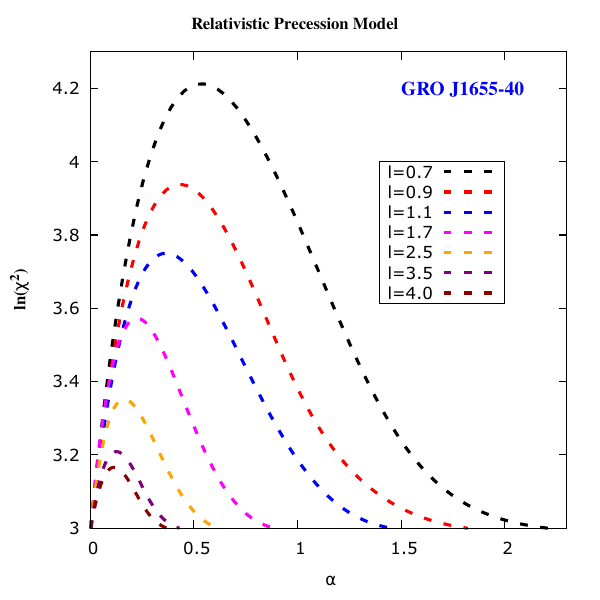}\label{RPMBH1All2}}
		\caption{The above figure shows the variation of ln$\chi^2$ with $\alpha$ for different choices of $l$ when the RPM model is used to explain HFQPOs and LFQPOs of GRO J1655-40. The red, blue, and green dotted lines are associated with the 1-$\sigma$, 2-$\sigma$, and 3-$\sigma$ confidence intervals, which correspond to $\chi^2_{min}+1$, $\chi^2_{min}+2.71$, and $\chi^2_{min}+6.63$ respectively.}
        \label{RPMGRO}
\end{figure}

In \autoref{RPMBH1All1} and \autoref{RPMBH1All2} we show the variation of ln$\chi^2$ with $\alpha$ for different choices of $l$ when the RPM is used to explain the twin-peak HFQPOs and the LFQPO of the source GRO J1655-40. We note that for $\alpha=0$ and $\alpha=\alpha_{max}=2/l$, the errors are the same, as both represent the Kerr case. While we have shown only certain values of $l$ in \ref{RPMBH1All1} and \ref{RPMBH1All2}, we have considered $0.1\leq l \leq 4$ in intervals of 0.2 for our analysis. 
For each $l$ we have varied $\alpha$ between $\alpha_{0}\leq\alpha\leq  \alpha_{max}=2/l$ in intervals of 0.1 when $l\geq 1.7$ and in intervals of 0.2 when $0.1\leq l <1.7$. This applies for all the BHs considered here and for both the models.
Note that, $\alpha_0\sim \alpha_{crit}$ for $l<0.3$ and $\sim 0$ otherwise (see \autoref{S2}). We note from \autoref{RPMGRO} that ln$\chi^2$ attains a minima near $l\sim 0.1, ~\alpha\sim 12$. The red, blue and green dotted lines are associated with the 1-$\sigma$, 2-$\sigma$ and 3-$\sigma$ confidence intervals, which correspond to $\chi^2_{min}+1$, $\chi^2_{min}+2.71$ and $\chi^2_{min}+6.63$ respectively \cite{avni1976energy}.
It is to be noted that for $l=0.1$ and $l=e^{-2}\approx 0.13533$, we have not started from $\alpha=0$ (as shown in \autoref{RPMBH1All1}), because SEC is satisfied from $\alpha=\alpha_{crit}\approx 11.4$ (for $l=0.1$) and $\alpha=\alpha_{crit}\approx 7.4$ (for $l=e^{-2}$) respectively, to $\alpha=\alpha_{max}\approx 2/l$ (see \autoref{S2}).
From \autoref{RPMBH1All1}, we notice a sharp decrease in ln$\chi^2$ for $l=0.1$ which attains a minima, increases slightly and then saturates.
For $l=0$, SEC is not satisfied for any $\alpha$, therefore we have excluded this from our consideration.
Since $\alpha\sim 0$ and $\alpha l\sim 2$ results in the Kerr metric, the figure reveals that the Kerr scenario is ruled out outside 2-$\sigma$ by the RPM model. Although it appears from \autoref{RPMGRO} that GR is allowed within the 3-$\sigma$ interval, if we had taken more points near $l\sim0.1~\alpha\sim 12$, then we would have noticed that the magnitude of $\chi^2_{min}\sim 0$, i.e., near  $l\sim0.1~\alpha\sim 12$, $\chi^2$ attains a very sharp minima. Thus, even when the 3-$\sigma$ interval is considered GR is excluded. 
Thus, if RPM is used to explain the QPOs of GRO J1655-40, the hairy Kerr BH scenario, particularly with $l\sim 0.1, ~\alpha\sim 12$ seems to be observationally favored compared to the general relativistic scenario. 
We have further confirmed this by Markov Chain Monte Carlo (MCMC) simulation in the next section.  However, this result is model dependent and should not be considered as a confirmatory signature of hairy Kerr scenario for this source.\\
\\
\textbf{2.\underline{\ XTE J1550-564}}\\
\begin{figure}[hbt!]
		\centering
		\subfloat[]{\includegraphics[width=0.41\linewidth]{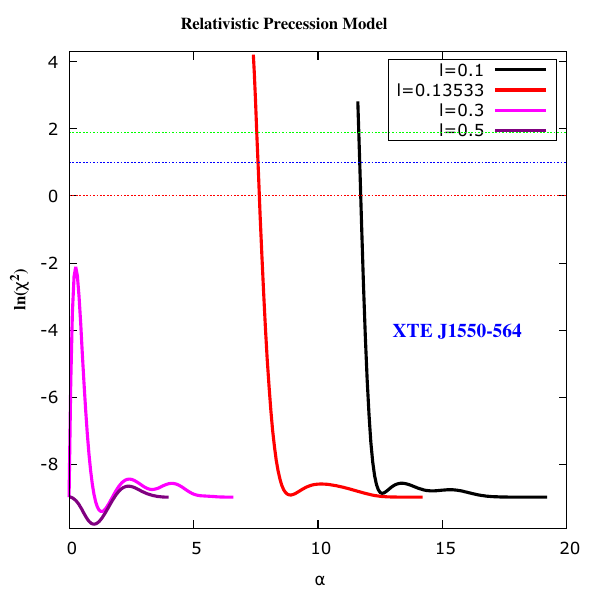}\label{RPMBH2All1}}
		\hfil
		\subfloat[]{\includegraphics[width=0.41\linewidth]{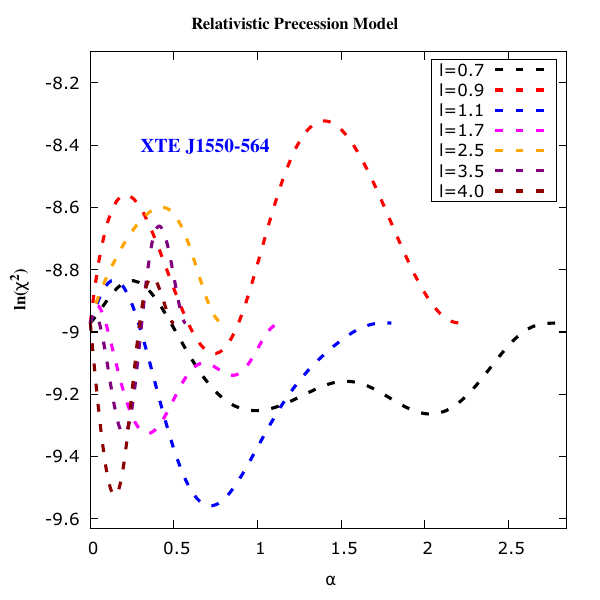}\label{RPMBH2All2}}
		\caption{The above figure shows the variation of ln$\chi^2$ with $\alpha$ for different choices of $l$ when the RPM model is used to explain the HFQPOs of XTE J1550-564}
        \label{XTERPM}
\end{figure}

\vspace*{0.2cm}
When RPM is used to explain the twin-peak HFQPO data of XTE J1550-564, from \autoref{XTERPM}, we note that the $\chi^2$ assumes very high values for $l\leq e^{-2}$ and $\alpha\simeq \alpha_{crit}$. Thus, the hairy BH scenario with such values of $l$, $\alpha$ are ruled out outside 3-$\sigma$ for the source XTE J1550-564. Once again, the red, blue and green dotted lines represent the 1-$\sigma$, 2-$\sigma$, and 3-$\sigma$ confidence intervals, respectively, from $\chi^2_{min}$. The $\chi^2$ does not attain a sharp minima for any particular $l,\alpha$, but is generally $\sim 0$ for most choices of $l, \alpha$ except the ones mentioned above.
Thus, the HFQPO data of XTE J1550-564 cannot distinguish the Kerr scenario from the hairy-Kerr scenario with $l>e^{-2}$ or $l\leq e^{-2}$ but $\alpha>\alpha_{crit}$. 
\\
\\
\textbf{3. \underline{\ \ GRS 1915+105 \& H 1743+322}}\\
\begin{figure}[hbt!]
		\centering
		\subfloat[]{\includegraphics[width=0.41\linewidth]{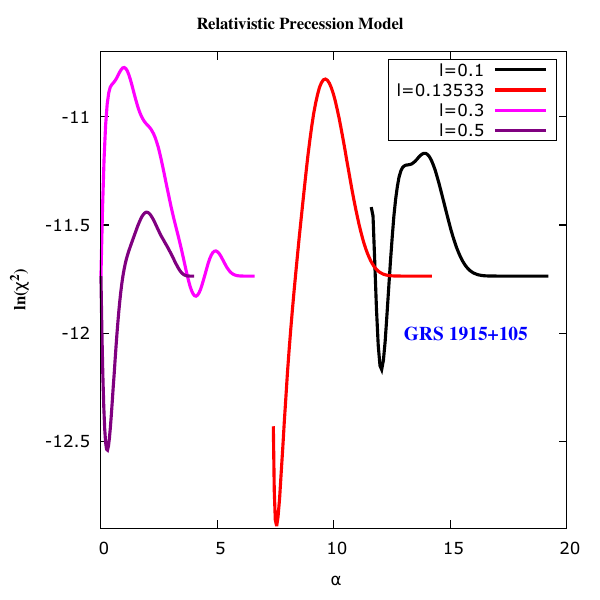}\label{RPMBH3All1}}
		\hfil
		\subfloat[]{\includegraphics[width=0.41\linewidth]{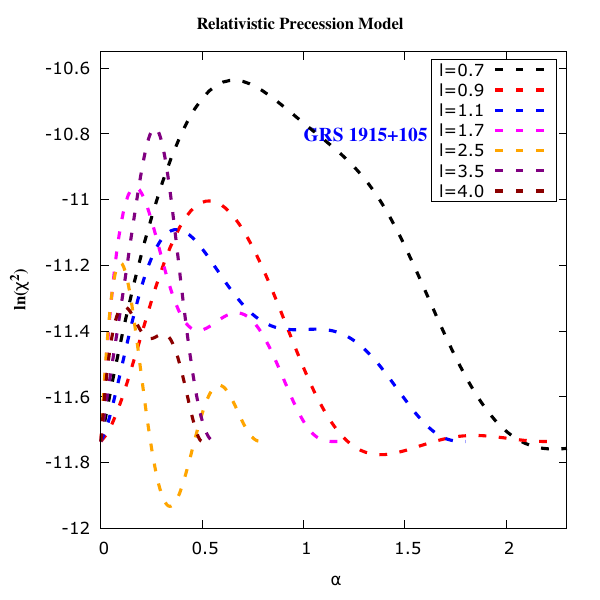}\label{RPMBH3All2}}
		\caption{The above figure shows the variation of ln$\chi^2$ with $\alpha$ for different choices of $l$ when the RPM model is used to explain the HFQPOs of GRS 1915+105. }
        \label{GRSRPM}
\end{figure}

\begin{figure}[hbt!]
		\centering
		\subfloat[]{\includegraphics[width=0.41\linewidth]{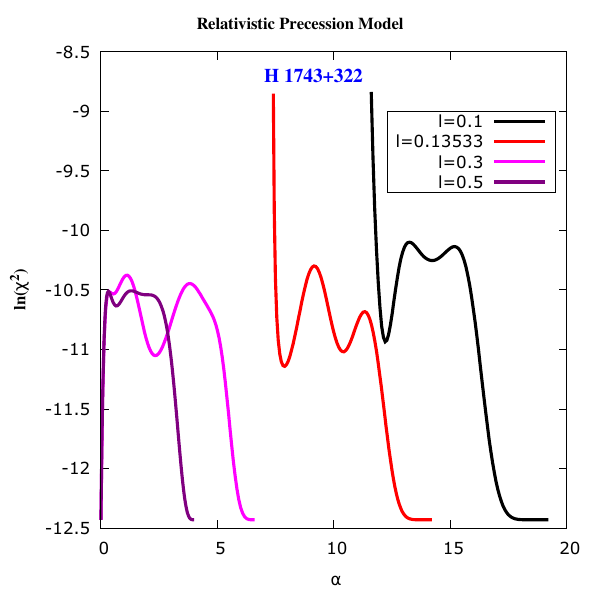}\label{RPMBH4All1}}
		\hfil
		\subfloat[]{\includegraphics[width=0.41\linewidth]{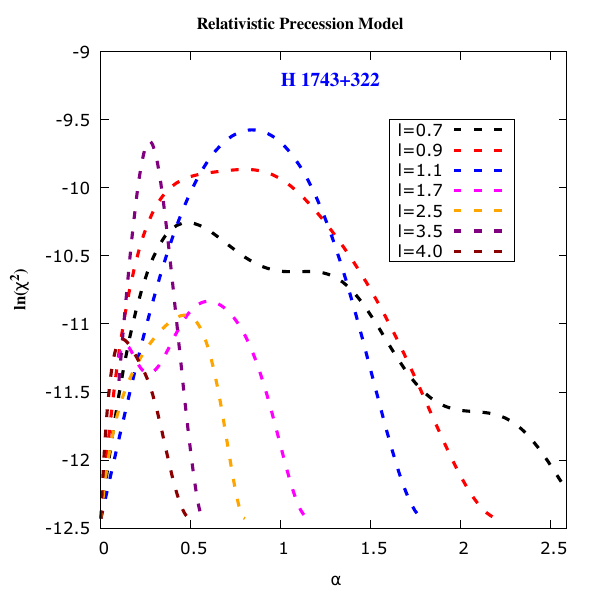}\label{RPMBH4All2}}
		\caption{The above figure shows the variation of ln$\chi^2$ with $\alpha$ for different choices of $l$ when the RPM model is used to explain the HFQPOs of H 1743-322}
        \label{HRPM}
\end{figure}

When RPM is used to explain the twin-peak HFQPO data of GRS 1915+105 and H 1743+322, from \autoref{GRSRPM} and \autoref{HRPM}, we note that $\chi^2\sim 0$ for all allowed values of $l, \alpha$. Once again the errors remain essentially the same at $\alpha = 0$ and $\alpha = \alpha_{\max}$, with only minor variations occurring in the intermediate region. These variations are very weakly sensitive to $l, \alpha$, and therefore do not significantly constrain the parameter space. Thus the HFQPO data of GRS 1915+105 and H 1743-322 fail to differentiate between the Kerr and the hairy Kerr scenario.
\vspace{1cm}
\\
\\
\textbf{4. \underline{Sgr A*}}\\
\begin{figure}[hbt!]
		\centering
		\subfloat[]{\includegraphics[width=0.41\linewidth]{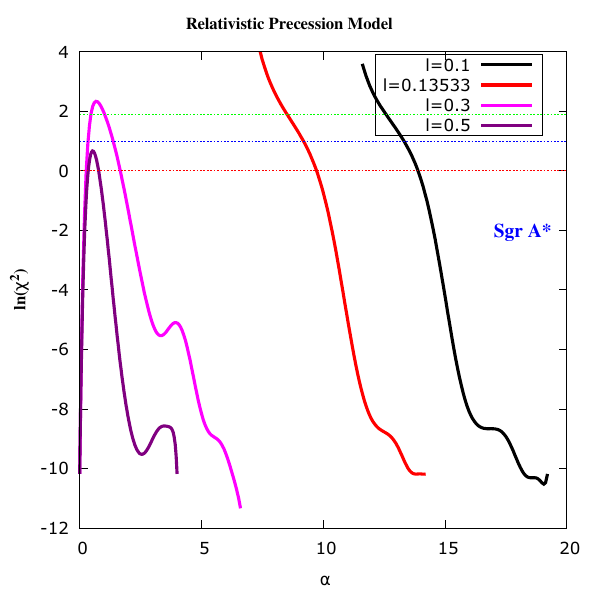}\label{RPMBH5All1}}
		\hfil
		\subfloat[]{\includegraphics[width=0.41\linewidth]{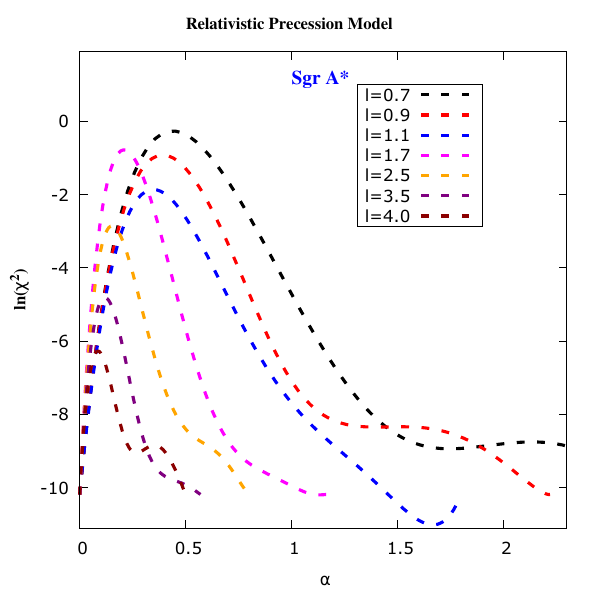}\label{RPMBH5All2}}
		\caption{The above figure shows the variation of ln$\chi^2$ with $\alpha$ for different choices of $l$ when the RPM model is used to explain the HFQPOs of Sgr A*. The red, blue and green dotted lines are associated with the 1-$\sigma$, 2-$\sigma$ and 3-$\sigma$ confidence intervals, which correspond to $\chi^2_{min}+1$, $\chi^2_{min}+2.71$ and $\chi^2_{min}+6.63$ respectively. Here $\chi_{min} \sim 0$. }
        \label{RPMSgr}
\end{figure}

\vspace{0.42cm}
Next, we discuss our result for supermassive BH Sgr A*. From \ref{RPMBH5All1} and \autoref{RPMBH5All2} we note that $\chi^2_{min}\sim 0$. For $l\lesssim e^{-2}$, $\alpha$ close to $\alpha_{crit}$ are ruled out outside 3-$\sigma$ (green dotted line in \autoref{RPMSgr}). For $l\sim e^{-2}$ and $l\sim 0.1$, the allowed range of $\alpha$ within 1-$\sigma$ corresponds to $10\lesssim \alpha \lesssim \alpha_{max}=2/l$ and  $15\lesssim \alpha \lesssim \alpha_{max}=2/l$, respectively. For $0.3\lesssim l \lesssim 0.5$, intermediate values of $\alpha$ are observationally less favored compared to $\alpha\to 0$ or $\alpha\to \alpha_{max}$ (\ref{RPMBH5All1}). For $l\gtrsim 0.7$, all values of $\alpha$ between $0\lesssim \alpha \lesssim \alpha_{max}$ are allowed within 1-$\sigma$. Thus, in the context of the RPM model, the HFQPO data of Sgr A* are best described either by the Kerr model or by a hairy Kerr model with relatively large length parameter ($l \gtrsim 0.7$ for any allowed $\alpha$), or alternatively with smaller $l$ provided the combined deformation satisfies $\alpha l \gtrsim 1$.
\\
\\
\textbf{5. \underline{\ XTE J1859+226}}\\
\begin{figure}[hbt!]
		\centering
		\subfloat[]{\includegraphics[width=0.41\linewidth]{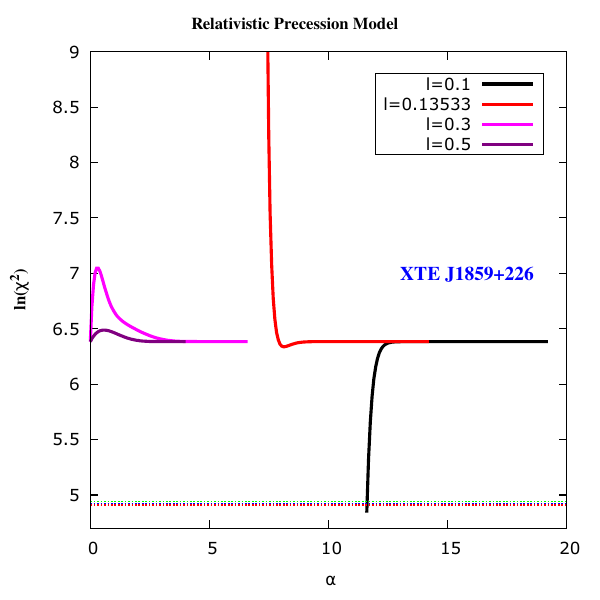}\label{RPMBH6All1}}
		\hfil
		\subfloat[]{\includegraphics[width=0.41\linewidth]{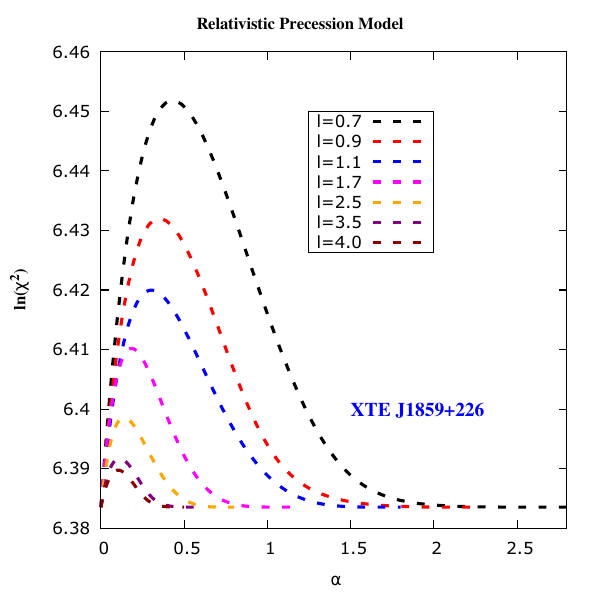}\label{RPMBH6All2}}
		\caption{The above figure shows the variation of ln$\chi^2$ with $\alpha$ for different choices of $l$ when the RPM model is used to explain the HFQPO data of XTE J1859+226. The red, blue and green dotted lines are associated with the 1-$\sigma$, 2-$\sigma$ and 3-$\sigma$ confidence intervals, which correspond to $\chi^2_{min}+1$, $\chi^2_{min}+2.71$ and $\chi^2_{min}+6.63$ respectively. Since the variation of $\chi^2$ with respect to $\alpha$ is substantial, the 3 confidence contours seem to overlap when plotted in the log scale. }
        \label{RPMXTE2}
\end{figure}

Now, we discuss our result related to XTE J1859+226 BH. For $l\geq 0.3$, $\chi^2$ assumes the same value for $\alpha \to 0$ or $\alpha\to \alpha_{max}$ as both represent the Kerr case.  
From \ref{RPMBH6All1} we note that $\chi^2$ attains a global minimum at $l\simeq 0.1$ and $\alpha\simeq \alpha_{crit}\approx 12$. 
Even the Kerr scenario is ruled out outside 3-$\sigma$ if RPM is used to explain the QPO data of XTE J1859+226.
For this black hole, the hairy BH scenario is more favored, the parameter space is highly constrained and  we will confirm this global minima by MCMC in the next section.  Again, this should not be considered as a confirmatory signature of hairy Kerr scenario for this source as this result is model dependent.

Therefore, considering the RPM model, GRO J1655-40 and XTE J1859-226 seem to favor the hairy Kerr scenario with $l\sim 0.1$ and $\alpha\sim 12$. For XTE J1550-564 and Sgr A*, lower $l$ with $0 <\alpha l\lesssim 1$ are ruled out, while for GRS 1915+105 and H 1743+322 the HFQPO data cannot distinguish the Kerr and the hairy Kerr scenario.

\subsubsection{Parameter Constraints from MCMC Analysis}
\label{MCMC}
In our earlier analysis, the model parameters were determined through a direct grid based search. In that approach, we scanned over the allowed ranges of the length parameter $l$, deformation parameter $\alpha$, the spin $a$, the black hole mass $M$, and the emission radius $r_{\rm cm}$, computing the corresponding $\chi^{2}$ for every point in the grid. The set of parameters yielding the lowest value of $\chi^{2}$ was taken as the best fit. A crucial requirement for this technique is that the grid must be fine enough to ensure that the global minimum is not overlooked. To confirm that our grid resolution is adequate, we compare these results with those from a Bayesian inference scheme that uses Markov Chain Monte Carlo (MCMC) sampling.

The posterior probability distribution for the parameter vector $\theta \equiv \{l,\alpha, a, M, r_{\rm cm}\}$ is written as
\begin{align}
P(\theta|D)=\frac{P(D|\theta)P(\theta)}{P(D)}    
\end{align}
where $P(\theta)$ is the prior on the parameters, $P(D|\theta)$ corresponds to the likelihood, and $P(D)$ is the evidence, which can be taken as unity since the data set is fixed \cite{Verde:2009tu}. For the parameters $l$, $\alpha$, $a$, and $r_{\rm cm}$, we adopt uniform priors across their allowed domains. For the mass parameter $M$, we choose Gaussian priors for GRO~J1655--40 and XTE~J1859+226, considering earlier independent measurements (\autoref{t1}).  As we could strongly constrain the parameter space for GRO J1655-40 and XTE J1859+226 from the grid search, here we present the corner plots for only these two black holes. The likelihood incorporates the contributions from the upper and lower HFQPOs and, where relevant, the LFQPO as well -- for example, GRO~J1655-40 displays a pair of HFQPOs along with a LFQPO, all of which are accounted for within the RPM framework. This likelihood is given as,
\begin{eqnarray}
\log \mathcal{L}=\log \mathcal{L}_{U_1} + \log \mathcal{L}_{U_2} + \log \mathcal{L}_{L}\,,
\label{4}
\end{eqnarray}
where 
\begin{eqnarray}
\log \mathcal{L}_{U_1} =-\frac{1}{2} \frac{\lbrace f_{\textrm{up1}{,i}}-f_1(l,\alpha,a,M,r_{\rm em}) \rbrace ^2}{\sigma_{up_{\rm 1},i}^2}\,,
\label{5}
\end{eqnarray}
\begin{eqnarray}
\log \mathcal{L}_{U_2} =-\frac{1}{2} \frac{\lbrace f_{\textrm{up2}{,i}}-f_2(l,\alpha,a,M,r_{\rm em}) \rbrace ^2}{\sigma_{up_{\rm 2},i}^2}\,,
\label{6}
\end{eqnarray}
\begin{eqnarray}
\log \mathcal{L}_{L} =-\frac{1}{2} \frac{\lbrace f_{\textrm{up3}{,i}}-f_3(l,\alpha,a,M,r_{\rm em}) \rbrace ^2}{\sigma_{up_{\rm 3},i}^2}\,.
\label{6}
\end{eqnarray}
\begin{figure}[hbt!]
		\centering
        \hspace{-1.4cm}
		\subfloat[]{\includegraphics[width=0.53\linewidth]{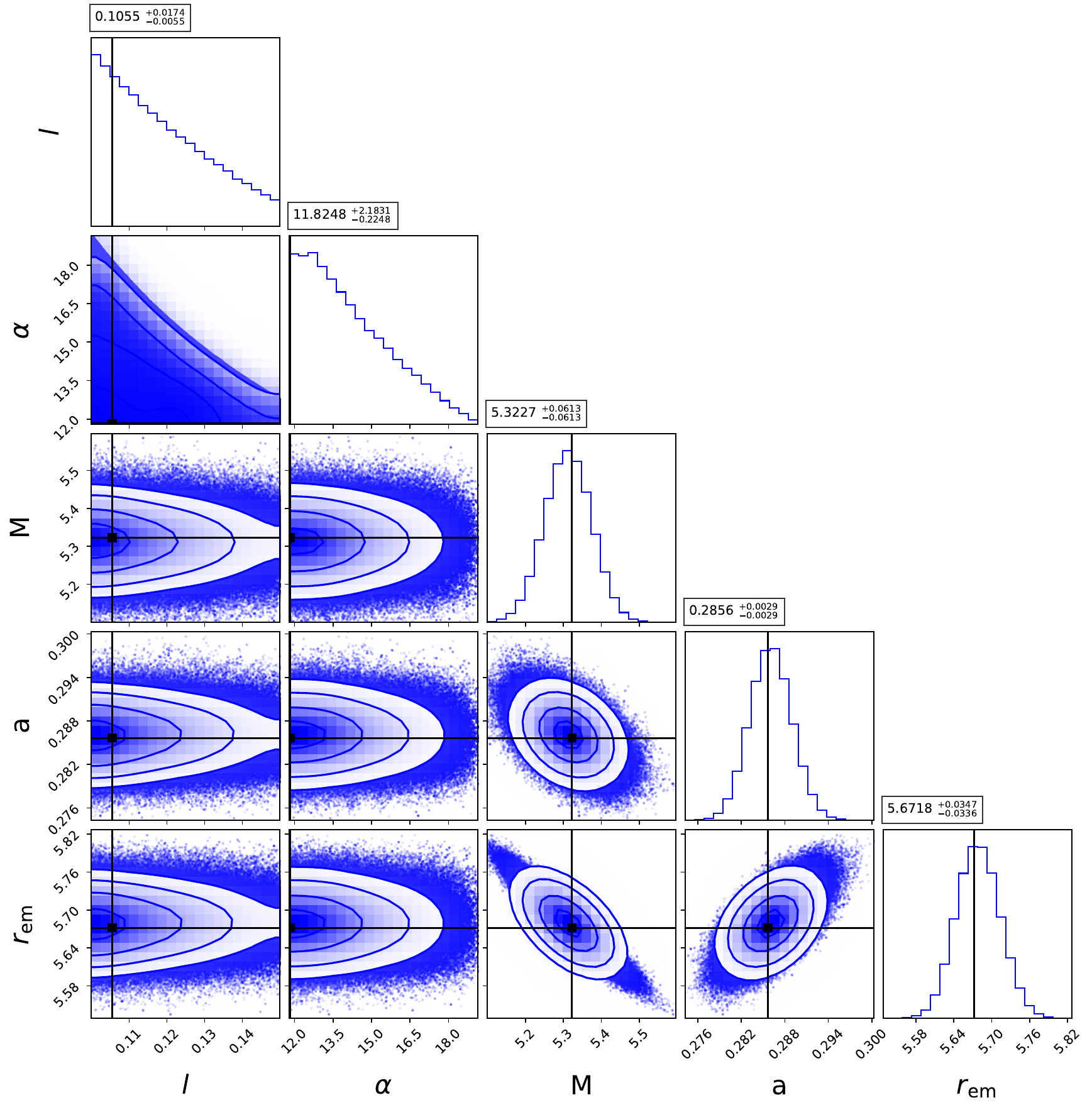}\label{GRORPM}}
		\hfil
		\subfloat[]{\includegraphics[width=0.53\linewidth]{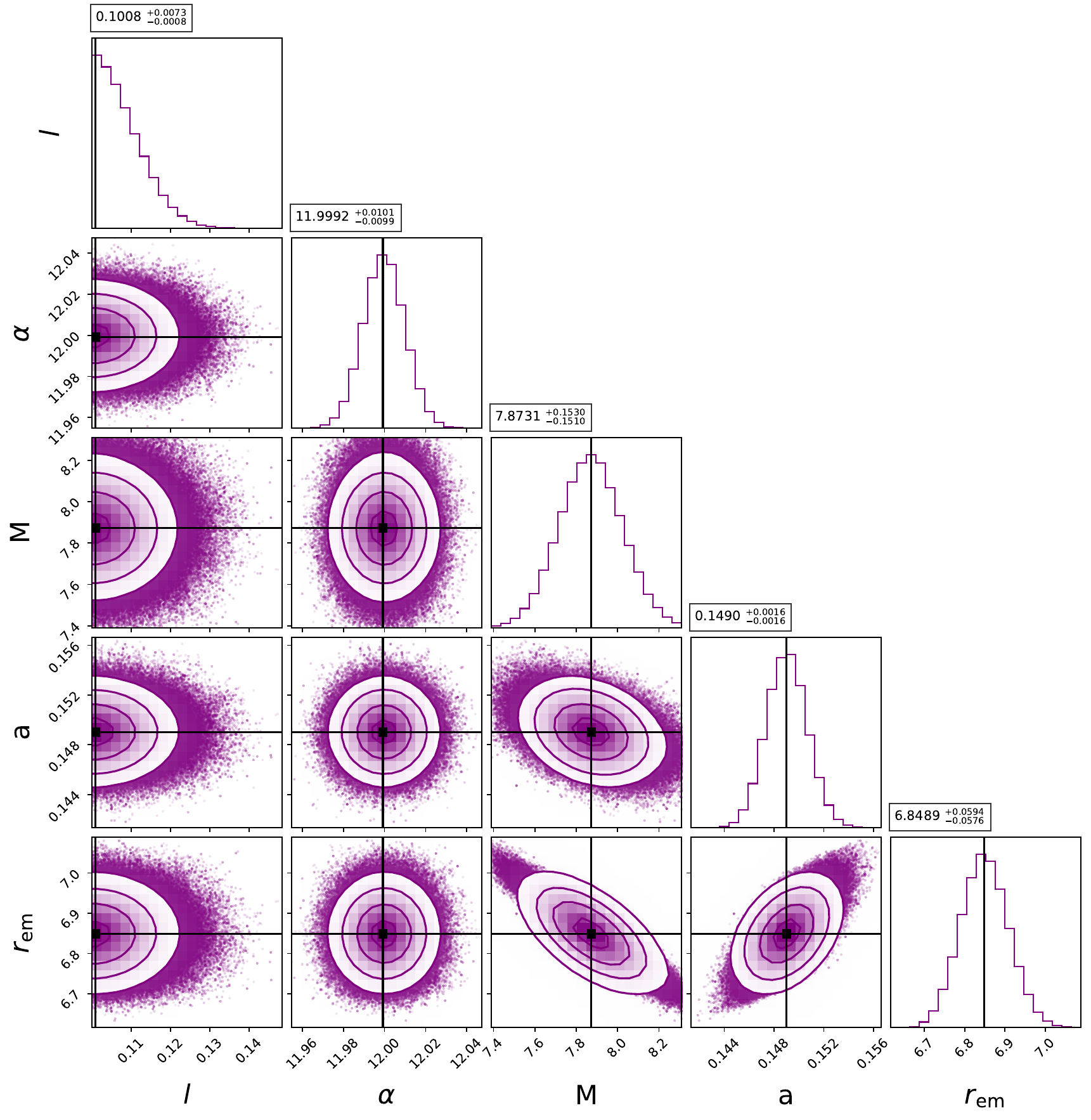}\label{XTE2RPM}}
		\caption{Constraints on the model parameters using the QPO data of GRO J1655-40 (blue) and XTE J1859+226 (purple) based on the RPM model. 
        }
        \label{cornerplot}
\end{figure}
From \autoref{GRORPM} and \autoref{XTE2RPM}, we confirm that for GRO J1655-40 and XTE J1859+226, the parameter estimates based on the grid-search and the MCMC methods are in agreement. However, $\chi_{min}\sim 0.0146$ for GRO J1655-40 and $\chi_{min}\sim 0.016$ for XTE J1859+226.

\subsection{Tidal Disruption Model}
\label{S4.1}
In the Tidal Disruption Model (TDM), the origin of HFQPOs is attributed to the interaction between the black hole’s tidal gravitational field and small, overdense patches of plasma that appear within the accretion disk\cite{Cadez:2008iv,Kostic:2009hp,Germana:2009ce}. These patches, or clumps, form because the disk is not perfectly uniform, and as they drift inward, the rapidly changing gravitational pull across their extent begins to stretch and tear them apart. During this process the clumps are elongated along their orbit, eventually taking on a ring-like shape. Once this structure forms, it undergoes characteristic oscillations influenced both by the geometry of the spacetime around the black hole and the dynamics of accretion flow.

Within this picture, the higher-frequency QPO is interpreted as the sum of the orbital motion and the radial oscillation of the distorted ring, represented as $f_{1} = f_{\phi} + f_{r}$. The lower HFQPO corresponds simply to the orbital motion, $f_{2} = f_{\phi}$. Because these frequencies are tied to the black hole mass, spin, charge, and the radius at which the emission originates ($r_{\mathrm{cm}}$), they can in principle be used to infer the properties of the system. As the clumps are disrupted, the infalling material heats up through viscous effects and the release of gravitational binding energy, giving rise to strong X-ray emission. In this way, the model connects the observed HFQPOs with the basic orbital dynamics of matter in the disk and with the tidal forces operating near the black hole.
\\
\\
\textbf{1. \underline{\ GRO J1655-40, XTE J1550-564, GRS 1915+105 \& H 1743-322}}\\
\begin{figure}[hbt!]
		\centering
		\subfloat[]{\includegraphics[width=0.41\linewidth]{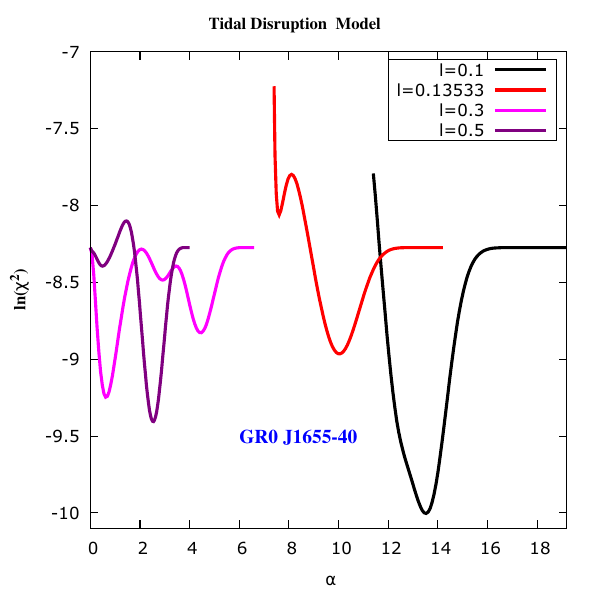}\label{TDMBH1All1}}
		\hfil
		\subfloat[]{\includegraphics[width=0.41\linewidth]{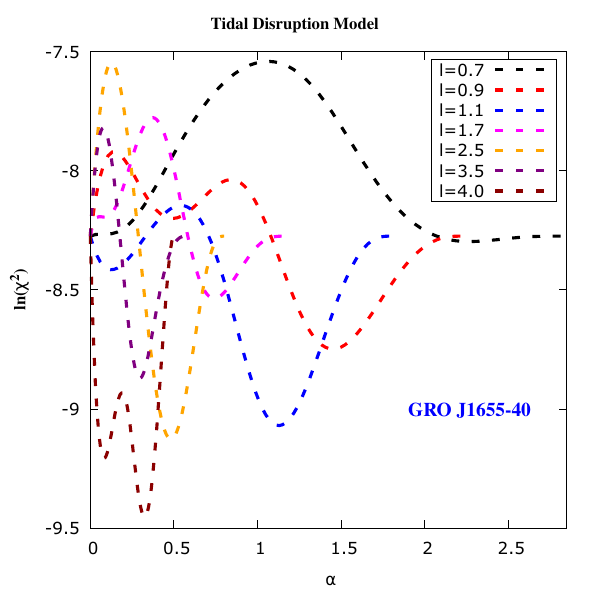}\label{TDMBH1All2}}
		\caption{The above figure shows the variation of ln$\chi^2$ with $\alpha$ for different choices of $l$ when the TDM model is used to explain the HFQPOs of GRO J1655-40}
        \label{TDMGRO}
\end{figure}

\vspace{0.15cm}

\begin{figure}[t!]
		\centering
		\subfloat[]{\includegraphics[width=0.41\linewidth]{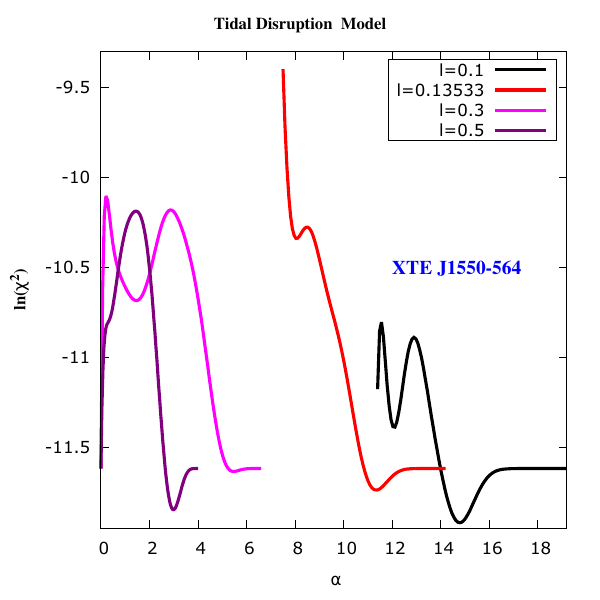}\label{TDMBH2All1}}
		\hfil
		\subfloat[]{\includegraphics[width=0.41\linewidth]{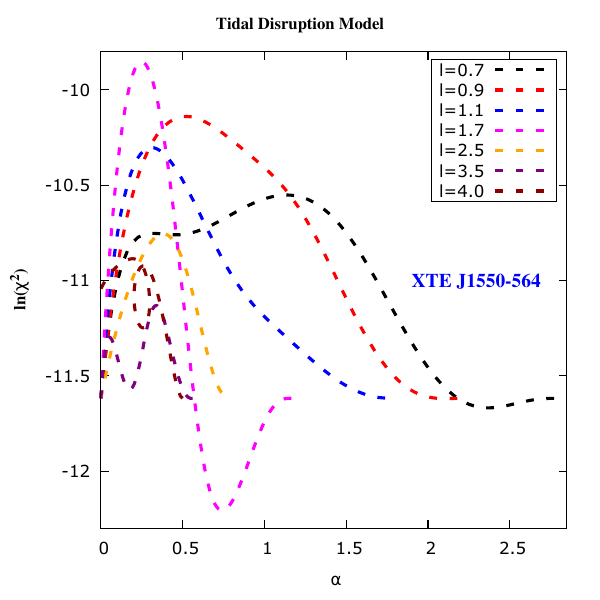}\label{TDMBH2All2}}
		\caption{The above figure shows the variation of ln$\chi^2$ with $\alpha$ for different choices of $l$ when the TDM model is used to explain the HFQPOs of XTE J1550-564}
        \label{TDMXTE}
\end{figure}

\begin{figure}[t!]
		\centering
		\subfloat[]{\includegraphics[width=0.41\linewidth]{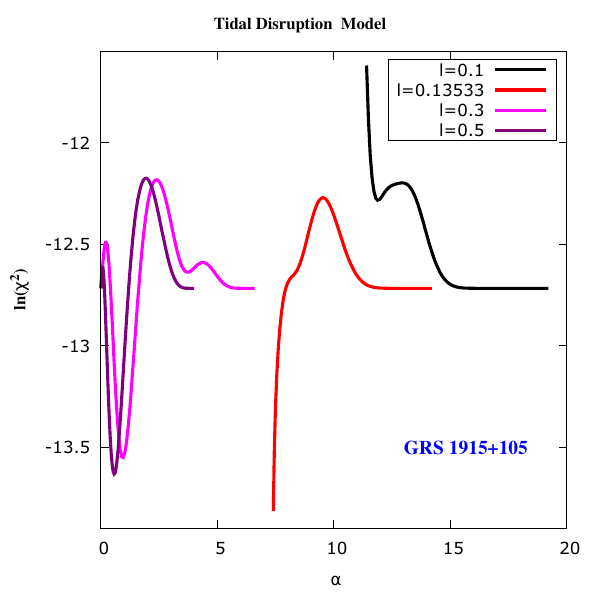}\label{TDMBH3All1}}
		\hfil
		\subfloat[]{\includegraphics[width=0.41\linewidth]{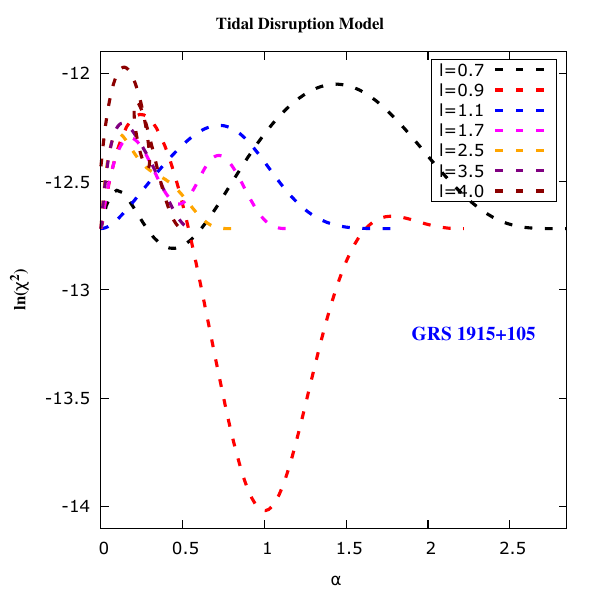}\label{TDMBH3All2}}
		\caption{The above figure shows the variation of ln$\chi^2$ with $\alpha$ for different choices of $l$ when the TDM model is used to explain the HFQPOs of GRS 1915+105.}
        \label{TDMGRS}
\end{figure}
\begin{figure}[t!]
		\centering
		\subfloat[]{\includegraphics[width=0.41\linewidth]{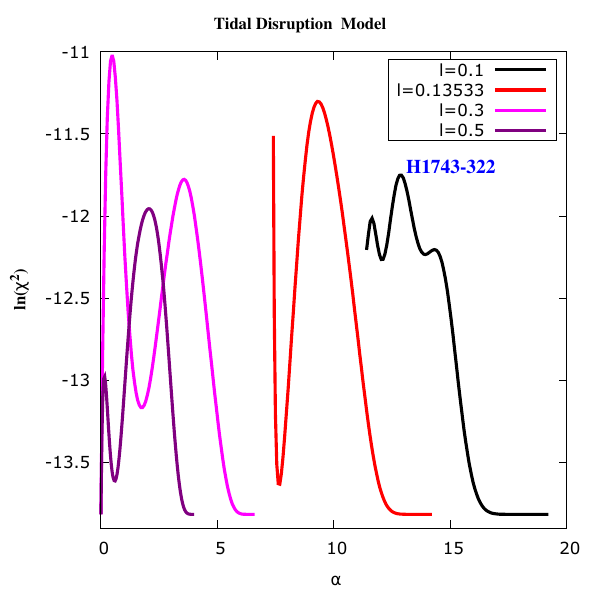}\label{TDMBH4All1}}
		\hfil
		\subfloat[]{\includegraphics[width=0.41\linewidth]{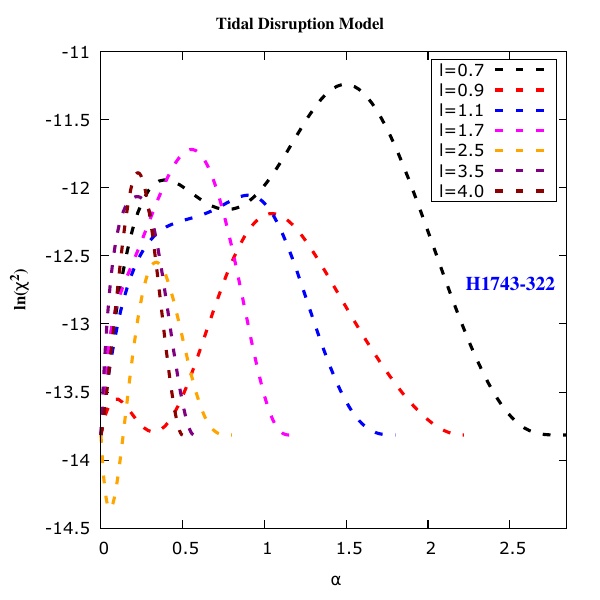}\label{TDMBH4All2}}
		\caption{The above figure shows the variation of ln$\chi^2$ with $\alpha$ for different choices of $l$ when the TDM model is used to explain the HFQPOs of H 1743+322}
        \label{TDMH}
\end{figure}

When we compare the observed QPO frequencies of GRO J1655-40 with the TDM model we note that the $\chi^2$ values are nearly zero irrespective of whether we consider the Kerr case or the hairy Kerr scenario.
From \autoref{TDMBH1All1}, it can be seen that for $l = 0.1$ and $l = 0.13533$, the behaviour of the error shows a 
non-monotonic trend: as $\alpha$ increases, the error initially declines, but beyond a particular value of $\alpha$ -- around $\alpha = 10$ for $l = 0.13533$ and $\alpha = 13.9$ for $l = 0.1$ -- the error reverses its trend, rises again, and eventually approaches the value corresponding to the Kerr limit ($\alpha = 0$). For the other choices of $l$ that appear in \autoref{TDMBH1All1} and \autoref{TDMBH1All2}, the situation is different. 
In these cases, the errors at $\alpha = 0$ and at $\alpha = \alpha_{\max}$ are practically indistinguishable (since both correspond to the Kerr scenario), and the intermediate values of $\alpha$ produce only minor deviations. The same can be observed for higher values of $l$ in \ref{TDMBH1All2}. 
Consequently, imposing $1\sigma$, $2\sigma$, or $3\sigma$ bounds does not meaningfully restrict the parameter space, as the error curves barely respond to changes 
in $\alpha$ or $l$. Similar conclusions can also be achieved for other BH sources like 
XTE J1550-564, GRS 1915+105, and H 1743-322, which is evident from \autoref{TDMXTE}, \autoref{TDMGRS} and \autoref{TDMH}.\\

Therefore, the TDM framework does not provide any robust constraints on the hairy parameters for the BH sources GRO J1655-40, XTE J1550-564, GRS 1915+105, and H 1743-322, indicating that the QPO data associated with these BHs cannot distinguish between the Kerr and the hairy Kerr scenario. 
\vspace{1cm}\\
\\
\textbf{2. \underline{Sgr A*}}
\\
\begin{figure}[t!]
		\centering
		\subfloat[]{\includegraphics[width=0.41\linewidth]{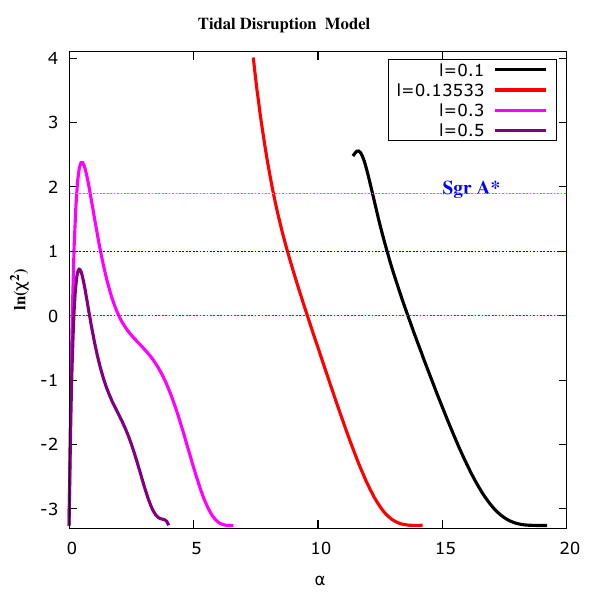}\label{TDMBH5All1}}
		\hfil
		\subfloat[]{\includegraphics[width=0.41\linewidth]{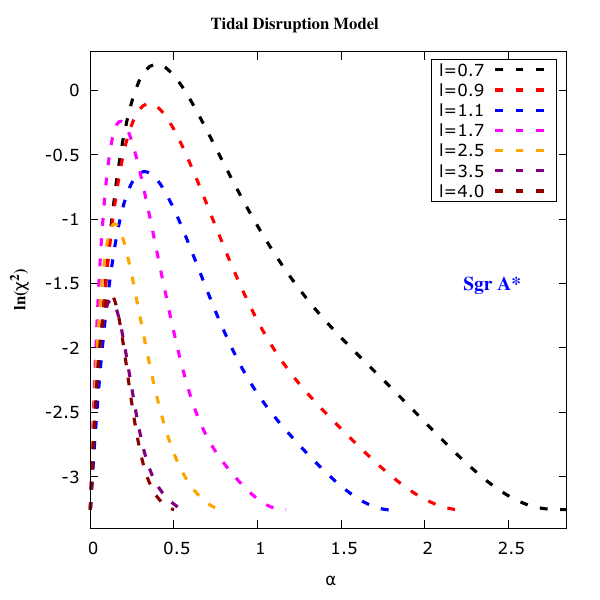}\label{TDMBH5All2}}
		\caption{The above figure shows the variation of ln$\chi^2$ with $\alpha$ for different choices of $l$ when the RPM model is used to explain the HFQPO data of Sgr A*. The red, blue and green dotted lines are associated with the 1-$\sigma$, 2-$\sigma$ and 3-$\sigma$ confidence intervals, which correspond to $\chi^2_{min}+1$, $\chi^2_{min}+2.71$ and $\chi^2_{min}+6.63$ respectively.}
        \label{TDMSgr}
\end{figure}

When TDM is used to fit the HFQPO data of Sgr A*, we note from \ref{TDMBH5All1} that for $l=0.1$ and $l=e^{-2}$ (i.e., $l\leq e^{-2}$), the $\chi^2$ increases substantially near $\alpha_{crit}$. Thus, for $l\sim 0.1$, $\alpha_{crit} \lesssim \alpha \lesssim 15 $ are ruled out outside 1-$\sigma$ while $\alpha_{crit}\lesssim \alpha \lesssim 13 $ are ruled out outside 3-$\sigma$. Similarly, for $l\sim e^{-2}$, $\alpha_{crit} \lesssim \alpha \lesssim 10 $ are ruled out outside 1-$\sigma$ while $\alpha_{crit}\lesssim \alpha \lesssim 8 $ are ruled out outside 3-$\sigma$. For $0.3\lesssim l \lesssim 0.7$, the Kerr scenario is preferred and the intermediate values of $\alpha$ are ruled out (refer \ref{TDMBH5All1}). For higher values of $l$ (see \ref{TDMBH5All2}), for a given $l$, the $\chi^2$ maximizes for intermediate $\alpha$, but the $\chi^2$ values are not high enough, such that all values of $l$ and $\alpha$ are allowed within 1-$\sigma$. Thus, certain parameter space of the hairy Kerr scenario are less preferred compared to the Kerr scenario for Sgr A* within the domain of this model.\\
\\
\textbf{3. \underline{\ XTE J1859+226}}\\
\begin{figure}[hbt!]
		\centering
		\subfloat[]{\includegraphics[width=0.41\linewidth]{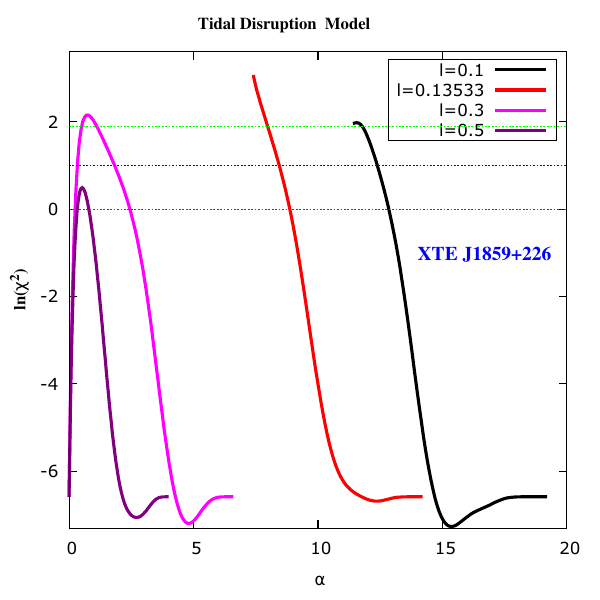}\label{TDMBH6All1}}
		\hfil
		\subfloat[]{\includegraphics[width=0.41\linewidth]{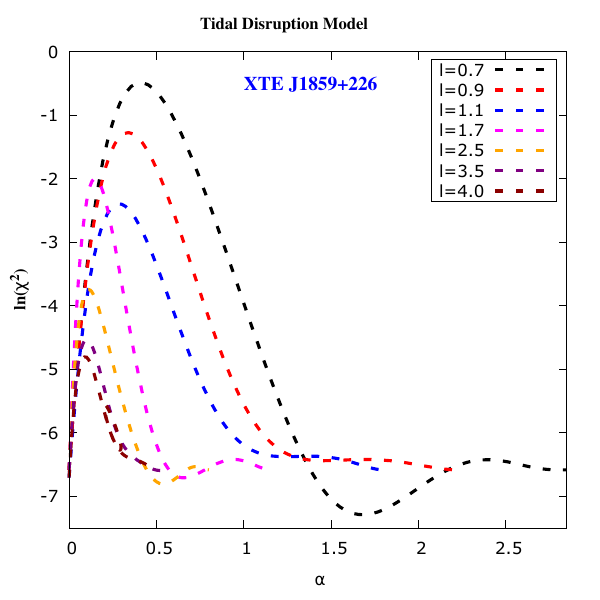}\label{TDMBH6All2}}
		\caption{The above figure shows the variation of ln$\chi^2$ with $\alpha$ for different choices of $l$ when the RPM model is used to explain the HFQPO data of XTE J1859+226. The red, blue and green dotted lines are associated with the 1-$\sigma$, 2-$\sigma$ and 3-$\sigma$ confidence intervals, which correspond to $\chi^2_{min}+1$, $\chi^2_{min}+2.71$ and $\chi^2_{min}+6.63$ respectively.}
         \label{TDMXTE2}
\end{figure}

When the TDM is applied to fit the HFQPO data of XTE J1859+226, we observe from \ref{TDMBH6All1} that for $l = 0.1$ and $l = e^{-2}$ (i.e., $l \leq e^{-2}$ ), the corresponding $\chi^2$ values rise sharply near $\alpha_{\mathrm{crit}}$. Consequently, for $l \sim 0.1$, the range $\alpha_{\mathrm{crit}} \lesssim \alpha \lesssim 14$ is excluded at the 1-$\sigma$ level, while $\alpha_{\mathrm{crit}} \lesssim \alpha \lesssim 12$ is excluded at the 3-$\sigma$ level. Likewise, for $l \sim e^{-2}$, the interval $\alpha_{\mathrm{crit}} \lesssim \alpha \lesssim 10$ is ruled out at 1-$\sigma$, and $\alpha_{\mathrm{crit}} \lesssim \alpha \lesssim 8$ at 3-$\sigma$.
For intermediate values of $l$ ($0.3 \lesssim l \lesssim 0.7$), the Kerr limit is favored, and the corresponding intermediate values of $\alpha$ are disfavored (see \ref{TDMBH6All1}). For larger values of $l$ (\ref{TDMBH6All2}), although the $\chi^2$ tends to peak at intermediate $\alpha$ for a fixed $l$, the peak values are not sufficiently large to exclude any portion of the $l,\alpha$ parameter space at the 1-$\sigma$ level. Overall, within the framework of this model, parts of the hairy Kerr parameter spacetime are less preferred than the Kerr solution for XTE J1859+226.

\subsection{Robustness of our results}
In this section we discuss the sensitivity of our results to different factors, e.g., grid resolution, parameter prior ranges, etc:

\begin{itemize}
    \item Regarding the grid resolution, we mention that in our analysis, we have carefully explored the  entire theoretically allowed parameter space using very fine gridding. Specifically, we consider $0.1 \leq l \leq 4$ (since $l=0$ is excluded as it violates the strong energy condition), with a step size of $\Delta l = 0.2$.  
    We have verified that our conclusions are not sensitive to the chosen upper bound $l = 4$. From the behavior observed in \autoref{RPMGRO}-\autoref{RPMXTE2} and \autoref{TDMGRO}-\autoref{TDMXTE2}, increasing $l$ does not introduce qualitatively new features in $\chi^2$ profiles. In fact, as $l$ increases, the deviation from the Kerr case diminishes, and the corresponding $\chi^2$ values decrease smoothly. Well before $l = 4$, the deviation from the observational constraints falls below the $1\sigma$ level for all sources, indicating that extending the parameter space to larger $l$ does not yield additional physically relevant information.
For each fixed value of $l$, the parameter $\alpha$ is varied within the physically allowed range $\alpha_{crit} \leq \alpha \leq \alpha_{\text{max}} = 2/l$, as dictated by the strong energy conditions (SECs) and the asymptotic flatness condition. We have noted that, if $l>4$, $\alpha_{max}<0.5$, which makes the allowed parameter space of $\alpha$ very restrictive. 
Hence, we restrict our parameter choice between $0.1 \lesssim l \lesssim 4$. As a consequence of this choice, smaller values of $l$ allow for substantially larger values of $\alpha$ (e.g., for $l \sim 0.1$, $\alpha_{max} \sim 20$), which we take into account.  

\item As demonstrated in \autoref{S2}, the parameter space can be naturally divided into three distinct regions: (i) $l < e^{-2}$, (ii) $l = e^{-2}$, and (iii) $l > e^{-2}$. Identifying these three distinct regimes is important, since based on the strong energy conditions (SECs), for a given $l$, not all values of $\alpha$ between $0\lesssim \alpha \lesssim 2/l$ are allowed. 
Only when $\alpha_{crit}<0$ for a given $l$, the entire allowed range of $\alpha$ from 
$0\lesssim \alpha \lesssim 2/l$ is consistent with the SECs and hence can be considered.
Moreover, the SECs further restrict the allowed range of spin for a given $l,\alpha$, i.e. for a given $l$, when $\alpha$ is between $\alpha_{crit}\lesssim \alpha \lesssim \alpha_{coin}$, the maximum allowed BH spin is smaller than $a_{max}$,
which corresponds to the spin of the extremal BH.
These factors are taken into account while choosing our allowed parameter ranges in the grid-search. 
The resolution in $\alpha$ is chosen adaptively: we use $\Delta \alpha = 0.2$ for $0.1 \leq l < 1.7$ and $\Delta \alpha = 0.1$ for $l \geq 1.7$, ensuring adequate sampling of the parameter space, particularly in regions where the allowed range of $\alpha$ decreases.
\item For each $(l,\alpha)$, the $\chi^2$ is calculated for spin in the range $-a_{max}\lesssim a\lesssim a_{max}$/ $-a_{SEC}\lesssim a\lesssim a_{SEC}$ in steps of 0.1. As $a\to a_{max}/a_{SEC}$, the stepsize of $\Delta a$ is further reduced since the radius of the marginally stable circular orbit $r_{ms}$ decreases more rapidly as $a\to a_{max}$, which may lead to a rapid variation of the model dependent QPO frequencies.

\item Apart from $(l,\alpha,a)$ the model dependent QPO frequencies are also sensitive to $M$ and $r_{cm}$. For stellar mass BHs, $\Delta M=0.01M_\odot$ is considered while for supermassive BHs like Sgr A*, $\Delta M=10^{-5}M_\odot$ is taken. On the other hand, $r_{cm}$ is varied between $r_{ms}\lesssim r_{cm}\lesssim r_{ms}+20 R_g$ ($R_g=GM/c^2$ being the gravitational radius) in steps of $\Delta r_{cm}=0.1 R_g$ when $r_{cm}>r_{ms}+5 R_g$ while $\Delta r_{cm}=10^{-3} R_g$ when $r_{ms}\lesssim r_{cm}\lesssim r_{ms}+5 R_g$. Thus, we consider sufficient fine-gridding of our parameter space such that the global minima of $\chi^2$ is not missed. For most BHs, the $\chi^2$ function does not have a sharp minima, i.e., the minimum $\chi^2\sim 0$ is achieved for a broad parameter range,
which is confirmed from the ln$\chi^2$ vs $\alpha$ plots for most sources using the RPM and the TDM in \autoref{S4.2} and \autoref{S4.1}. 

\item Although  the $\chi^2$ landscape is relatively flat across a broad region of the parameter space for the majority of sources, there are exceptions. These are GRO J1655--40 and XTE J1859+226 when tested with the RPM, where a distinct minima appear. These features are independently confirmed through the MCMC analysis, which provides a complementary method to validate our findings (\autoref{cornerplot}).

\item Regarding the choice of priors in the MCMC, we have adopted ranges that are consistent with physically allowed domains of the model parameters as discussed above. We have noted that 
the best-fit values of the model parameters are not sensitive to moderate variations in the prior ranges, although they can influence the shape of the posterior distributions, including the degree of asymmetry observed in some cases. 
Thus, we report that by changing prior ranges of our MCMC moderately or by increasing the grid resolution, our results would not get altered. 

\item However, our results are sensitive to the specific identification of a QPO triplet. 
In our analysis, we find that relatively tight constraints on the model parameters can be obtained only for the sources GRO~J1655$-$40 and XTE~J1859+226 within the framework of the relativistic precession model (RPM). Notably, these two sources are among the few systems where a full QPO triplet have been observed. Since RPM provides a natural interpretation of such triplets in terms of orbital and epicyclic frequencies at a common radius, the availability of three simultaneously identified QPO frequencies significantly enhances the constraining power of the model. Thus, the full potential of the RPM can be realized only when the BHs exhibit QPO triplets in their power spectrum. 
This suggests that the robustness of parameter estimation is closely tied to the identification of a complete QPO triplet. In contrast, for sources where only doublets are observed, the reduced number of observational constraints lead to broader allowed regions in parameter space. This highlights the importance of QPO identification in shaping the inferred properties of the underlying spacetime, and further motivates the need for more precise and comprehensive observations.

\item Furthermore, our results are sensitive to observational uncertainties. The constraints on the model parameters are expected to significantly improve if the error bars in the observed QPO frequencies reduce and if more sources exhibiting a QPO triplet are observed in future.

\end{itemize}

 Before concluding this section we would like to mention that previously Liu et al. ~\cite{Liu:2023ggze} performed a similar analysis and reported constraints on the hairy Kerr parameters. Our analysis differs from that of ~\cite{Liu:2023ggze} on the following grounds:
\begin{itemize}
   \item  We consider both the available kinematic models of QPOs, namely, the Relativistic Precession Model (RPM) and the Tidal Disruption Model (TDM) in our study, while Liu et al. ~\cite{Liu:2023ggze} considers only the Relativistic Precession Model. 
   
    \item In Ref.~\cite{Liu:2023ggze}, the authors adopt relatively restricted priors for the hairy parameters since they have taken $\alpha l \in [0,2)$ and $\alpha \in [0,5]$.
    However, we report that based on the strong energy condition (SEC) $l = 0$ is not allowed which we exclude from our analysis.
    Moreover,  we consider a broader range of $l$
    namely, $0.1\lesssim l \lesssim 4$ since the $\chi^2$ values do not exhibit any notable features for larger values of $l$, e.g. $l>2$. 
    Furthermore, the requirement of asymptotic flatness imposes the condition $l\alpha \lesssim 2$, which introduces a nontrivial correlation between the parameters. As a consequence of this choice, smaller values of $l$ allow for substantially larger values of $\alpha$ (e.g., for $l \sim 0.1$, $\alpha_{max} \sim 20$), which has not been considered in ~\cite{Liu:2023ggze}.

    \item  In this work, we systematically explored the full parameter space, ranging from $\alpha = 0$ up to $l\alpha \lesssim 2$. As demonstrated in \autoref{S2} of the manuscript, this parameter space can be naturally divided into three distinct regions: (i) $l < e^{-2}$, (ii) $l = e^{-2}$, and (iii) $l > e^{-2}$, which we report for the first time. Identifying these three distinct regimes is important, since based on the strong energy conditions (SECs), for a given $l$, not all values of $\alpha$ between $0\lesssim \alpha \lesssim 2/l$ are allowed. Moreover, SEC further restricts the allowed range of spin for a given $l,\alpha$.
    One needs to keep these factors in mind while setting the priors for the MCMC or exploring the parameter space using grid-search method.
    Such comprehensive exploration of the parameter space has not been carried out in 
    \cite{Liu:2023ggze} or in any previous work related to hairy Kerr BHs.
    \item Due to the limited prior range $\alpha \in [0,5]$ adopted in Ref.~\cite{Liu:2023ggze}, a significant portion of the physically allowed parameter space is not explored in that study. As a result, regions that we find to be observationally preferred---such as the sharp minimum around $l \sim 0.1$ and $\alpha \sim 12$ in the case of GRO J1655--40 when tested with the RPM---are not captured in their analysis. For the same reason, for each source, the constraints on the model parameters  reported by us and Liu et al. ~\cite{Liu:2023ggze} significantly differ.
    
    \end{itemize}

\subsection{Model Comparison}
In order to understand the relative importance of RPM or TDM in explaining the observed HFQPO frequencies of the six BH sources, we use the Akaike Information Criteria (AIC) \cite{1100705} defined as:
\begin{align}
    AIC= -2\ln \mathcal{L}_{max} + 2p\,,
    \label{AIC}
\end{align}
where, $\mathcal{L}_{max}$ represents the maximum likelihood and $p$ the number of parameters in the model. The optimal model is the one that yields the lowest $AIC$ value. In general, models with too few parameters fit the data poorly and therefore have lower log-likelihoods, whereas models with too many parameters are penalized by the second term in the $AIC$ expression. Note that $-2\ln \mathcal{L}_{max}$ corresponds to $\chi^2_{min}$ and therefore based on our previous analysis we can compare between RPM and TDM for each of the six BH sources. For both models, the number of parameters is $p=5$, corresponding to   
$l$, $\alpha$, $a$, $M$, and $r_{cm}$. Since both models have the same number of parameters $AIC$ is purely determined by $\chi^2_{min}$.

The $AIC$ of a single model is not meaningful on its own; what matters is the comparison of $AIC$ values across different models. Consequently, the model with the smallest $AIC$ denoted as $AIC_{min}$ is regarded as the best choice while the degree of evidential support for each model, relative to the others, can be determined with $\Delta_i=AIC_i - AIC_{min}$, where $AIC_i$ represents the $AIC$ of the $i^{th}$ model. The model selection criteria corresponds to \cite{Yang:2024mro}:\\\\
1. $0< \Delta_i\leq 2$: The $i^{th}$ model and the most optimal model receives the same level of support.\\
2. $4< \Delta_i\leq 10$: The $i^{th}$ model has significantly less support compared to the optimal model. \\
3. $\Delta_i> 10$: Model $i$ is essentially unsupported by the observational data.\\

 We now compare the performance of the RPM and the TDM in explaining the QPO data of each of the six BH sources, based on AIC and also their consistency with previous spin estimates from other independent methods (apart from QPO based data). 
Since the underlying physical mechanism responsible for QPOs is not yet conclusively established, it remains unclear which model  provides the most appropriate description for a given source. For this reason, we consider multiple models and identify the suitable models based on the best-fit model parameters.
In this regard it is important to note that the previous spin estimates of these sources were obtained assuming the sources to be Kerr BHs. Thus, when we derive the most favored spin for each of these sources assuming the RPM and the TDM we consider the $a_{min}$ corresponding to the $\alpha\simeq 0$ case, such that the comparison with earlier estimates is meaningful. 

From previous discussion we note that for GRO J1655-40, the magnitude of $\chi_{min}^2 \sim 0$ for both the RPM and the TDM model. This becomes more explicit from the MCMC corner plot (\autoref{cornerplot}). Thus, $AIC_{\rm TDM}=AIC_{\rm RPM}\sim 10$ and based on AIC both models are equally preferred by the source GRO J1655-40. 
 We next compare the spin of GRO J1655-40 predicted by the RPM and the TDM with that of previous independent spin estimates of the source. Assuming the Kerr geometry, we report that the spin of GRO J1655-40 predicted from the RPM is $a_{min}\sim 0.3$ and from the TDM, $a_{min}\sim 0.1$. None of these estimates
 are consistent with the spin of this source determined previously by other independent methods, e.g. the Continuum-Fitting method predicts $ a \sim 0.65-0.75$ \cite{Shafee_2005} while the Fe-line method predicts $a \sim 0.94-0.98$ \cite{Miller:2009cw}. If at least one of the previous estimates is correct, then neither RPM nor TDM is suitable for this source.
 Therefore, one needs to analyze other available HFQPO models to check the consistency with earlier estimates, which we leave for a future work. In particular, if a QPO model predicts a spin in agreement with at least one of the previous estimates (assuming Kerr geometry), then we consider that model to be observationally preferred. 
 It is interesting to note however, that the two previous spin estimates of GRO J1655-40  \cite{Shafee_2005, Miller:2009cw} assuming the Kerr scenario are also not in mutual agreement. In this context, it may be worthwhile to recall that the RPM model strongly favors the hairy Kerr scenario with $l\sim 0.1, \alpha\sim 12$ and rules out the Kerr model outside 3-$\sigma$ (although this cannot be considered as a confirmatory signature of the hairy Kerr spacetime as the correct QPO model for GRO J1655-40 is not yet known), while using the TDM the Kerr and the hairy Kerr scenario are equally favored for this source. 
 This when coupled to the largely discrepant previous spin estimates might motivate one to revisit the spin estimates of GRO J1655-40 using the Continuum-Fitting and the Fe-line method assuming a non-Kerr background which might yield consistent values of spin and the hair parameter(s) from different methods.

For the source XTE J1550-564, \autoref{XTERPM} and \autoref{TDMXTE} indicate that for both the RPM and the TDM $\chi^2_{min} \sim 0$ and hence $AIC\sim 10$. Thus, based on AIC, both models seem to be equally favored by the HFQPO data for this source. Assuming the Kerr background, the observationally favored spin of this source corresponds to $a\sim 0.4$ (based on RPM) and $a\sim 0.2$ (based on TDM). Independent approaches—namely the Continuum-Fitting method, the Fe-line analysis and the jet power—have earlier yielded spin estimates of $a \sim 0.34$ (with an error bar of $-0.11<a<0.71$), 
$a = 0.55^{+0.15}_{-0.22}$ \cite{Steiner:2010bt}, and $0.3\lesssim a \lesssim 0.6$ \cite{Banerjee:2020ubc} respectively. Thus, the RPM predicts a spin
in better agreement with the previous estimates while the spin predicted by the TDM is in agreement only with the Continuum-Fitting method when the error bars are considered (which is however very large). Thus, RPM seems to be a better model for this source compared to TDM.

For the source GRS 1915+105, from \autoref{GRSRPM} and \autoref{TDMGRS} we note that $\chi^2_{min} \sim 0$ and hence $AIC\sim 10$ for both the RPM and the TDM, indicating that both models are equally favored.  Considering the Kerr background, the spin of GRS 1915+105 reported previously using different independent methods exhibit substantial variation. For example, the Fe-line analysis indicates a range of $a \sim 0.6-0.98$ \cite{Blum:2009ez}, while the Continuum-Fitting method yields both intermediate values near $a \sim 0.7$ \cite{Middleton:2006kj} and maximal spin at $a \sim 0.98$ \cite{McClintock:2006xd}. More recent studies, which incorporate updated measurements of the mass and inclination of GRS 1915+105, place the spin in the interval $a \sim 0.4-0.98$ \cite{Mills:2021dxs}. Spin values inferred from jet-power observations further suggest a range of $a \sim 0.6-0.9$ \cite{Banerjee:2020ubc}. For $\alpha\sim 0$ in the hairy Kerr model, both RPM and TDM predict $|a|\lesssim 0.1$, which is not in agreement with any of the previous estimates based on independent methods. Thus, if at least one of the previous spin estimates is correct, then neither RPM nor TDM is suitable for this source.
This again requires one to test other available HFQPO models. Also, assuming both the models the HFQPO data cannot distinguish between the Kerr and the hairy Kerr scenario (\autoref{GRSRPM} and \autoref{TDMGRS}).
This, when coupled with the huge disparity in the previous spin estimates may require revisiting the previous spin estimates of  GRS 1915+105 by the Continuum-Fitting or the Fe-line method assuming a non-Kerr background.

For the source H1743-322 when we compare \autoref{HRPM} and \autoref{TDMH}, we note that $\chi^2_{min} \sim 0$ and $\chi^2$ is insensitive to the variation in $l,\alpha$ such that the data fails to differentiate between the Kerr and the hairy Kerr scenario.
For this source, assuming the Kerr geometry, RPM predicts $a_{min}\sim 0.6$ while TDM predicts $a_{min}\sim 0.1$. 
Using the Continuum-Fitting method, the spin of this source has been inferred to be $a = 0.2 \pm 0.3$ at the 68\% confidence level, with a corresponding upper bound of $a < 0.92$ at 99.7\% confidence \cite{Steiner:2011kd}. An independent estimate derived from the observed jet power constrains the spin to lie in the range $0.25 \lesssim a \lesssim 0.5$ \cite{Banerjee:2020ubc}, in reasonable agreement with the Continuum-Fitting measurements reported in \cite{Steiner:2011kd}. Since the previous spin estimates are more or less consistent and the spin predicted by the TDM is in better agreement with \cite{Steiner:2011kd}, TDM seems to provide a better explanation of the HFQPO data of 
H1743-322 compared to RPM.

Although the $\chi^2$ exhibits substantial enhancement for smaller $l$ and intermediate $\alpha$, $\chi^2_{\min}\sim 0$ for the source Sgr A* assuming both RPM and TDM (\autoref{RPMSgr} and \autoref{TDMSgr}). Thus both models predict $AIC\sim 10$
and hence seem to be equally favored. Earlier determinations of the spin of this source display considerable disparity, with some analyses supporting a nearly vanishing spin $\rm a \lesssim 0.1 $ \cite{Fragione:2020khu}, others indicating an intermediate to low value ($\rm a\sim 0.52 $, $ \rm a\sim 0.5$, $ \rm a\sim 0.22$) \cite{Genzel:2003as,Shcherbakov:2010ki,Belanger:2006gm}, and still others favoring a high to near-extremal spin ($\rm a= 0.9\pm 0.06$, $\rm a\sim 0.9$, $\rm 0.4 \lesssim a \lesssim 1$) \cite{Daly:2023axh,Moscibrodzka:2009gw,Meyer:2006fd}. These spins were determined assuming Sgr A* to be a Kerr BH. Here we assumed Sgr A* to be a hairy Kerr BH and based on RPM and TDM, the most favorable spins (assuming $\alpha\sim 0$) are $a_{min}\sim 0.999$ for both the models, roughly in agreement with \cite{Meyer:2006fd,Daly:2023axh}. Thus, the present data cannot distinguish between the RPM and the TDM, but the huge disparity in the earlier spin estimates possibly requires revisiting the spin estimates of this source by earlier independent methods assuming a non-Kerr geometry, e.g. the hairy Kerr scenario considered here. 

Finally, when we compare \autoref{RPMXTE2} and \autoref{TDMXTE2} for XTE J1859+226, we note that $\chi^2_{min}\sim 0$ for both the RPM and the TDM, but for RPM $\chi^2$ varies very sharply near the minima. This becomes more explicit from the MCMC analysis, which yields $\chi^2_{min}\sim 0$ for this source assuming RPM (see Sec. \ref{MCMC}). Thus, assuming the RPM this source strongly favors the hairy Kerr scenario with $l\sim 0.1$ and $\alpha\sim 12$ and rules out the Kerr scenario outside 3-$\sigma$. However, this result is model dependent and not a conclusive signature of the hairy Kerr scenario, since the correct QPO model for each of these sources is not yet known.
Moreover, since $\chi^2_{min}\sim 0$ for both the RPM and the TDM, both models are equally favored, based on $AIC$.
The spin of this object has been estimated using the relativistic precession model (RPM), yielding a low value of $ a = 0.149 \pm 0.005 $ \cite{Motta:2022rku}. This estimate is based on QPO data similar to the present analysis, and is consistent with the spin obtained in our RPM-based study. In contrast, spin measurements derived from other independent methods suggest a near-maximal value ($a = 0.987 \pm 0.003$) \cite{Mall:2023nab}. 
The spin of XTE J1859+226 predicted by the TDM in the present analysis assuming the Kerr geometry corresponds to $a_{min}\sim 0.9$ in better agreement with \cite{Mall:2023nab}. Thus, TDM seems to be a more plausible model when compared to RPM in order to explain the HFQPO data of XTE J1859+226. But, one needs to test the validity of the other HFQPO models for this source, which we leave for a future work.

 Here, we would like to mention that the small number of BH sources exhibiting HFQPOs is an intrinsic limitation of current observations rather than a shortcoming of the present analysis. As discussed in standard reviews \cite{Remillard:2006fc,INGRAM2019101524}, HFQPOs are rare and typically only two high-frequency QPO peaks are detected for a given black hole source. Thus, we do not have the provision to aim for large-sample statistical inference, but rather be limited to testing the consistency of theoretical models with the observed QPO frequencies on a source-by-source basis. Indeed, we encounter degeneracies and in most cases we cannot provide definitive parameter estimations.
Interestingly, even in presence of such degeneracies, our analyis allows us to rule out certain theoretically allowed portions of the parameter space for each source (based on agiven model), which we report in \autoref{S4.1} and \autoref{S4.2}. Whenever, the source exhibits a LFQPO and the model has the provision to include a third QPO frequency, the constraints are much more tighter (e.g. GRO J1655-40 \& XTE J1859+226 using the RPM model). Thus, similar approaches have been adopted widely in the literature \cite{Motta:2013wgae,Awal:2025dxs,Hazarika:2025axz,Ghorani:2024ufk,Jumaniyozov:2025wcs,Wu:2025ccc,Allahyari:2021bsq,Jiang:2021ajk,Bambi:2012pa,Bambi:2013fea,Dasgupta:2025qwq}.
We also emphasize that our conclusions are intentionally cautious. In several cases, we explicitly find that the available data do not allow us to distinguish between the Kerr and the hairy Kerr scenarios, reflecting the limited constraining power of current HFQPO based studies. Therefore, our results should be interpreted as preliminary constraints on the parameter space rather than statistically definitive conclusions.
We expect the constraints to improve with the availability of more precise data.

 Before concluding this section we  present the key results of our analysis in tabular form, on a source by source basis.

\begin{table}[H]
\centering
\setlength{\tabcolsep}{5pt}               
\renewcommand{\arraystretch}{1.6}         
\footnotesize
\begin{adjustbox}{max width=\textwidth}
\begin{tabular}{|l|l|l|l|l|l|l|l|}
\hline
\multicolumn{8}{|c|}{\textbf{GRO J1655-40}} \\ \hline
\multirow{2}{*}{\textbf{Previous constrains}}     & \multicolumn{3}{c|}{\textbf{Spin}} & \multicolumn{4}{c|}{\textbf{Mass $(M_{\odot})$} }\\  \cline{2-8} 
& {$\rm a\sim 0.65-0.75 $ \cite{Shafee_2005}}  & {$\rm a\sim 0.94-0.98$ \cite{Miller:2009cw}} & {$ \rm a=0.29\pm 0.003  $ \cite{Motta:2013wgae}}   &  \multicolumn{4}{|c|}{\textbf{$\rm 5.4\pm 0.3$} \cite{Beer:2001cg} } \\

   \hline \hline                 
\multirow{1}{*}{\textbf{Model}} & \multicolumn{7}{c|}{\textbf{}} \\ \cline{2-8} 
& \textbf{ $\mathbf{Best-fit ~hairy ~parameters}$} & \textbf{ $\mathbf{\chi^2_{min}}\sim$} & AIC &\textbf{Whether Kerr allowed within 1-$\sigma$} & \textbf{Whether Kerr allowed within 3-$\sigma$} & \textbf{$a_{min}$ ~($\alpha\sim 0$)} & \textbf{$M_{min}$ ~($\alpha\sim 0$)} \\\hline
RPM   & $l \sim 0.1$, $\alpha \sim 12$    &  0.0146 & 10 & No & No & $\sim 0.3$ &  5.12\\ \hline
TDM   & -  & 0   & 10  & Yes & Yes & $\sim 0.1$ & 5.32\\ 
\hline
\end{tabular}
\end{adjustbox}
\caption{Table summarizing the best-fit hairy parameters for GRO J1655-40 and comparing the spin derived from QPO related studies with previous independent estimates.
}
\label{Tab2}
\end{table}
\begin{table}[H]
\centering
\setlength{\tabcolsep}{5pt}               
\renewcommand{\arraystretch}{1.6}         
\footnotesize
\begin{adjustbox}{max width=\textwidth}
\begin{tabular}{|l|l|l|l|l|l|l|l|}
\hline
\multicolumn{8}{|c|}{\textbf{XTE J1550-564}} \\ \hline
\multirow{2}{*}{\textbf{Previous constrains}}     & \multicolumn{3}{c|}{\textbf{Spin}} & \multicolumn{4}{c|}{\textbf{Mass $(M_{\odot})$} }\\  \cline{2-8} 
& {$-0.11<a<0.71 $ \cite{Steiner:2010bt}}  & {$a=0.55^{+0.15}_{-0.22}$ \cite{Steiner:2010bt}} & $0.3\lesssim a \lesssim 0.6$ \cite{Banerjee:2020ubc}  &  \multicolumn{4}{|c|}{\textbf{$\rm 9.1\pm 0.61$} \cite{Orosz:2011ki} } \\

   \hline \hline                 
\multirow{2}{*}{\textbf{Model}} & \multicolumn{7}{c|}{\textbf{}} \\ \cline{2-8} 
& \textbf{ $\mathbf{Best-fit~hairy ~parameters}$} & \textbf{ $\mathbf{\chi^2_{min}}\sim$} & AIC &\textbf{Whether Kerr allowed within 1-$\sigma$} & \textbf{Whether Kerr allowed within 3-$\sigma$} & \textbf{$a_{min}$ ~($\alpha\sim 0$)} & \textbf{$M_{min}$~($\alpha\sim 0$)}\\ \hline
RPM   & -    & 0 & 10 & Yes & Yes & $\sim 0.4$ &   9.29\\ \hline
TDM   & -  & 0   & 10  & Yes & Yes & $\sim 0.2$ &  8.91\\ 
\hline
\end{tabular}
\end{adjustbox}
\caption{Table summarizing the best-fit hairy parameters for XTE J1550-564  and comparing the spin derived from QPO related studies with previous independent estimates.}
\label{Tab3}
\end{table}
\begin{table}[H]
\centering
\setlength{\tabcolsep}{5pt}               
\renewcommand{\arraystretch}{1.6}         
\footnotesize
\begin{adjustbox}{max width=\textwidth}
\begin{tabular}{|l|l|l|l|l|l|l|l|}
\hline
\multicolumn{8}{|c|}{\textbf{GRS 1915+105}} \\ \hline
\multirow{2}{*}{\textbf{Previous constrains}}     & \multicolumn{5}{c|}{\textbf{Spin}} & \multicolumn{2}{c|}{\textbf{Mass $(M_{\odot})$} }\\  \cline{2-8} 
& {$ \rm a\sim 0.98$ \cite{McClintock:2006xd}} &  {$\rm a\sim 0.7$ \cite{Middleton:2006kj}} & {$ \rm a\sim 0.6-0.98  $ \cite{Blum:2009ez}}   &  $a \sim 0.4-0.98$ \cite{Mills:2021dxs} & $a \sim 0.6-0.9$ \cite{Banerjee:2020ubc} & \multicolumn{2}{|c|}{\textbf{$\rm 12.4^{+2.0}_{-1.8}$} \cite{Reid:2014ywa} } \\

   \hline \hline                 
\multirow{2}{*}{\textbf{Model}} & \multicolumn{7}{c|}{\textbf{}} \\ \cline{2-8} 
& \textbf{ $\mathbf{Best-fit ~hairy ~parameters}$} & \textbf{ $\mathbf{\chi^2_{min}}\sim$} & AIC &\textbf{Whether Kerr allowed within 1-$\sigma$} & \textbf{Whether Kerr allowed within 3-$\sigma$} & \textbf{$a_{min}$ ~($\alpha\sim 0$)} & \textbf{$M_{min}$~($\alpha\sim 0$)}\\ \hline
RPM   & -   & 0 & 10 & Yes & Yes & $\sim 0.0$  & 11.00\\ \hline
TDM   & -  & 0   & 10  & Yes & Yes & $\sim -0.1$ & 12.11\\ 
\hline
\end{tabular}
\end{adjustbox}
\caption{Table summarizing the best-fit hairy parameters for GRS 1915+105 and comparing the spin derived from QPO related studies with previous independent estimates.
}
\label{Tab4}
\end{table}
\begin{table}[H]
\centering
\setlength{\tabcolsep}{5pt}               
\renewcommand{\arraystretch}{1.6}         
\footnotesize
\begin{adjustbox}{max width=\textwidth}
\begin{tabular}{|l|l|l|l|l|l|l|l|}
\hline
\multicolumn{8}{|c|}{\textbf{H1743-322}} \\ \hline
\multirow{2}{*}{\textbf{Previous constrains}}     & \multicolumn{2}{c|}{\textbf{Spin}} & \multicolumn{5}{c|}{\textbf{Mass $(M_{\odot})$} }\\  \cline{2-8} 
& {$ a=0.2\pm0.3 $ \cite{Steiner:2011kd}} &  {$0.25\lesssim a\lesssim0.5$ \cite{Banerjee:2020ubc}}   &  \multicolumn{5}{|c|}{\textbf{$8-14.07$} 
\cite{Pei:2016kka,Bhattacharjee:2017rbl,Petri:2008jc} } \\

   \hline \hline                 
\multirow{2}{*}{\textbf{Model}} & \multicolumn{7}{c|}{\textbf{}} \\ \cline{2-8} 
& \textbf{ $\mathbf{Best-fit ~hairy ~parameters}$} & \textbf{ $\mathbf{\chi^2_{min}}\sim$} & AIC &\textbf{Whether Kerr allowed within 1-$\sigma$} & \textbf{Whether Kerr allowed within 3-$\sigma$} & \textbf{$a_{min}$ ~($\alpha\sim 0$)} & \textbf{$M_{min}$~($\alpha\sim 0$)} \\ \hline
RPM   & -   & 0 & 10 & Yes & Yes & $\sim 0.6$ &  13.59\\ \hline
TDM   & -  & 0   & 10  & Yes & Yes &  $\sim 0.1$ & 11.63\\ 
\hline
\end{tabular}
\end{adjustbox}
\caption{Table summarizing the best-fit hairy parameters for H1743-322 and comparing the spin derived from QPO related studies with previous independent estimates.}
\label{Tab5}
\end{table}
\begin{table}[H]
\centering
\setlength{\tabcolsep}{5pt}               
\renewcommand{\arraystretch}{1.6}         
\footnotesize
\begin{adjustbox}{max width=\textwidth}
\begin{tabular}{|l|l|l|l|l|l|l|l|}
\hline
\multicolumn{8}{|c|}{\textbf{Sgr A*}} \\ \hline
\multirow{2}{*}{\textbf{Previous constrains}}     & \multicolumn{6}{c|}{\textbf{Spin}} & \multicolumn{1}{c|}{\textbf{Mass $(M_{\odot})$} }\\  \cline{2-8} 
& { $\rm a= 0.9\pm 0.06$ \cite{Daly:2023axh}}& 
{$ \rm a\sim 0.9$ \cite{Moscibrodzka:2009gw}} &  {$\rm a\sim 0.52$ \cite{Genzel:2003as}} & {$\rm a\sim 0.22$ \cite{Belanger:2006gm}} & {$ a \lesssim 0.1 $ \cite{Fragione:2020khu}}   & $\rm 0.4 \lesssim a \lesssim 1$ \cite{Meyer:2006fd} &
\multicolumn{1}{|c|}{\textbf{$(3.5-4.9)*10^6$} \cite{Ghez:2008ms,Gillessen:2008qv} } \\
    
   \hline \hline                 
\multirow{2}{*}{\textbf{Model}} & \multicolumn{7}{c|}{\textbf{}} \\ \cline{2-8} 
& \textbf{ $\mathbf{Best-fit ~hairy ~parameters}$} & \textbf{ $\mathbf{\chi^2_{min}}\sim$} & AIC &\textbf{Whether Kerr allowed within 1-$\sigma$} & \textbf{Whether Kerr allowed within 3-$\sigma$} & \textbf{$a_{min}$ ~($\alpha\sim 0$)} & \textbf{$M_{min}$ ~($\alpha\sim 0$)}\\ \hline
RPM   & -   & 0 & 10 & Yes & Yes & $\sim 0.999$ & $4.4\times 10^6$ \\ \hline
TDM   & -  & 0   & 10  & Yes & Yes & $\sim 0.999$ & $3.5\times 10^6$\\ 
\hline
\end{tabular}
\end{adjustbox}
\caption{Table summarizing the best-fit hairy parameters for Sgr A* and comparing the spin derived from QPO related studies with previous independent estimates.
}
\label{Tab6}
\end{table}
\begin{table}[H]
\centering
\setlength{\tabcolsep}{5pt}               
\renewcommand{\arraystretch}{1.6}         
\footnotesize
\begin{adjustbox}{max width=\textwidth}
\begin{tabular}{|l|l|l|l|l|l|l|l|}
\hline
\multicolumn{8}{|c|}{\textbf{XTE J1859+226}} \\ \hline
\multirow{2}{*}{\textbf{Previous constrains}}     & \multicolumn{2}{c|}{\textbf{Spin}} & \multicolumn{5}{c|}{\textbf{Mass $(M_{\odot})$} }\\  \cline{2-8} 
& {$ 0.149\pm0.005 $ \cite{Motta:2022rku}} &  {$0.987\pm0.003$ \cite{Mall:2023nab}}   &  \multicolumn{5}{|c|}{\textbf{$\rm 7.85\pm 0.46$} \cite{motta2022black} } \\
\hline \hline                 
\multirow{2}{*}{\textbf{Model}} & \multicolumn{7}{c|}{\textbf{}} \\ \cline{2-8} 
& \textbf{ $\mathbf{Best-fit ~hairy ~parameters}$} & \textbf{ $\mathbf{\chi^2_{min}}\sim$} & AIC &\textbf{Whether Kerr allowed within 1-$\sigma$} & \textbf{Whether Kerr allowed within 3-$\sigma$} & \textbf{$a_{min}$ ~($\alpha\sim 0$)} & \textbf{$M_{min}$ ~($\alpha\sim 0$)}\\ \hline
RPM   &  $l \sim 0.1$, $\alpha \sim 12$   & 0.016 & 10 & No & No & $\sim 0.2$  & 7.39\\ \hline
TDM   & -  & 0   & 10  & Yes & Yes & $\sim 0.9$ & 7.49\\ 
\hline
\end{tabular}
\end{adjustbox}
\caption{Table summarizing the best-fit hairy parameters for XTE J1859+226  and comparing the spin derived from QPO related studies with previous independent estimates.
}
\label{Tab7}
\end{table}

\section{Conclusion}
In this work, we investigate the role of the hairy Kerr scenario in explaining the observed high frequency quasi-periodic oscillations (HFQPOs) in BH sources. Hairy black holes have garnered considerable interest, as they naturally arise in modified gravity theories, effective field frameworks, or within Einstein gravity coupled to additional self-interacting fields. Such BH solutions have been obtained by the method of extended gravitational decoupling \cite{Ovalle:2020kpd} and are described by the deviation parameters $\alpha$ and $l$. The Kerr scenario is recovered either when $\alpha\to 0$ or when $l\alpha \to 2$. The rotating hairy black hole solution considered here satisfies the Einstein field equations with a conserved energy–momentum tensor that obeys the strong energy conditions (SECs). 
We investigate the horizon structure of the hairy Kerr BHs and report that the ($\alpha, l$) parameter space can be classified into 3 domains: (a) $l\geq e^{-2}$, (b) $l=e^{-2}$ and (c) $l\leq e^{-2}$, manifesting certain interesting and unusual features, compared to the standard Kerr scenario. For certain choices of $\alpha$ and $l$ there can be more than two horizons, for a given $l$ the SECs are satisfied only above a critical $\alpha$, and for any $l$, $\alpha$ can go only as high as $\alpha_{max}=2/l$ to ensure asymptotic flatness. As $\alpha\to \alpha_{max}$, the Kerr limit is recovered.
When $l=e^{-2}$ and $\alpha\simeq \alpha_{crit}$, the non-rotating BH is also the extremal BH and the horizon radius corresponds to the inflection point of the function $\Delta(l,\alpha,a, r)$ appearing in the metric. For $l>e^{-2}$ and $\alpha_{crit}\leq \alpha\leq \alpha_{coin}=1/l$ the SEC is satisfied only in the spin range $0\leq |a|\leq a_{SEC}<a_{max}$, while when $\alpha_{coin}<\alpha\leq \alpha_{max}$ all allowed spins satisfy the SEC. Based on the above classification we systematically study the radial variation of the three fundamental frequencies associated with the hairy Kerr spacetime. We also explore the sensitivity of the fundamental frequencies on the the deviation parameters ($\alpha, l$) and also the BH spin and compare them with the Kerr scenario. This is important as we aim to test the viability of this solution in explaining the observed HFQPOs in BH sources, where the model dependent HFQPO frequencies depend on the three fundamental frequencies. This gives us an opportunity to constrain the parameter space of the hairy Kerr BH based on the available HFQPO observations. 
 In this regard, we would like to emphasize that 
from our analysis we try to comment on two issues for each source (a) whether the HFQPO data of each source prefers the Kerr or the hairy Kerr spacetime, and (b) the most favoured HFQPO model for each source.
Several theoretical models are proposed in the literature to explain the observed HFQPOs. 
Since the underlying physical mechanism
responsible for QPOs is not yet conclusively established, it remains unclear which model
provides the most appropriate description for a given source. For this reason, we consider
multiple models and attempt to identify the suitable model(s) for each source.
Here, we consider the viability of the kinematic models of HFQPOs, namely, the Relativistic Precession Model (RPM) and the Tidal Disruption Model (TDM). 

We compare the model dependent QPO frequencies with the observed QPO frequencies of six available BH sources. Using $\chi^2$ analysis, the parameter space of $l$ and $\alpha$ is constrained based on the RPM and the TDM for each of the six BH sources.
For most sources, the HFQPO data with the current level of precision cannot distinguish between the Kerr and the hairy Kerr scenario. Only for GRO J1655-40 and XTE J1859+226, the QPO data exhibits a strong preference towards the hairy Kerr scenario with $l\simeq 0.1$ and $\alpha\simeq 12$, ruling out the Kerr scenario at 3-$\sigma$, when the RPM model is used. This result is also supported from the MCMC analyses (\autoref{MCMC}). 
The observation of a sharp minima of $\chi^2$ for the two aforesaid sources may be attributed to an identification of a QPO triplet for these two sources and the provision to explain a LFQPO (in addition to the two HFQPOs) by the RPM model. When QPO triplets are observed the constraining power of models like the RPM is substantially enhanced. 
Thus, this result is model dependent and is not yet a definitive signature of the hairy Kerr spacetime. This is because the correct QPO model for each of the sources is unknown and hence we do not know for sure whether RPM is the most suitable model for GRO J1655-40 and XTE J1859+226.

In order to comment on the most favoured HFQPO model for each source, we compare the Akaike Information Criterion (AIC) associated with the RPM and the TDM for each source. If the difference between the AICs denoted by $\Delta_i>4$, then the model with lower AIC gets more observational support. 
However, we find that we cannot distinguish between the performance of the RPM and the TDM for each source based on AIC, as both have nearly the same magnitude $\sim10$. Thus, we attempt to distinguish between the two models based on the consistency of the spin predicted from the present study with that of previous independent estimates. It is worth noting that the spin estimates of most sources derived from independent methods, such as the Continuum-Fitting and the Fe-line spectroscopy, often exhibit noticeable discrepancies. In this regard, we mention that the previous spin estimates of these sources were obtained assuming the sources to be Kerr BHs. 
Therefore, when we derive the most favored spin for each of these sources assuming the RPM and the TDM we consider the $a_{min}$ corresponding to the $\alpha\simeq 0$ case, such that the comparison with earlier estimates is meaningful.
In \autoref{Tab2}-\autoref{Tab7} we report the spin of the six BH sources obtained from the present study and also the previous independent estimates. We also report the $\chi^2_{min}$ and the AIC in each of these tables.
For most BHs, $\chi^2_{min}\sim 0$ is achieved over a broad parameter range, hence we refrain from reporting the best-fit hairy parameters in such cases.

We assume that for a given source at least one of the previous estimate is correct. The model which predicts a spin for a given source in agreement with at least one of the previous estimate is considered to be a preferred model for that source. Since, the previous spin estimates are not consistent, then there can be multiple suitable models for a given source, which the present data cannot distinguish.
However, if a model predicted spin is not in agreement with any of the previous spin estimates, then that model is considered to be unsuitable for that source.
Based on this argument, for XTE J1550-564, the RPM
predicts a spin in better agreement with the previous estimates while the spin predicted by the TDM is in
agreement only with the Continuum-Fitting method when the error bars are considered (which is however pretty large). Thus, the RPM seems to
be a better model for this source when compared to the TDM. For H 1743+322, on the other hand the TDM seems to be a better model. The HFQPO data for Sgr A* cannot distinguish between the RPM and the TDM while for XTE J1859+226,  the TDM seems to be a more plausible model.
 For the sources GRO J1655-40 and GRS 1915+105, the predicted spins from both the RPM and the TDM are much lower than all of the earlier independent estimates. Therefore, both models seem to be unsuitable for both these sources.
 Therefore, we need to analyze and compare other available HFQPO models to understand which is the most suitable model for each of these sources. We leave this for a future work which will be reported soon. 

Alternatively, the fact that the previous spin estimates are often mutually inconsistent may raise a question on the Kerr paradigm. 
This,
when coupled with the fact that the data often fails to distinguish between the Kerr and the hairy Kerr scenario and often exhibits a preference towards an alternative Kerr scenario (e.g. the hairy Kerr spacetime here) when QPO triplets are observed, might
indicate revisiting the previous spin measurements of these sources by the Continuum-Fitting or the Fe-line
method assuming a non-Kerr background.

We emphasize that our conclusions are intentionally cautious. In several cases, we explicitly find that the available data do not allow us to distinguish between the Kerr and the hairy Kerr scenarios, reflecting the limited constraining power of current HFQPO based studies. This however, is an intrinsic limitation of current observations rather than a shortcoming of the present analysis. 
Therefore, our results should be interpreted as preliminary constraints on the parameter space rather than statistically definitive conclusions.
The present analysis is limited by poor statistics, as only a few black hole sources exhibit HFQPOs, and by the lack of precise data. These limitations are expected to improve significantly with the launch of the ESA X-ray mission LOFT, enabling much tighter constraints on black hole spin and additional hairs from QPO based observations.


\section*{Acknowledgments}
Research of I.B. is funded by the Start-Up
Research Grant from SERB, DST, Government of India
(Reg. No. SRG/2021/000418). The work of S. M. is supported by the core research
grant from the Science and Engineering Research Board,
a statutory body under the Department of Science and
Technology, Government of India, under Grant
Agreement No. CRG/2023/007670.\\

\label{S6}

\bibliography{mybib.bib,Re(1),QPO,Re,new-ref,revision,revision2}
\bibliographystyle{unsrt}

\end{document}